\documentclass[fleqn,10pt]{wlscirep}
\usepackage[utf8]{inputenc}
\usepackage[T1]{fontenc}
\usepackage{subfigure}
\usepackage{caption}
\usepackage{float}
\usepackage{multirow}
\usepackage{multicol}
\usepackage{rotating}
\usepackage{lineno}
\usepackage{ragged2e}
\usepackage{xcolor}
\definecolor{fullred}{rgb}{1 0 0}
\usepackage{bm}
\usepackage{soul}

\newenvironment{suppfigure}
{\begin{figure}}
{\end{figure}}

\newenvironment{supptable}
{\begin{table}}
{\end{table}}

\title{Interpretable Material Spatial Intelligence for Discovery of Governing Microstructural Features}

\author[1,*]{Mathieu Calvat}
\author[2]{Gregory Sparks}
\author[1]{Dhruv Anjaria}
\author[1]{Chris Bean}
\author[3]{Haoren Wang}
\author[4]{Paul Gradl}
\author[5]{Timothy M. Smith}
\author[6]{Allison M. Beese}
\author[4]{Gabriel Demeneghi}
\author[3]{Kenneth Vecchio}
\author[7,8]{Morad Behandish}
\author[1,*]{J.C. Stinville}

\affil[1]{Materials Science and Engineering Department, University of Illinois Urbana-Champaign, Urbana, IL, USA}
\affil[2]{University of Dayton Research Institute, Dayton, OH, USA}
\affil[3]{Department of NanoEngineering, University of California San Diego, San Diego La Jolla, CA, USA}
\affil[4]{NASA Marshall Space Flight Center, Huntsville, AL, USA}
\affil[5]{NASA Glenn Research Center, Cleveland, OH, USA}
\affil[6]{Department of Materials Science and Engineering, Pennsylvania State University, University Park, PA, USA}
\affil[7]{School of Mechanical, Aerospace, and Manufacturing Engineering, University of Connecticut, Storrs, CT, USA}
\affil[8]{AIMETRA, San Mateo, CA, USA}

\affil[*]{e-mail: mcalvat@illinois.edu; jcstinv@illinois.edu}

\begin{abstract}

    Many material systems exhibit complex spatial and temporal interactions across multiple length scales and modalities that govern macroscopic behavior. Although Machine Learning (ML) is widely used in materials science to predict this behavior, most approaches still rely on handcrafted descriptors or aggregated representations that overlook spatial organization, limiting insight into governing mechanisms. We introduce Materials Spatial Intelligence (MSI), a framework inspired by spatial intelligence that learns directly from multimodal spatial observations of material systems. MSI encodes high-resolution microstructural and deformation data into shared latent representations that preserve spatial relationships while supporting property prediction, interpretation, and optimization. By combining multimodal representation learning, MSI identifies the key features governing mechanical behavior and property trade-offs in structural alloys. Beyond prediction, MSI enables feature-driven microstructure optimization and mechanism discovery. More broadly, MSI establishes a foundation for applying spatial intelligence to materials science, leveraging interpretable ML systems to accelerate scientific discovery and materiel design.

\end{abstract}

\begin{document}

\flushbottom
\maketitle

\thispagestyle{empty}

\justify Material properties, particularly those of metallic materials, are governed by their (micro)structure across multiple length scales, from the atomic to the macroscopic scale, and by the associated spatial heterogeneity \cite{Olson1997}. The structure cannot be described through a single data modality because it comprises complex combinations of coupled chemical, crystallographic, phase, and defect heterogeneity. Material structure and local behavior also evolve continuously under external stimuli such as mechanical deformation, thermal exposure, irradiation, and corrosion. Materials therefore constitute hierarchical spatial and temporal physical systems whose behavior emerges from interactions among constituent features and their evolution over stimuli. Modern AI tools are used extensively in material science; for example, Large Language Models (LLMs) have demonstrated impressive capabilities in scientific knowledge extraction, information retrieval and specific reasoning tasks \cite{Miret2025,Tang2026,Ahlawat2026}, meanwhile, diffusion models have attracted considerable interest due to their capacity to generate realistic synthetic data based on contextual information \cite{Bastek2023,Igashov2024,Li2024}. However, they remain limited in understanding and reasoning over physical systems governed by evolving spatial and temporal interactions. Recently, spatial intelligence and world-model approaches \cite{Hafner2025,Xie2021,Liu2022} have emerged as powerful frameworks for learning, representing, and reasoning over spatial and temporal information \cite{Wang2026}. The key challenge lies not only in defining the optimal ML approach, but also in identifying the relevant data modalities that are predictive of the desired material behavior and macroscopic properties. By explicitly preserving spatial relationships throughout prediction, such frameworks also offer unique opportunities for interpretability \cite{Rudin2019,Karniadakis2021}, enabling identification of the mechanisms governing material behavior and property evolution.

\justify Here, we introduce a Materials Spatial Intelligence (MSI) framework for discovering microstructural features governing mechanical properties. Using metallic materials as a model system, we demonstrate how MSI learns directly from multimodal spatial measurements to identify governing microstructural features and predict material performance. Beyond prediction, MSI enables feature-driven microstructure optimization and establishes a new paradigm for microstructure-based materials research, transforming how microstructure--property relationships are discovered, interpreted, and engineered.

\justify Mechanical property trade-offs in structural alloys, such as strength--ductility \cite{Wei2014} and fatigue ratio (efficiency)--strength \cite{Pang2013,Stinville2022}, remain a major challenge in designing materials for complex thermomechanical loading conditions \cite{Charkaluk2002,Cassagne2025}. Significant progress has been made to link microstructure to mechanical behavior through experiments and physically informed models, providing key insights into deformation mechanisms and the role of microstructural features \cite{Prouteau2022,Anjaria2024,Texier2024,Yvinec2026}. However, these approaches are time-consuming and typically focus on a single loading condition, microstructural state, or alloy, making it difficult to systematically identify features governing behavior across alloy systems and novel materials \cite{GIANOLA2023101090}. This limitation is further amplified by advanced processing methods, including additive manufacturing, powder metallurgy, and automated synthesis \cite{Salzbrenner2017}, which generate increasingly complex microstructures and alloy systems at an accelerating pace \cite{Gu2021,Ren2022,Gao2023,Zhu2026}.

\justify Conventional approaches typically analyze microstructure and deformation separately or through localized correlations, limiting their ability to capture collective behavior and its relationship to mechanical properties. Conversely, data-driven approaches often rely on averaged or simplified microstructural descriptors \cite{Borg2020,Lee2021,Hu2022,Deng2022,Gu2024,Sohail2025}, neglecting the local states, deformation heterogeneity, and interactions that govern macroscopic behavior. Bridging local microstructure, deformation, and mechanical behavior therefore requires a paradigm shift: materials must be represented as spatially organized systems in which macroscopic properties emerge from the distribution and interaction of local microstructural and deformation states.

\justify Recent advances in high-fidelity experimental techniques are transforming how microstructure--property relationships can be learned in structural alloys. High-resolution, large-area characterization methods \cite{Black2023,Deronja2024}, combined with rapid data acquisition, now enable spatially resolved measurements of microstructural and deformation states at unprecedented throughput. These developments allow direct learning from microstructural heterogeneity, including grain-scale features, defect distributions, and their influence on deformation across multiple length scales. Combined with advances in spatial intelligence \cite{Xie2021,Liu2022,Calvat2026}, namely frameworks designed to learn and exploit spatial relationships in structured data, such datasets can establish quantitative links between microstructural features, deformation states, and macroscopic behavior.

\begin{figure}
    \centering
    \includegraphics[width=1\linewidth]{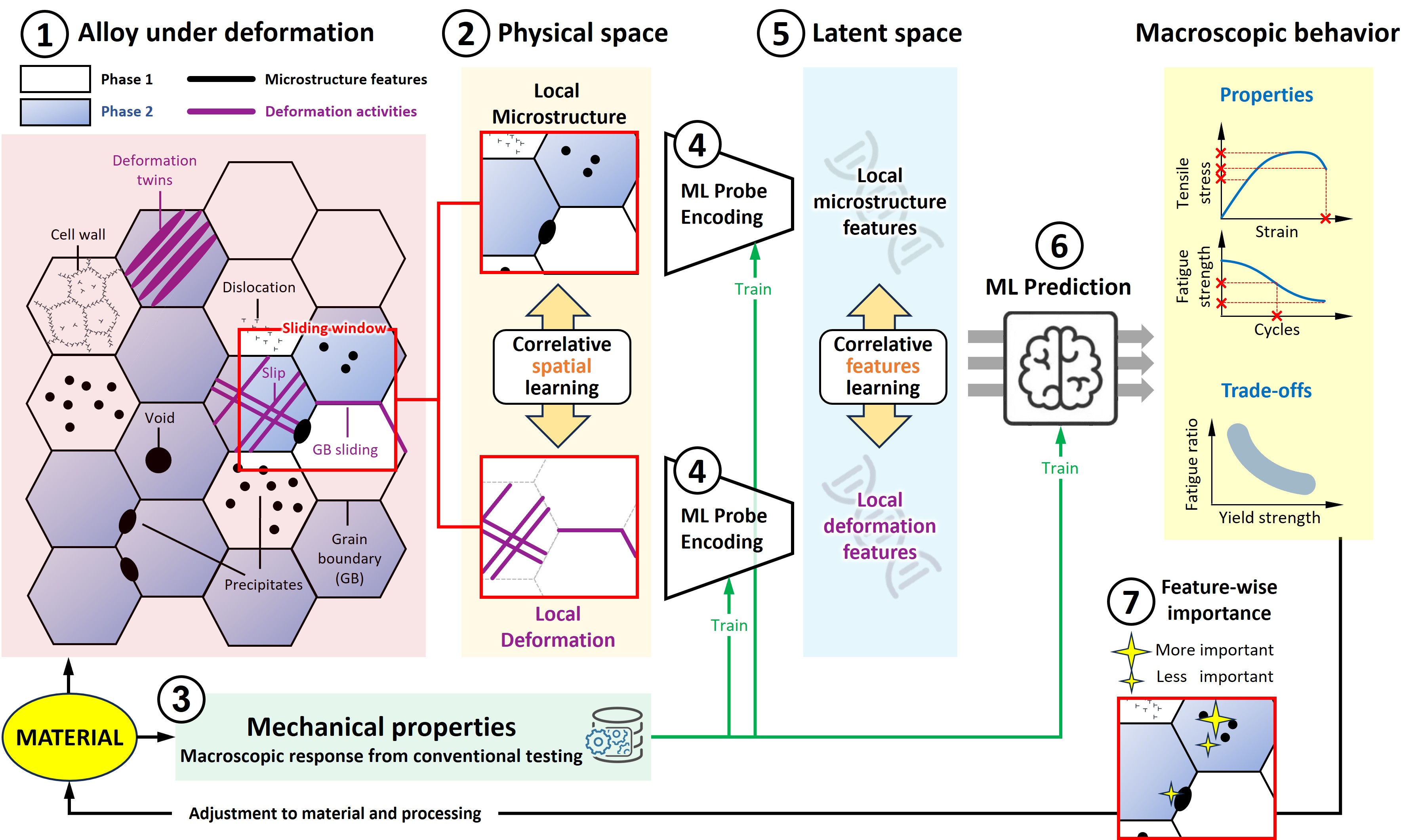}
    \caption{
        \textbf{Material spatial intelligence (MSI) for identification of microstructural features governing macroscopic properties.} 
        (1) Alloy macroscopic mechanical properties are governed by the local microstructural state and deformation response under external loading, as well as their \emph{spatial heterogeneity}. 
        (2) These local states, their associated deformation responses, and their spatial heterogeneity can be captured using scanning probe electron microscopy techniques. 
        (3) Macroscopic properties, which emerge from the collective response of an ensemble of local states, can be captured by conventional testing to train ML-based predictive models. 
        (4) ML-based encoding can be performed on the ensemble of a reduced region of the microstructural and deformation states to enable feature learning within latent spaces. We refer to this as ML-based probe encoding and it is performed with the knowledge of macroscopic properties to identify relevant microstructural and deformation features. 
        (5) We generate latent spaces that describe the essential features of material microstructure and deformation states. 
        (6) Such datasets enable training ML models to predict mechanical properties from local states. 
        (7) We leverage microstructure feature-wise importance to achieve a data-driven identification of the fundamental features that control mechanical properties. 
    }
    \label{Figure1}
\end{figure}

\justify We schematically illustrate the conceptual foundation of the proposed MSI framework in Figure \ref{Figure1}. MSI represents materials as spatially resolved systems in which microstructure and deformation are described as fields (\textit{i.e.}, maps), capturing their heterogeneity (Fig. \ref{Figure1}(1)). Each material is characterized by an ensemble of local microstructural states and their associated local deformation responses during loading (Fig. \ref{Figure1}(2)). These local states are linked to macroscopic mechanical properties (Fig. \ref{Figure1}(3)) for model training. Unlike conventional data-driven approaches that rely on predefined engineered descriptors based on assumed controlling features, we directly encode correlated microstructural and deformation states into latent representations (Fig. \ref{Figure1}(4)), preserving spatial relationships and feature correlations (Fig. \ref{Figure1}(5)). These representations are then used to predict macroscopic properties, enabling direct relationships between local material states and mechanical behavior (Fig. \ref{Figure1}(6)). Combined with ML interpretability tools (Fig. \ref{Figure1}(7)), MSI identifies the microstructural and deformation features controlling mechanical properties.

\section*{RESULTS}

\begin{figure}
    \centering
    \includegraphics[width=1\linewidth]{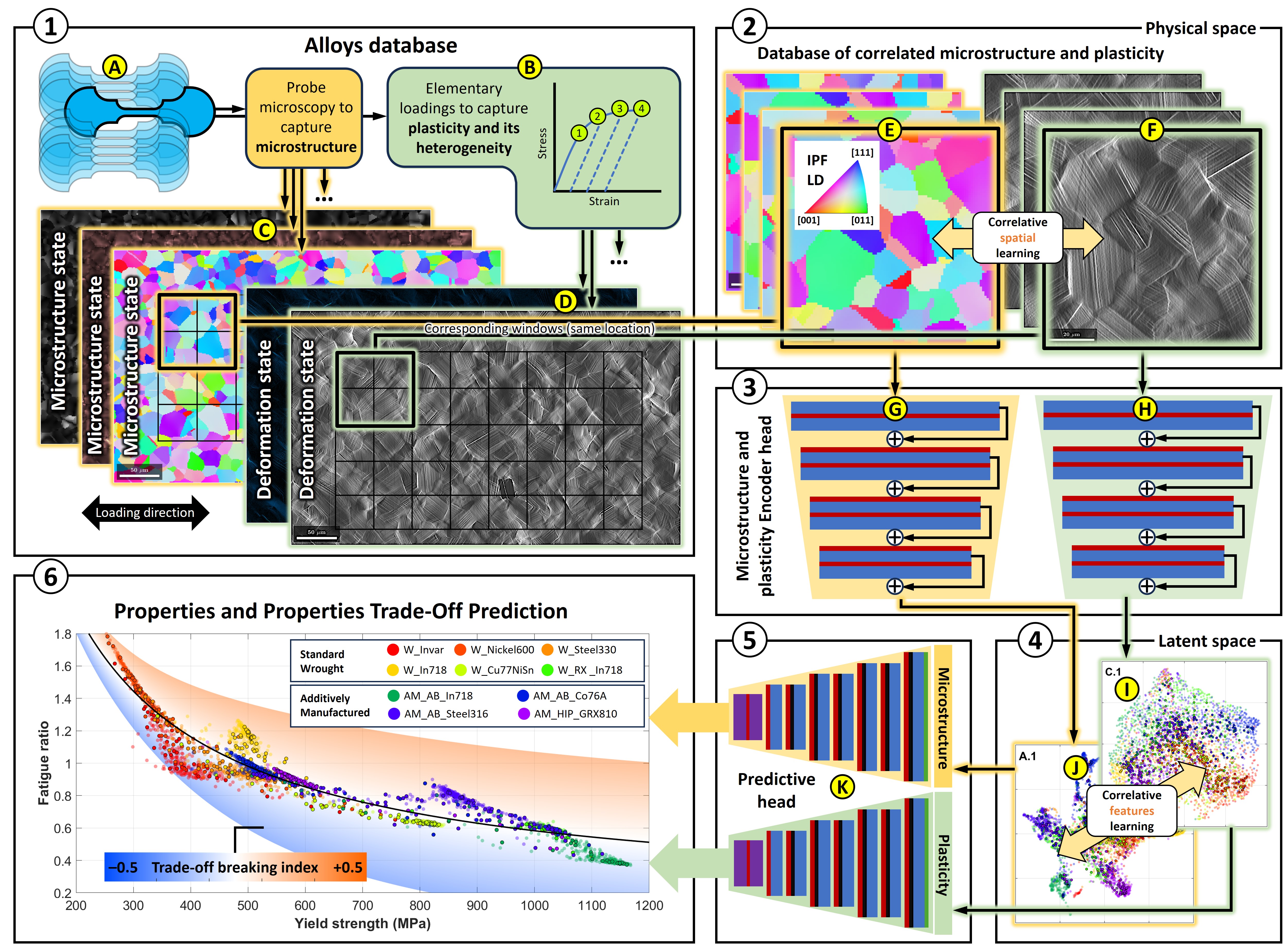}
    \caption{\textbf{Spatial data modalities and map-encoding scheme for microstructural and deformation states, enabling construction of alloy microstructure and deformation latent spaces for mechanical properties prediction.} (1) Elementary mechanical loadings (B) are applied to capture spatial deformation heterogeneity (D) under representative loading conditions. (C)(D) Large-field-of-view probe microscopy is used to characterize local microstructural states and their associated local deformation responses. (2) Microstructure and deformation maps are partitioned into corresponding local regions (E)(F), defining an ensemble of paired local microstructural and deformation states per alloy. (3) Separate convolutional-based encoder processes the (G) microstructural and (H) deformation information of the paired local states. (4) Uniform Mapping Approximation and Projection (UMAP) latent representation of all investigated local (I) microstructural and (J) deformation states. Colors correspond to different alloys. (5) Such an encoding scheme is used within a regression model to predict simultaneously a set of six mechanical properties through the (K) ML architecture . (6) Predicted fatigue ratio--yield strength trade-off for the validation dataset. Each alloy system is represented by a distinct color, while each point corresponds to an individual local microstructural state within that alloy. The color scale indicates the deviation from the average fatigue ratio--yield strength trade-off (solid black line), highlighting local microstructural states overcoming the trade-off.}
    \label{Figure2}
\end{figure}

\justify Building on the MSI framework conceptualized in Fig. \ref{Figure1}, Fig. \ref{Figure2} presents the spatial data modalities and map-encoding strategy used to construct latent representations of microstructural and deformation states and relate them to macroscopic mechanical properties. Figure \ref{Figure2}(1) shows representative microstructure and deformation maps for one investigated alloy (see Fig. S2 to Fig. S11 in the Supplementary Materials). These large-field measurements, obtained using Electron BackScatter Diffraction (EBSD) and High-Resolution Digital Image Correlation (HR-DIC) (see Methods), provide statistically representative, high-resolution descriptions of both material microstructure and deformation. Importantly, both are represented through a plurality of modalities that provide an in-depth description. Microstructural states include crystallographic orientation, phase distribution, and defect-related descriptors such as qualitative dislocation states \cite{Pantleon2008}, while deformation states include strain, lattice rotation \cite{STINVILLE2020110600}, and measures of deformation-event intensity \cite{Bourdin2018,Stinville2022}. We obtained material deformation responses under simplified loading conditions (\textit{i.e.}, tensile tests or limited cycling) to rapidly capture deformation states and their evolution during loading across various loading conditions (Fig. \ref{Figure2}(B)) \cite{Stinville2022}. Using large datasets across alloys and processing conditions, we learn general microstructure-deformation relationships rather than alloy specific ones that are not easily transferable from one alloy to another.

\justify Although these maps fully describe the microstructure and associated deformation response, their large dimensions prevent direct use. The maps are therefore partitioned into correlated reduced regions, generating a database of paired local microstructural and deformation states. Each region captures local microstructural features and their associated deformation response, forming the fundamental unit through which microstructure–deformation relationships are learned (Fig. \ref{Figure2}(E) and (F)). These local states are independently encoded using convolutional neural networks (Fig. \ref{Figure2}(G) and (H), see Fig. S18 in the Supplementary Materials), enabling latent spaces that capture the plurality and heterogeneity of microstructure and deformation across alloys and processing conditions. The encoded states are then projected into lower-dimensional manifolds organizing microstructural states according to processing routes, grain structure, crystallographic texture, defect-related density (see Fig. S18(A.1--A.4) in the Supplementary Materials) and deformation states according to classes of deformation mechanisms (Fig. \ref{Figure2}(4), see Fig. S18(C.1--C.4) in the Supplementary Materials).

\justify We designed the ML architecture shown in Fig. \ref{Figure2}(5) (see Fig. S13 in the Supplementary Materials) to infer six mechanical properties from latent representations: yield strength, ultimate tensile strength, two hardening parameters, ductility, and fatigue strength. We use a modular structure to enable direct assessment of the respective contributions of microstructure and deformation independently, while a fused representation captures their coupled effects.

\justify As we simultaneously predict multiple properties, we combine these predicted properties to investigate property trade-offs and how they are controlled by microstructure and deformation. Figure \ref{Figure2}(6) shows the fatigue ratio--strength trade-off obtained. As in the latent spaces of Fig. \ref{Figure2}(4), each point represents a local microstructural or deformation state, while a color designates a given alloy. Each alloy occupies a distinct region of the fatigue ratio--yield strength space, with the extent of this region reflecting microstructural heterogeneity effects learned form the ensemble of investigated alloys. Some alloys remain localized in this space, whereas others span broader regions due to microstructural heterogeneity. We learn the well-known fatigue ratio--yield strength trade-off \cite{Pang2013,Stinville2022} and establish relationships between microstructure and deformation states, and mechanical behavior.

\begin{figure}
    \centering
    \includegraphics[width=1\linewidth]{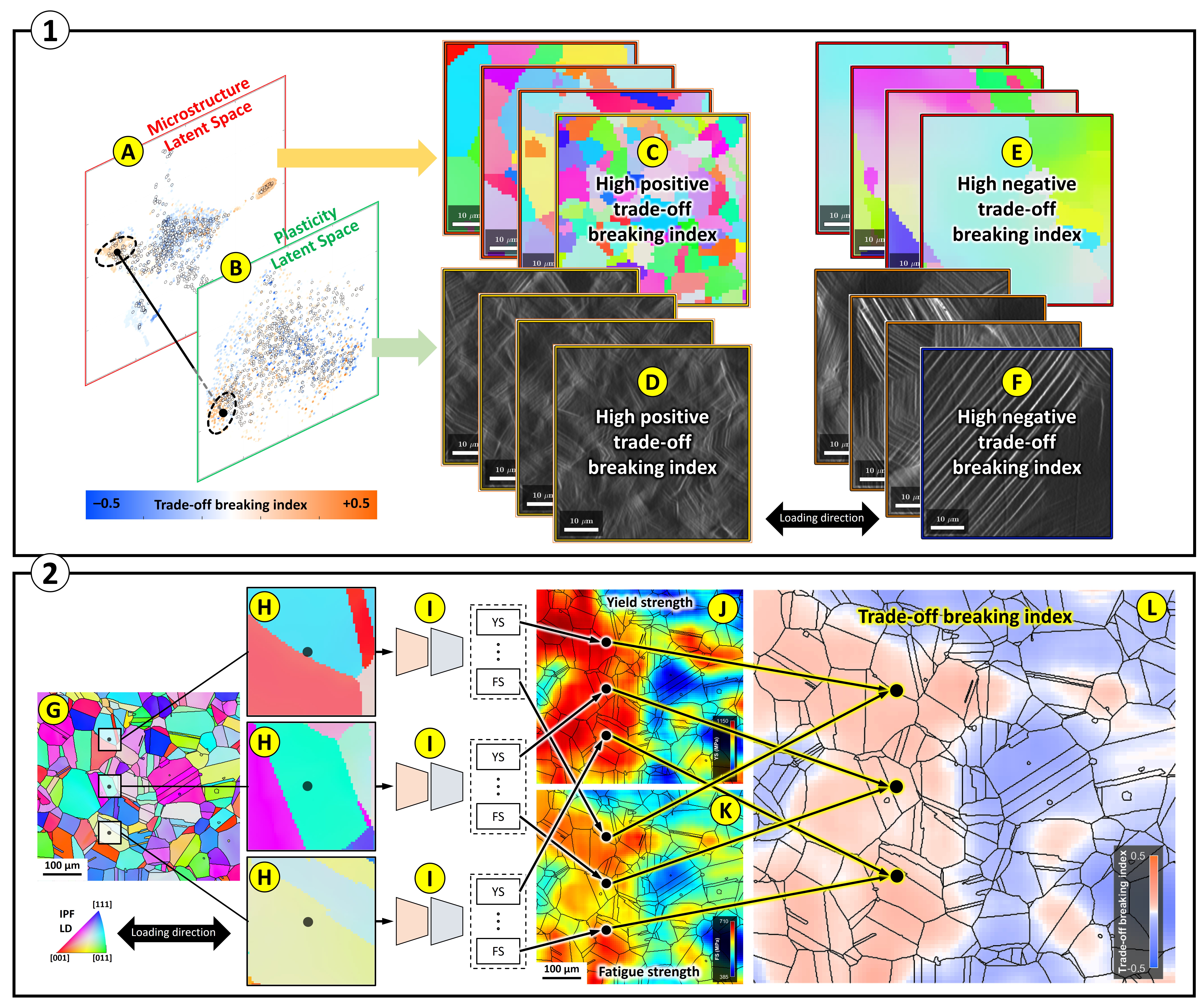}
    \caption{
        \textbf{Extraction of microstructural and deformation states that break macroscopic property trade-offs.} (1) A trade-off breaking index for fatigue ratio versus yield strength predicted based on either (A) microstructure or (B) deformation state  latent space. The trade-off breaking index is defined relative to the average trade-off behavior (black line in (6) of Fig. \ref{Figure2}) across all alloys. Representative local (C)(E) microstructural and (D)(F) deformation  states associated with the (C)(D) highest and (E)(F) lowest trade-off breaking index. (2) Reconstruction of spatially resolved mechanical property and trade-off maps from local states. (G) Representative microstructure used to characterize local microstructural states for the alloy W\_RX\_718. (H) Selected local regions are extracted and encoded to predict (I) macroscopic properties maps, such as (J) yield strength and (K) fatigue strength  maps. A sliding window approach is used to systematically extract local microstructural states and predict properties enabling the generation of property maps. (L) Reconstruction of the trade-off breaking index map derived from the predicted property maps.
    }
    \label{Figure3}
\end{figure}

\justify We leverage this unified representation to identify the local microstructural and deformation states that govern, favor, or mitigate macroscopic property trade-offs. Figure \ref{Figure2}(6) introduces a trade-off breaking index defined relative to the average fatigue ratio--yield strength relationship across all investigated alloys (solid black line), providing a quantitative measure of the ability of local states to exceed average trade-off behavior. This index is predicted using either microstructural or deformation states alone, enabling direct assessment of their respective contributions. The predicted trade-off breaking index is then overlaid onto the projected latent representations (Fig. \ref{Figure3}(A) and (B)), allowing identification of favorable (orange) and detrimental (blue) regions associated with microstructural or deformation states. Regions with large positive and negative trade-off breaking index values are shown in Fig. \ref{Figure3}(C,D) and (E,F), respectively. We recover, for instance, the beneficial effect of refined grain sizes on fatigue and yield strength (Fig. \ref{Figure3}(C) and (E)). Favorable states also exhibit limited deformation localization \cite{Anjaria2026} (Fig. \ref{Figure3}(D)), whereas detrimental states display intense localized deformation (Fig. \ref{Figure3}(F)). 

\justify To further improve interpretability, we combine MSI predictions with a sliding-window strategy (see Section 11 in the Supplementary Materials) that enables analysis over large fields of view. As summarized in Fig. \ref{Figure3}(2), full-field maps (Fig. \ref{Figure3}(G)) are divided into small regions (Fig. \ref{Figure3}(H)), encoded into latent representations, and used to predict local mechanical properties (Fig. \ref{Figure3}(I)). This process generates spatial maps of mechanical properties (Fig. \ref{Figure3}(J) and (K)) and trade-off breaking index map (Fig. \ref{Figure3}(L)), revealing the governing regions within microstructural or deformation states.

\begin{figure}
    \centering
    \includegraphics[width=0.8\linewidth]{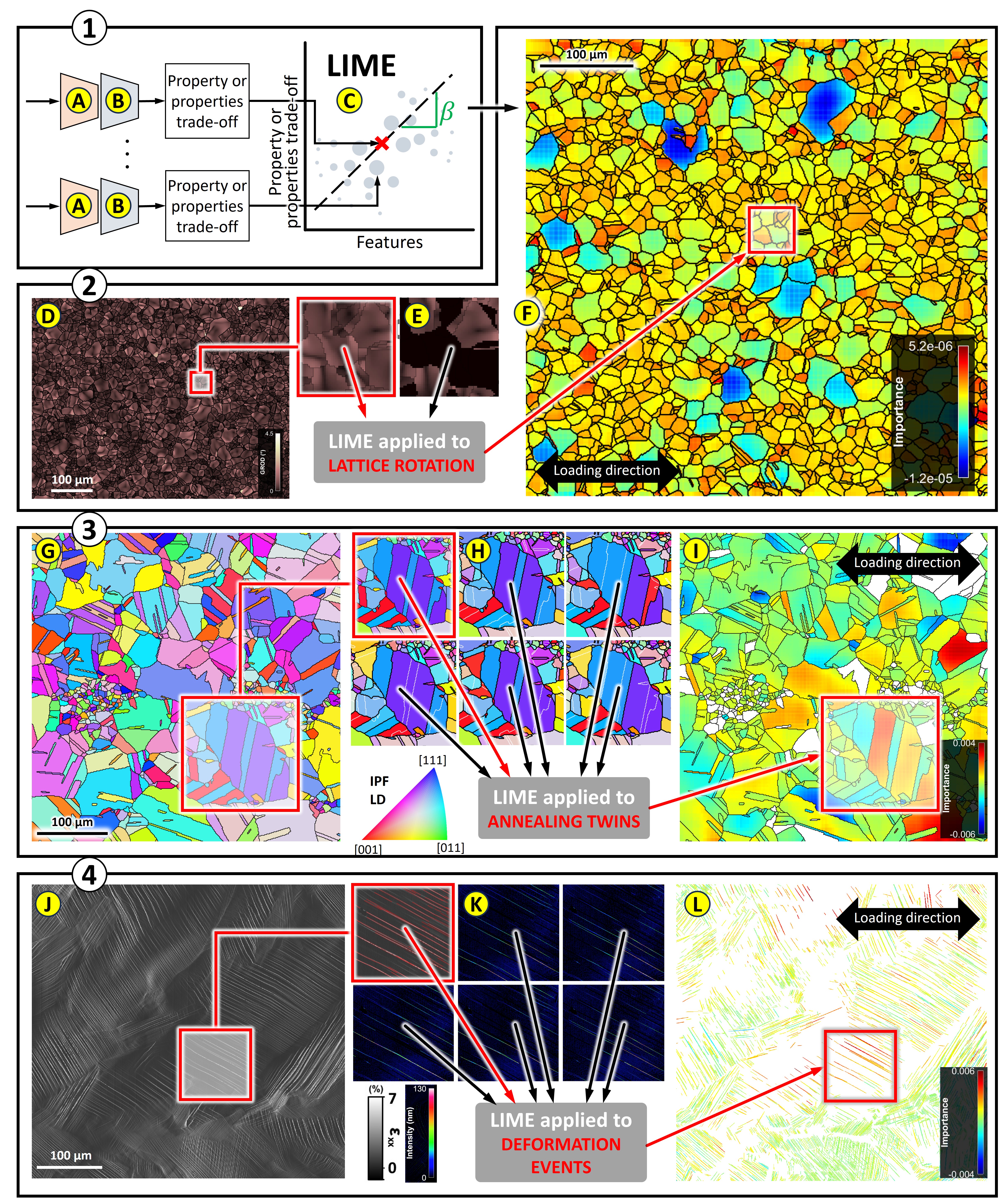}
    \caption{
        \textbf{Revealing microstructural and deformation features contributions to mechanical properties.} (1) Local interpretable model-agnostic explanations framework for identifying microstructural features importance on mechanical properties. (A)(B) Encoding of the initial state (red cross marker) and perturbed states (blue circle markers) to predict properties form these states and capture how the perturbation operation affects properties. (C) Illustration of the regression model employed in LIME relating feature perturbation to a mechanical property.
        (2) Generation of importance map from GROD perturbation within individual grains. (D) Representative GROD map for the W\_Cu77NiSn alloy. (E) Perturbation of a local microstructural state from a selected region (left), where the GROD values of selected grains were randomly set to zero (right). (F) Generated importance map from LIME illustrating the influence of GROD on the fatigue ratio--yield strength trade-off breaking index. (3) Generation of importance map from annealing twin perturbation. (G) Initial microstructure map of the W\_Steel330 alloy. (H) Random perturbation of a local microstructural state from a selected region (top left), where annealing twins are removed from parent grains. (I) Generated importance map from LIME illustrating the influence of annealing twins on the fatigue ratio--yield strength trade-off breaking index. (4) Generation of importance map from single deformation event perturbation. (J) Initial strain map showing deformation state of the AM\_AB\_Steel316 alloy. (K) Perturbation of the deformation state by removing individual deformation events identified using computer vision (red masks) and setting the corresponding deformation event intensity to zero. (L) Generated importance map illustrating the influence of single deformation event on the fatigue ratio--yield strength trade-off breaking index.}
    \label{Figure4}
\end{figure}

\justify To separate the influence of multiscale features within the governing regions in the microstructural state on mechanical behavior, we combine MSI with Local Interpretable Model-Agnostic Explanations (LIME) \cite{Ribeiro2016}. The LIME process (see Methods), illustrated in Fig. \ref{Figure4}(1), quantifies the importance of microstructural and deformation features by comparing predictions from original regions (local states) with randomly perturbed versions. For materials applications, we adapted LIME for regression using physically meaningful features such as grains, annealing twins, phases, defects-related features, and deformation events. These features are extracted using physics-based methods or computer-vision approaches (see Methods) \cite{VERMEIJ2023118502,HU2023103618,NI2024104119,Bean2025a}. Using predictions from the original and perturbed regions, a regression model (Fig. \ref{Figure4}(C)) is established to identify the importance of each feature to a given property or property trade-off. Predictions are weighted using a similarity metric, allowing regions closer to the original state to contribute more strongly. The resulting regression slopes ($\beta$ in Fig. \ref{Figure4}(C)) quantify how perturbing each feature affects the predicted properties.

.

\justify We extend LIME from single regions to large fields of view using the same sliding-window strategy employed for local property mapping. Instead of calculating feature importance over the entire microstructural or deformation states, a regression model is established locally at each position taken by the sliding window. Figure \ref{Figure4}(2) illustrates this process for Grain Reference Orientation Deviation (GROD), a metric to describe lattice rotation, using grains as the perturbed microstructural feature. Figure \ref{Figure4}(D) and (E) show the original and modified microstructure maps, where GROD values were randomly set to zero within grains, allowing to construct the feature-importance map shown in Fig. \ref{Figure4}(F), here for fatigue ratio--yield strength trade-off. This analysis confirms the detrimental effect of large grains, which mitigate the fatigue ratio-yield strength trade-off. Additionally, it reveals the significant influence of processing-induced lattice rotations within these grains.

\justify We further apply LIME to investigate the role of annealing twins, as shown in Fig. \ref{Figure4}(3). Their crystallographic orientation was replaced by that of the parent grain (see Methods) to evaluate their contribution. Figure \ref{Figure4}(G) and (H) show the original and modified microstructure maps, where several annealing twins were removed. The resulting feature-importance map (Fig. \ref{Figure4}(I)) reveals the detrimental effect of extended annealing twins within large grains, consistent with previous observations associating them with reduced fatigue strength \cite{Stein2014,Stinville2022}. Interestingly, the model also identifies alternating positive and negative feature importance in relation to the crystallographic orientation taken by the grain when the annealing twins are removed.

\justify Finally, we use LIME to investigate the influence of individual deformation events on the fatigue ratio--yield strength trade-off. The approach, illustrated in Fig. \ref{Figure4}(4), uses deformation-event intensity maps (see Methods) and deformation-event masks extracted from computer vision \cite{Bean2025a}. Deformation events were randomly masked and their intensity values replaced by zero, as shown in Fig. \ref{Figure4}(K). The resulting feature-importance map (Fig. \ref{Figure4}(L)) reveals the critical role of intense and extended deformation events in controlling the trade-off, consistent with previous literature observations \cite{Zhang2008,Pang2013}.

\begin{figure}
    \centering
    \includegraphics[width=0.78\linewidth]{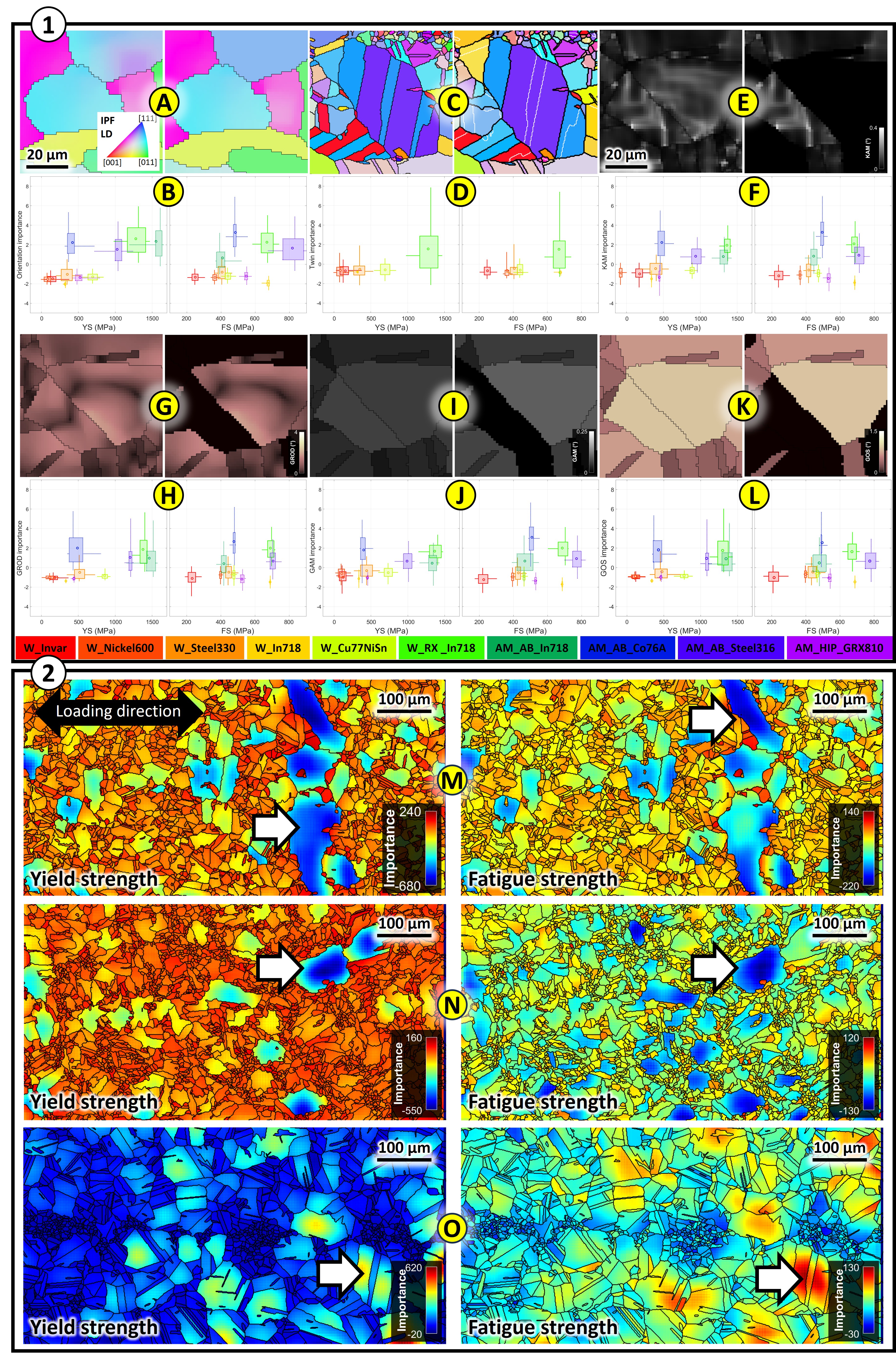}
    \caption{
        \textbf{Interpretability and spatial importance analysis of microstructural features governing mechanical properties.} (1) Representative examples illustrating the effect of modifying local microstructural and deformation states on the predicted mechanical properties, including here (left) Yield Strength (YS) and (right) Fatigue Strength (FS). Each row corresponds to a different modality extracted from experimentally measured datasets, including (A) crystallographic orientation (quaternion), (C) annealing twin, (E) kernel average misorientation, (I) grain reference orientation deviation, (G) grain average misorientation, and (K) grain orientation spread. For each feature, the left and right maps show the local state before and after perturbation, respectively. The associated box plots (B)(D)(F)(H)(J)(L) summarize the distribution of importance of the considered feature on (left) yield strength and (right) fatigue strength. The scatter per box in X represents the distribution in predicted properties from the predicted property maps. (2) Importance maps established from KAM and reconstructed for (left) yield strength and (right) fatigue strength for (M) W\_Invar, (N) AM\_HIP\_GRX810, and (O) W\_Steel330 alloys.
    }
    \label{Figure5} 
\end{figure}

\justify In Fig. \ref{Figure5}(1), we apply our LIME framework to all investigated alloys and multiple orientation-derived maps, including annealing twin substitution, crystallographic orientation substitution, GROD, Grain Average Misorientation (GAM), Kernel Average Misorientation (KAM), and Grain Orientation Spread (GOS). The corresponding box plots summarize both the predicted mechanical properties obtained from crystallographic orientation (quaternion), KAM, GROD, GAM and GOS maps, and the distributions of feature importance values extracted from the LIME analysis. Each box corresponds to a different alloy, with additively manufactured alloys represented using dashed lines. The boxes extend from the first to the third quartile (25\% to 75\% of the distribution), while whiskers indicate the extreme values spanning from the 5\textsuperscript{th} to the 95\textsuperscript{th} percentiles. In addition, markers represent the average predicted property values together with the average feature importance estimated from the spatial maps. Each line in Fig. \ref{Figure5}(1) therefore reflects the contribution of a given microstructural feature modality to the predicted mechanical properties shown in each column (with Fig. \ref{Figure5}(B--L)(left) and (right) being the Yield Strength (YS) and Fatigue Strength (FS), respectively).

\justify Such analysis provides a direct and quantitative evaluation of the relative importance of individual microstructural features with respect to the predicted mechanical properties. More importantly, the LIME framework enables these feature contributions to be spatially resolved and projected back onto the microstructure itself, as illustrated in Fig. \ref{Figure5}(2) for yield strength and fatigue strength, which together govern the fatigue ratio--yield strength trade-off. These importance maps reveal regions contributing either favorably (positive value) or detrimentally (negative value) to the predicted mechanical properties. 

\justify For instance, in the W\_Invar alloy (Fig. \ref{Figure5}(M)), the framework identifies extended partially recrystallized regions (see white arrows) that negatively influence both yield and fatigue strength. These regions, characterized by elevated lattice rotation and dislocation density (see Fig. S2 in the Supplementary Materials), demonstrate limited strengthening while affecting deformation localization during fatigue loading. Similar observations are obtained for the AM\_HIP\_GRX810 alloy (Fig. \ref{Figure5}(N)), where large grains (see white arrows) are identified as limiting the predicted mechanical properties due to their association with intense lattice rotation and high local dislocation density (see Fig. S11 in the Supplementary Materials). Interestingly, we also capture the known influence of specific annealing twin configurations on yield strength and fatigue performance \cite{Prouteau2022}, as indicated by the white arrow in Fig. \ref{Figure5}(O) in the W\_Steel330 alloy. 

\begin{figure}
    \centering
    \includegraphics[width=1\linewidth]{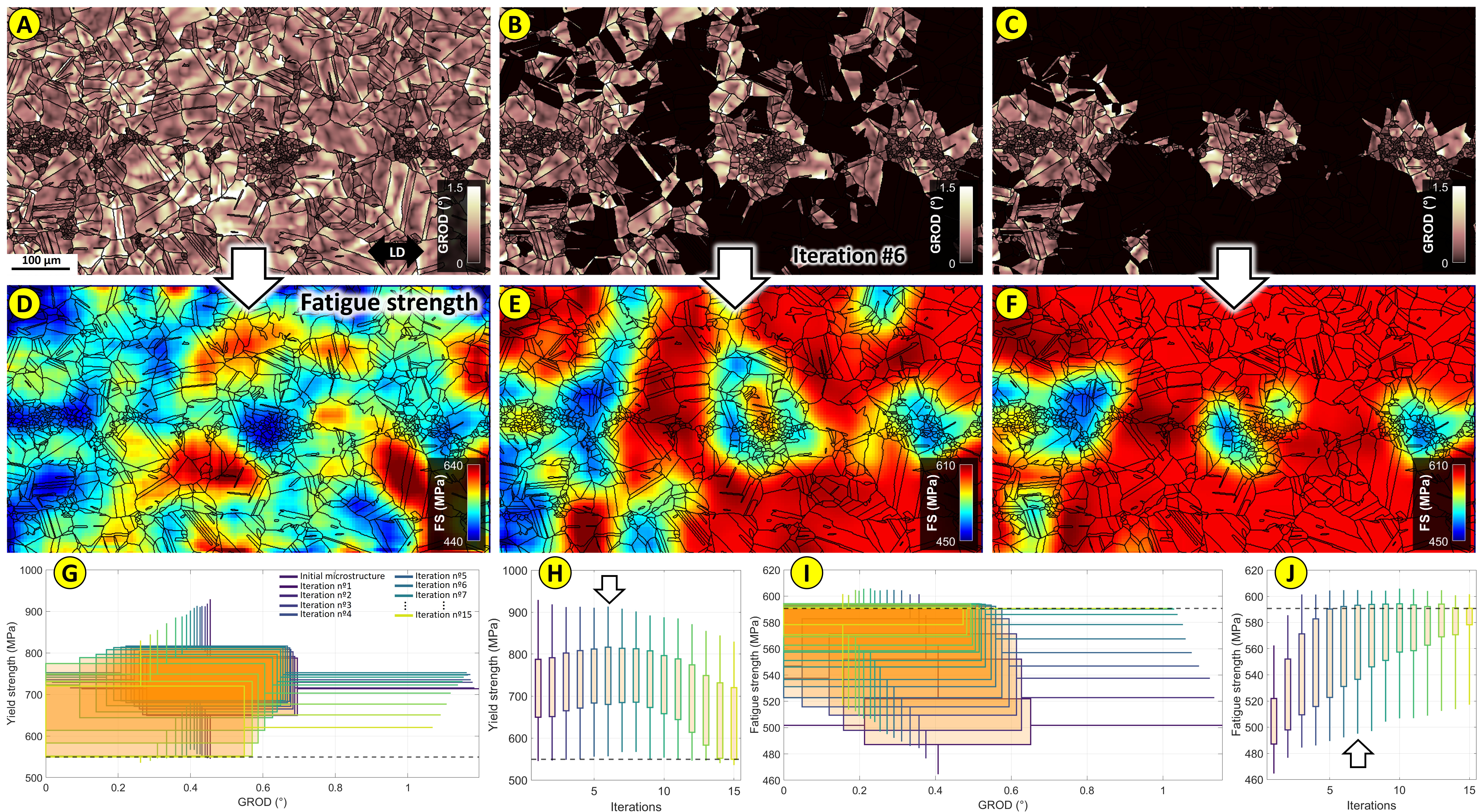}
    \caption{
        \textbf{An iterative synthetic microstructure optimization framework guided by the LIME.} The progressive modification of the spatial distribution of GROD where the most detrimental regions identified by LIME are iteratively (A to C) replaced by configurations with zero GROD values. The associated predicted fatigue strength spatial distributions are generated (D to F) during the optimization process for the W\_Steel330 alloy. Resulting improvement and their variability for (G)(H) yield strength and (I)(J) fatigue strength distributions for dedicated optimization process and derived from the predicted property maps across all optimization iterations, during which the most detrimental grains are progressively replaced by configurations with zero GROD values. The x-axis in (G, I) represents the distribution of GROD values within the microstructure after each iteration.
    }
    \label{Figure6}
\end{figure}

\justify We also leverage our framework to investigate the role of spatial microstructural configurations and enable synthetic microstructure design. Rather than randomly replacing features, we iteratively modify the 5\% most detrimental grains identified by their LIME importance values, replacing unfavorable features with favorable ones to generate microstructural states with improved properties. As illustrated for the W\_Steel330 alloy in Fig. \ref{Figure6}, the initial microstructural state (A) and associated predicted fatigue strength map (D) are progressively modified by removing processing-induced lattice rotations. This is represented by setting the GROD values to zero in selected grains (B--C) along with the results predicted maps shown in (E--F). The evolution of yield strength and fatigue strength over many iterations is summarized in Fig. \ref{Figure6}(G--J). The framework identifies microstructures exhibiting a slight increase in yield strength through the formation of a heterostructure containing grains with and without lattice rotations, whereas further removal of this feature reduces the associated strengthening effect and decreases yield strength. In contrast, fatigue strength progressively improves owing to the reduction in dislocation density. These results demonstrate the ability of MSI to identify spatial distributions of microstructural features that simultaneously optimize multiple mechanical properties and capture property trade-offs. This behavior differs from uniformly removing the feature across the entire microstructure map (dashed lines in Fig. \ref{Figure6}(G--J)), where spatial distribution effects are not exploited.

\section*{DISCUSSION}


\justify Spatial intelligence marks a paradigm shift in mechanical metallurgy, shifting the focus from averaged material descriptors to spatially resolved representations of microstructure and deformation. Optical, electron, and X-ray microscopy enable detailed observation of these features and underlying mechanisms, while their correlation with mechanical behavior is predominantly established until recently through human interpretation or through physics-based simulations \cite{Meric1991,Berbenni2007,Zhou2010,Zepeda2017}. Physics-based approaches, including crystal plasticity, phase-field models, discrete dislocation dynamics, and molecular dynamics, has remained the predominant and most effective approach for describing the influence of microstructure and deformation mechanisms on mechanical behavior. However, these approaches often require numerous parameters \cite{Meric1991,Cormier2010,Castelluccio2017} and substantial computational resources \cite{Zepeda2017,Zhu2023}, while relying on simplified representations of microstructure or mean-field approximations \cite{Liu2017} or being restricted to small material volumes \cite{Healy2014,Zepeda2017,Calvat2025a}. More importantly, their application to new alloy systems relies heavily on re-parameterization, slowing discovery in metallic materials and hindering materials design. In parallel, many data-driven approaches still rely on simplified materials descriptors such as average chemical composition \cite{Lee2021,Hu2022,Sohail2025}, phase fractions \cite{Borg2020,Deng2022}, grain size \cite{Gu2024}, texture \cite{Gu2024}, or simplified simulation outputs, limiting predictive accuracy and access to a fundamental understanding at the microstructural scale. Instead, our MSI framework exploits high-resolution, spatially resolved experimental measurements over large fields of view to characterize both microstructure and deformation together with their heterogeneity. These measurements provide statistically representative and spatially explicit descriptions of material states, shifting data-driven materials science from averaged descriptors to spatially resolved maps and enabling spatial intelligence.


\justify Mechanism-aware prediction from microstructural and deformation states provides a framework for linking microstructural features and small-scale deformation mechanisms to macroscopic mechanical properties, that. Microstructure alone is insufficient for prediction, as identical microstructures can exhibit different responses under varying loading conditions \cite{Aubin2003,Kumar2016}. In our MSI framework, materials are described through both their microstructural and deformation states under simplified (elementary) loadings for rapid measurement, enabling prediction of macroscopic properties under various loading conditions. This approach is strongly supported by mechanical metallurgy, as mechanical properties are known to correlate with microstructural features, deformation processes, as well as their characteristics. Yield strength correlates with deformation event intensity and length, cross-slip activity, and event connectivity (\textit{i.e.}, slip transmission) \cite{HULL2011171, Andani2020, ANDANI2022117613,doi:https://doi.org/10.1002/9781119296126.ch173, doi:10.1126/sciadv.abo5735, CLEVERINGA1999837, CHAKRAVARTHY2010625, refId0}. Slip event length is linked to the dislocation mean free path, governed by internal length scales such as grain size in wrought alloys, , consistent with the Hall--Petch relationship \cite{Hall1951,Petch1953} or dislocation cell size in additively manufactured alloys \cite{Bean2025a}. Strain hardening depends on the evolution of strain localization, including the number, intensity, and connectivity of deformation events, as well as the activation of slip systems or deformation modes such as twinning \cite{Salem2006,Bean2022,Yvinec2026b}. Ultimate tensile strength and ductility are governed by strain accumulation. Lastly, fatigue performance is controlled by slip irreversibility \cite{Mughrabi2009}, which drives crack initiation and governs fatigue strength \cite{Mughrabi2009,Stinville2022}. Slip irreversibility is strongly influenced by plastic localization \cite{Stinville2022,Mughrabi2009,STINVILLE2020172,Pan2026}, with intense localization sites correlating with crack nucleation \cite{Stinville2022, ANDANI2022117613,STINVILLE201816}. Importantly, alloy properties depend not only on microstructural and deformation features at the nanometer-to-micron scale but also on their spatial arrangement across the macroscopic scale.This creates an intrinsic experimental challenge that requires characterization methods capable of resolving small-scale mechanisms over millimeter-scale fields of view. Recent advances in detector technology, electron optics, and automation now enable rapid acquisition of large-area datasets with nanometer resolution \cite{Black2023, annurev:/content/journals/10.1146/annurev-matsci-080921-102621}. In this work, deformation events and deformation processes are quantified using high-throughput, high-resolution digital image correlation combined with electron microscopy \cite{Bourdin2018,Vermeij2023,Stinville2022,Texier2024,Mornout2025}. This approach shifts deformation analysis from site-specific investigations toward large fields of view measurements. It also enables the characterization of the nature (slip, deformation twinning or grain boundary sliding), morphology, connectivity, and spatial distribution of these deformation events to be captured \cite{POLAK20031,MUGHRABI20131197,Mughrabi2009,CHARPAGNE2021117037,HONG19901581,Bean2025a,CHARPAGNE2022143291,Anjaria2026}. During training of our MSI framework, the latent space evolves to capture the most informative features \cite{Goodfellow2016,Pei2021,Calvat2025b,Calvat2026,Islam2026}, improving predictive capability while preserving the diversity and complexity of the experimental dataset. The resulting representation extracts the key features required to describe material states and mechanical behavior. Importantly, the MSI framework is designed to operate across multiple material systems. This enables the learning of both general microstructural and deformation features (for example, grains and slip events) and specific feature–property relationships. 



\justify A data-driven route to fundamental materials understanding emerges from the analysis and interpretation of the MSI latent space. Latent space projection using UMAP \cite{Mcinnes2018,Becht2019} reveals the structure of the latent space which captures grain sizes, shapes, and local misorientations, as well as their nature (\textit{e.g.}, partially recrystallized microstructures, intense plasticity localization, or additive manufacturing processing as shown in Fig. S18 in the Supplementary Materials). Interestingly, due to the highly heterogeneous microstructural or deformation states of alloys, some alloys are spread widely throughout the latent space giving an opportunity to learn and capture the effect of local states on properties. Since our approach is based on local measurements, we can also construct maps of macroscopic properties, as shown in Fig. \ref{Figure3}(2), that illustrate regions (microstructure or deformation states) that promote or hinder mechanical properties or their trade-offs. However, this analysis remains patch-based, with each patch containing potentially correlated combinations of microstructural and deformation features. To improve interpretability, MSI was combined with LIME \cite{Ribeiro2016} to explain predictions and quantify feature importance. By applying controlled physics-based perturbations to microstructural and deformation states, LIME probes the local model response and identifies the features and spatial regions that most strongly influence predictions (Fig. \ref{Figure4}). Importantly, this feature-wise importance analysis captured that material properties are governed not only by the magnitude of individual features, but also by their spatial organization and interactions. 


\justify Materials Spatial Intelligence provides a practical tool for materials design by enabling rapid property prediction, qualification (see Section 12 in the Supplementary Materials) and physically interpretable microstructure optimization. Although MSI provides limited time savings for monotonic properties, it offers substantial advantages for advanced properties such as fatigue performance, where conventional testing is extremely time-consuming. Fatigue properties can typically be estimated 30- to 100-fold faster. More importantly, since predictions are derived from local microstructural and deformation states, the framework can also be applied to small coupons and spatially heterogeneous materials, including functionally graded and multi-material systems \cite{Bobbio2017,BEAN2025114115,Jullien2025}. Our MSI framework can also be integrated into material design loops. As a first implementation, MSI can be combined with an experimental loop to enable rapid property prediction, providing the opportunity to evaluate a large number of alloys. In parallel, the LIME-based microstructure feature-wise importance approach identifies the influence of specific microstructural and deformation features, together with their spatial organization, on mechanical properties. This quantification can be used to virtually modify the governing features to improve performance or overcome property trade-offs, as illustrated in Fig. \ref{Figure6}. Although the resulting optimized microstructures may not always be experimentally attainable, this strategy retains a key advantage over purely generative ML approaches by focusing on physically identifiable features and spatial arrangements rather than unconstrained synthetic structures. Existing processing knowledge can subsequently guide adjustments of processing parameters to target these desired features and configurations \cite{Gu2018,Keller2021}.

\justify This MSI approach transforms microstructure-based research by enabling data-driven and physically interpretable identification of governing microstructural features and deformation mechanisms from experimental full-field measurements, while providing a pathway toward data-driven microstructure-based alloy design.

\newpage
\section*{METHODS}

\subsection*{Materials}

\justify A total of ten FCC materials, including conventionally wrought and additively manufactured materials, were considered in this study. These alloys are referred to as \textbf{W}, \textbf{AM}, \textbf{RX}, \textbf{AB} and \textbf{HIP}, corresponding to Wrought, Additively Manufactured, Recrystallized, As-Built, and Hot Isostatic Pressing, respectively. The alloys' compositions are listed in Table S1 of the Supplementary Materials. Additionally, Fig. S1 in the Supplementary Materials provides an overview of the investigated materials' microstructure and the associated deformation. Figures S2 through S11 present the microstructure of each material (Inverse Pole Figure (IPF)-colored orientation maps, Kernel Average Misorientation (KAM), and Grain Reference Orientation Deviation (GROD)), the longitudinal strain fields showing plasticity localization and evolution throughout four consecutive deformation steps, and macroscopic tensile behavior and fatigue results.

\subsection*{Sample preparation}

\justify The various specimen geometries employed for HR-DIC, optical DIC and conventional fatigue testing are provided in Fig. S12 in the Supplementary Materials. Both HR-DIC and optical DIC specimens were machined by EDM (Electrical Discharge Machining). HR-DIC specimens were first mechanically polished with abrasive papers down to 1200 grit followed by $3\:\mathrm{\mu m}$ diamond suspensions. The specimens were finished with a mechanical/chemical polishing using a $50\:\mathrm{nm}$ colloidal silica suspension, either manually for 20 minutes or automatically using a Buehler Vibromet 2 for 24 hours. Two layers of $3\:\mathrm{nm}$ of Titanium and $10\:\mathrm{nm}$ of Silver were then deposited on the surface of the samples using an AJA Sputter Coater. The Silver layer was reconfigured into speckles by immersing the samples into a 1\% salt water solution for 2 hours (see Fig. S12(B) in the Supplementary Materials). Additional details concerning the reconfiguration process can be found in Ref. \cite{Kammers_2013a, Montgomery2019}. AM\_HIP\_GRX810 was heat treated at $275^{\circ}\mathrm{C}$ for 90 minutes under Argon flow. The same heat treatment was applied to W\_Cu77NiSn, however under vacuum ($10^{-4}\:\mathrm{Torr}$). Optical DIC specimens were mechanically polished with abrasive papers down to 320 grit. A speckle pattern was then produced by laser engraving (see Fig. S12(D) in the Supplementary Materials) using an OMTech $30\:\mathrm{W}$ laser with a wavelength of $1064\:\mathrm{nm}$. Conventional fatigue specimens were machined by CNC, mechanically polished with abrasive papers down to 1200 grit followed by $3\:\mathrm{\mu m}$ diamond suspensions. 

\subsection*{Mechanical Testing}

\justify HR-DIC and optical DIC testing was performed using a $5\:\mathrm{kN}$ NewTec MT1000 under quasi-static loading conditions (force rate of $2\:\mathrm{N\cdot s^{-1}}$ corresponding to $\dot{\varepsilon}$ between $10^{-5}$ and $10^{-4}\:\mathrm{s^{-1}}$). For HR-DIC, mechanical testing was interrupted four times to collect images and characterize the deformation processes. For optical DIC tensile testing, images were captured every 5 seconds using a Pixelink PL-D7912MU-T camera. Specimens were deformed up to fracture to estimate the mechanical monotonic properties listed in Table S2 in the Supplementary Materials. Yield Strength (YS) and Ultimate Tensile Strength (UTS) were directly identified from the tensile curves. Strain at Failure (SF) was identified from images taken during mechanical testing. The corresponding strain was directly estimated using the first image and the last image taken before failure of the specimen. Following optical DIC computations, longitudinal strain was averaged over the gauge length and Young's modulus was identified from the corresponding stress strain curve. Using the identified Young's modulus and a default Poisson ratio $\nu$ of 0.3, isotropic nonlinear hardening was modeled using Eq. \ref{eq:hardening} according to monotonic loading in the elastoplastic regime\cite{Besson2009}. The hardening parameters $Q$ and $b$ were then manually identified and are listed in Table S2 in the Supplementary Materials.

\begin{equation}
    \sigma = \sigma_y + Q(1 - e^{-b\varepsilon^p})
    \label{eq:hardening}
\end{equation}

\justify For fatigue testing, each specimen underwent testing for 500,000 cycles at a stress ratio $R_{\sigma}$ of 0.1 and a frequency of $2\:\mathrm{Hz}$ on an Instron 8500 or an Instron 8800 with $100\:\mathrm{kN}$ force cells. Fatigue testing was carried out incrementally using 50\% of the yield strength as the initial stress. If no fracture occurred before run-out, the stress was increased by 10\% of the yield strength and this procedure was repeated until fracture of the specimen. The Fatigue Strength (FS) was identified as the highest stress of run-out before failure and is listed in Table S2 in the Supplementary Materials.

\subsection*{Electron microscopy}

\justify The high-resolution imaging was performed on a Thermo Fisher Scios 2 Dual Beam SEM/FIB. SEM images were acquired ex-situ (in an unloaded state) at different macroscopic plastic deformation states listed in Table S2 in the Supplementary Materials. For each specimen and deformation state, fields of view of $840 \times 630\:\mathrm{\mu m}$ were acquired consisting of a 7 by 7 grid of secondary electron images with a resolution of 6144 by 4096 pixels (with a $15\%$ overlap between neighboring images). According to the guidance given in Ref. \cite{Kammers_2013b,Stinville_2015a,Mello2017} to mitigate distortions, the images were collected with the following acquisition parameters: an accelerating voltage of $10\:\mathrm{kV}$, a current of $0.8\:\mathrm{nA}$, a dwell time of $10\:\mathrm{\mu s}$ per pixel and at a working distance of $5\:\mathrm{mm}$. Imaging was automated using Thermo Fisher's AutoScript software and a custom Python code that automated focus, astigmatism, contrast and brightness, as well as aligning subsequent images with respect to their undeformed state. The corresponding Python routine has been made available on GitHub (\url{https://github.com/cmbean2/Automated-SEM-Procedure}).

\subsection*{High-resolution digital image correlation}

\justify Digital Image Correlation (DIC) was performed using the Heaviside-DIC (H-DIC) formulation which is a discontinuity-tolerant DIC formulation as detailed in Ref. \cite{Valle2017}. This method allows the quantification of in-plane displacements associated with irreversible deformation mechanisms found in metallic materials such as dislocation slip \cite{Bourdin2018,Anjaria2024}, grain boundary sliding \cite{Texier2024} or twinning \cite{Anjaria2024}. This method has a typical resolution of 0.2 to 0.3 pixels corresponding here to 7 to $10\:\mathrm{nm}$. In this study, all modalities were computed with respect to the initial state using a subset size of $35 \times 35$ or $40 \times 40$ pixels and a step size of 3 pixels. The computations were performed using a proprietary software, XCorrel, on a workstation equipped with NVIDIA Geforce RTX 4090 and NVIDIA Geforce RTX 3090.

\subsection*{Computer vision}

\justify Following HR-DIC computations, deformation events were identified using a computer vision framework developed in our previous study \cite{Bean2025a}. This model was trained using a collection of manually annotated slip deformation events from Inconel 718, Invar, pure Niobium, and Ti-6Al-4V using the CVAT.ai online annotation tool \cite{CVAT2022}. For this application, Ultralytic's pre-trained Yolov8 base was retrained to specifically identify slip deformation events using plastic deformation intensity maps. Deformation events are identified from reduced regions of $1024\times1024$ pixels ($69 \times 69\:\mathrm{\mu m}$) as masks (polygons) around the events. These masks are  finally post-processed to combine different masks representing the same deformation event. Additional details can be found in Ref. \cite{Bean2025a}. In this study, the slip deformations events were identified from the plasticity deformation maps warped into the EBSD coordinates.

\subsection*{Electron backscatter diffraction}

\justify After deformation, all the specimens were manually repolished for 20 minutes using only colloidal silica. EBSD measurements were then performed on a Thermo Fisher Scientific Apreo SEM quipped with an Oxford Symmetry EBSD detector with a square grid collection, a $1\:\mathrm{\mu m}$ step size, an accelerating voltage of $30\:\mathrm{kV}$, a current of $6.5\:\mathrm{nA}$, an exposure time of $6.5\:\mathrm{ms}$, $10\:\mathrm{mm}$ of working distance and $70^{\circ}$ tilt. The diffraction patterns were indexed on-the-fly using a Hough-based transform and the resulting EBSD data were processed using MTEX \cite{Bachmann2010}. The grains were first reconstructed with a minimum misorientation criterion of $5^{\circ}$ and a minimum of 10 pixels per grain. All pixels considered as noise were corrected using a half quadratic filter \cite{Bergmann2015}. Twins boundaries were identified from the $\Sigma 3$ crystallographic orientation relationship and allowing a maximum misorientation of $5^{\circ}$. EBSD maps were finally processed to extract Kernel Average Misorientations (KAM), Grain Average Misorientation (GAM), Grain Reference Orientation Deviation (GROD), Grain Orientation Spread (GOS) and Geometrically Necessary Dislocations (GND) \cite{Pantleon2008}. EBSD measurements were performed on slightly larger regions to facilitate the alignments with HR-DIC results. In addition to orientation-derived maps, conventional microstructure characteristics have been computed over the entire region investigated by EBSD and are provided in Table S3 in the Supplementary Materials.

\subsection*{CNN architectures}

\justify The architecture used for property predictions in this study has been adapted from Ref. \cite{Calvat2026} and is illustrated in Fig. S13 in the Supplementary Materials. First, it includes two encoding heads, shown in Fig. S13(A) and Fig. S13(B), which extract relevant features from HR-DIC or EBSD maps, respectively. These encoders have been developed to be compatible with various resolutions: $846 \times 846$, $1410 \times 1410$, $1974 \times 1974$ or $2538 \times 2538$ for HR-DIC, and $57 \times 57$, $95 \times 95$, $133 \times 133$ or $171 \times 171$ for EBSD maps. The employed convolution kernels remain identical across the investigated resolutions by adjusting padding dimensions to fit the different resolutions. Depending on the input resolution, the encoding heads produce $9 \times 9$, $18 \times 18$, $27 \times 27$, or $36 \times 36$ activation maps, which are transformed into $9 \times 9$ maps using a max pooling operation with strides of 2, 3, or 4. These activation maps are then used as inputs for the prediction head, the structure of which is detailed in Fig. S13(C). The same predicting head architecture is used for activation maps produced from either HR-DIC or EBSD. Due to the limited amount of data, the prediction head includes 15\% dropout layers to prevent overfitting during training. The prediction head finally outputs six values from a fully connected layer without any activation, enabling the simultaneous prediction of six mechanical properties, including Yield Strength (YS), two hardening parameters (Q and b), Ultimate Tensile Strength (UTS), Strain at Failure (SF), and Fatigue Strength (FS). Unlike the model developed in Ref. \cite{Calvat2026}, conditional macroscopic states (stress and strain states) have not been employed to force the model in identifying relevant features with respect to mechanical properties. Including both the encoding and prediction heads, the resulting HR-DIC and EBSD models have 12.1M and 11.3M parameters, respectively. All these architectures were implemented and trained on MATLAB using the Deep Learning Toolbox.

\subsection*{Training data and preprocessing} 

\justify Prior to any CNN training, microstructure and plasticity maps were preprocessed. In total, four different HR-DIC maps ($\varepsilon_{xx}$, $\varepsilon_{eq}$, plastic deformation intensity $\|\tau\|$ and lattice rotation $\gamma$) and six EBSD maps (quaternions, KAM, GAM, GROD, GOS and GND) were considered. The HR-DIC maps were interpolated at 1\% macroscopic plastic strain using a linear pixel-wise interpolation method derived from the four different states (see Table S2 in the Supplementary Materials). Strain components and lattice rotation were then preprocessed with Gaussian blur of standard deviation of 3 and 10, respectively. Following HR-DIC/EBSD maps alignment and warping, these maps were divided into two regions along the horizontal direction for training and validation with respective proportions of 75\% and 25\%. Each region was then divided into smaller tiles while allowing an overlap of $50\%$ between horizontal and vertical tiles but no overlap between training and validation regions. Four different tile dimensions were considered in this study: $57 \times 57\:\mathrm{\mu m}$, $95 \times 95\:\mathrm{\mu m}$, $133 \times 133\:\mathrm{\mu m}$ or $171 \times 171\:\mathrm{\mu m}$ corresponding to $846 \times 846\:\mathrm{px.}$, $1410 \times 1410\:\mathrm{px.}$, $1974 \times 1974\:\mathrm{px.}$ or $2538 \times 2538\:\mathrm{px.}$ for HR-DIC maps. The resulting number of HR-DIC tiles per material is listed in Table S5 in the Supplementary Materials. Since HR-DIC maps were interpolated at 1\% macroscopic plastic strain, the considered number of EBSD and HR-DIC tiles are identical.

\subsection*{CNN trainings}

\justify The different CNN (Convolutional Neural Network) architectures were trained to predict all 6 mechanical properties in a simultaneous manner from a single microstructure or deformation patch. To improve stability during training and ensure similar accuracy among mechanical properties, they were standardized, ensuring a consistent range for all properties. The network was then trained using a conventional $L_2$ loss established between the predicted and experimentally obtained values. All weights were initialized using the Glorot initializer \cite{Glorot2010}. All models were trained using the same hyperparameters for 60 epochs with the Adam updater \cite{Kingma2014} using a gradient decay of $0.9$, a squared gradient decay of $0.999$, batches of 32 and a constant learning rate of $1 \cdot 10^{-4}$. All trainings were conducted on an NVIDIA Geforce RTX 4090. Additional results regarding the various trainings are provided in Section 7 of the Supplementary Materials.

\subsection*{LIME}

\justify To investigate the importance of features within local regions that control mechanical properties, we have combined our prediction framework with LIME (Local Interpretable Model-Agnostic Explanations) \cite{Ribeiro2016}. While the original LIME implementation identifies features from superpixels \cite{Ribeiro2016}, we have modified LIME to employ physical, microstructure- or deformation-relevant features such as grains, annealing twins, phases, defects and deformation events. These features are identified from microstructure or deformation measurements. These features were extracted using physics-based approaches such as grain and annealing twin identification based on misorientation criteria or crystallographic orientation relationships, or through computer vision-based segmentation approaches for deformation events\cite{Bean2025a}. A large number of randomly perturbed configurations are generated by randomly modifying each feature using a perturbation operation (depending on the modality considered). The CNN model then processes these configurations, assigning a weight to each according to its similarity with the original configuration. A regression model is then established from these predictions to identify feature importance  that reveals how each feature affects the prediction statistically, as schematized by the slope $\beta$ (Fig. \ref{Figure4}(1)). For KAM, GAM, GROD and GOS, as well as for the deformation intensity, the features are replaces by zeros as perturbation operation. Analogously, to investigate the influence of the distribution of crystallographic orientations within grains, grains orientation are replaced by a constant value corresponding to the average orientation. LIME was also employed to study the importance of annealing twins. These twins were initially identified using the $\Sigma 3$ crystallographic orientation relationship and, in principle, can be replaced by the parent grain's orientation as a perturbation operation. However, identifying the parent grain may prove difficult. When two grains are tied to a CSL crystallographic orientation relationship and do not share any other similar crystallographic orientation relationship with other grains, the orientation of grain 1 can be replaced by the orientation of grain 2 (or vice versa). This results in two distinct, mutually exclusive modifications. Conversely, feature values were uniquely modified for orientation-derived maps (\textit{i.e.}, set to zero). Since removing a CSL boundary can cause a grain to adopt various crystallographic orientations, the perturbed state corresponds to any orientation that differs from the original. The grains depicted in white in Fig. \ref{Figure4}(I) have not been modified in the perturbation process; either they do not share any CSL boundaries, or their configuration prevents perturbation. Feature importance was evaluated in this study through 100 randomly perturbed samples of the considered microstructure and plasticity maps. Each sample corresponds to a full perturbed map, and the mechanical properties were predicted using a sliding window approach. Feature importance was estimated using ridge regression considering perturbed and original features with abscissa value of 0 and 1, respectively (Fig. \ref{Figure4}(C)). All the predictions were weighted using radial basis functions and cosine similarity at each position taken by the sliding window. 


\bibliography{sample} 

\begin{thebibliography}{100}
\urlstyle{rm}
\expandafter\ifx\csname url\endcsname\relax
  \def\url#1{\texttt{#1}}\fi
\expandafter\ifx\csname urlprefix\endcsname\relax\def\urlprefix{URL }\fi
\expandafter\ifx\csname doiprefix\endcsname\relax\def\doiprefix{DOI: }\fi
\providecommand{\bibinfo}[2]{#2}
\providecommand{\eprint}[2][]{\url{#2}}

\bibitem{Olson1997}
\bibinfo{author}{Olson, G.~B.}
\newblock \bibinfo{journal}{\bibinfo{title}{Computational design of hierarchically structured materials}}.
\newblock {\emph{\JournalTitle{Science}}} \textbf{\bibinfo{volume}{277}}, \bibinfo{pages}{1237–1242}, \doiprefix\url{10.1126/science.277.5330.1237} (\bibinfo{year}{1997}).

\bibitem{Miret2025}
\bibinfo{author}{Miret, S.} \& \bibinfo{author}{Krishnan, N.~A.}
\newblock \bibinfo{journal}{\bibinfo{title}{Enabling large language models for real-world materials discovery}}.
\newblock {\emph{\JournalTitle{Nature Machine Intelligence}}} \textbf{\bibinfo{volume}{7}}, \bibinfo{pages}{991--998} (\bibinfo{year}{2025}).

\bibitem{Tang2026}
\bibinfo{author}{Tang, Y.} \emph{et~al.}
\newblock \bibinfo{journal}{\bibinfo{title}{A multimodal large language model for materials science}}.
\newblock {\emph{\JournalTitle{Nature Machine Intelligence}}} \bibinfo{pages}{1--14} (\bibinfo{year}{2026}).

\bibitem{Ahlawat2026}
\bibinfo{author}{Ahlawat, D.} \emph{et~al.}
\newblock \bibinfo{journal}{\bibinfo{title}{A family of large language models for materials research with insights into model adaptability in continued pretraining}}.
\newblock {\emph{\JournalTitle{Nature Machine Intelligence}}} \bibinfo{pages}{1--14} (\bibinfo{year}{2026}).

\bibitem{Bastek2023}
\bibinfo{author}{Bastek, J.-H.} \& \bibinfo{author}{Kochmann, D.~M.}
\newblock \bibinfo{journal}{\bibinfo{title}{Inverse design of nonlinear mechanical metamaterials via video denoising diffusion models}}.
\newblock {\emph{\JournalTitle{Nature Machine Intelligence}}} \textbf{\bibinfo{volume}{5}}, \bibinfo{pages}{1466--1475} (\bibinfo{year}{2023}).

\bibitem{Igashov2024}
\bibinfo{author}{Igashov, I.} \emph{et~al.}
\newblock \bibinfo{journal}{\bibinfo{title}{Equivariant 3d-conditional diffusion model for molecular linker design}}.
\newblock {\emph{\JournalTitle{Nature Machine Intelligence}}} \textbf{\bibinfo{volume}{6}}, \bibinfo{pages}{417--427} (\bibinfo{year}{2024}).

\bibitem{Li2024}
\bibinfo{author}{Li, T.}, \bibinfo{author}{Biferale, L.}, \bibinfo{author}{Bonaccorso, F.}, \bibinfo{author}{Scarpolini, M.~A.} \& \bibinfo{author}{Buzzicotti, M.}
\newblock \bibinfo{journal}{\bibinfo{title}{Synthetic lagrangian turbulence by generative diffusion models}}.
\newblock {\emph{\JournalTitle{Nature Machine Intelligence}}} \textbf{\bibinfo{volume}{6}}, \bibinfo{pages}{393--403} (\bibinfo{year}{2024}).

\bibitem{Hafner2025}
\bibinfo{author}{Hafner, D.}, \bibinfo{author}{Pasukonis, J.}, \bibinfo{author}{Ba, J.} \& \bibinfo{author}{Lillicrap, T.}
\newblock \bibinfo{journal}{\bibinfo{title}{Mastering diverse control tasks through world models}}.
\newblock {\emph{\JournalTitle{Nature}}} \textbf{\bibinfo{volume}{640}}, \bibinfo{pages}{647–653}, \doiprefix\url{10.1038/s41586-025-08744-2} (\bibinfo{year}{2025}).

\bibitem{Xie2021}
\bibinfo{author}{Xie, X.} \emph{et~al.}
\newblock \bibinfo{journal}{\bibinfo{title}{Mechanistic data-driven prediction of as-built mechanical properties in metal additive manufacturing}}.
\newblock {\emph{\JournalTitle{npj Computational Materials}}} \textbf{\bibinfo{volume}{7}}, \bibinfo{pages}{86} (\bibinfo{year}{2021}).

\bibitem{Liu2022}
\bibinfo{author}{Liu, Y.} \emph{et~al.}
\newblock \bibinfo{journal}{\bibinfo{title}{Experimental discovery of structure--property relationships in ferroelectric materials via active learning}}.
\newblock {\emph{\JournalTitle{Nature Machine Intelligence}}} \textbf{\bibinfo{volume}{4}}, \bibinfo{pages}{341--350} (\bibinfo{year}{2022}).

\bibitem{Wang2026}
\bibinfo{author}{Wang, X.} \emph{et~al.}
\newblock \bibinfo{journal}{\bibinfo{title}{Multimodal learning with next-token prediction for large multimodal models}}.
\newblock {\emph{\JournalTitle{Nature}}} \textbf{\bibinfo{volume}{650}}, \bibinfo{pages}{327–333}, \doiprefix\url{10.1038/s41586-025-10041-x} (\bibinfo{year}{2026}).

\bibitem{Rudin2019}
\bibinfo{author}{Rudin, C.}
\newblock \bibinfo{journal}{\bibinfo{title}{Stop explaining black box machine learning models for high stakes decisions and use interpretable models instead}}.
\newblock {\emph{\JournalTitle{Nature Machine Intelligence}}} \textbf{\bibinfo{volume}{1}}, \bibinfo{pages}{206–215}, \doiprefix\url{10.1038/s42256-019-0048-x} (\bibinfo{year}{2019}).

\bibitem{Karniadakis2021}
\bibinfo{author}{Karniadakis, G.~E.} \emph{et~al.}
\newblock \bibinfo{journal}{\bibinfo{title}{Physics-informed machine learning}}.
\newblock {\emph{\JournalTitle{Nature Reviews Physics}}} \textbf{\bibinfo{volume}{3}}, \bibinfo{pages}{422–440}, \doiprefix\url{10.1038/s42254-021-00314-5} (\bibinfo{year}{2021}).

\bibitem{Wei2014}
\bibinfo{author}{Wei, Y.} \emph{et~al.}
\newblock \bibinfo{journal}{\bibinfo{title}{Evading the strength--ductility trade-off dilemma in steel through gradient hierarchical nanotwins}}.
\newblock {\emph{\JournalTitle{Nature communications}}} \textbf{\bibinfo{volume}{5}}, \bibinfo{pages}{3580} (\bibinfo{year}{2014}).

\bibitem{Pang2013}
\bibinfo{author}{Pang, J.}, \bibinfo{author}{Li, S.}, \bibinfo{author}{Wang, Z.} \& \bibinfo{author}{Zhang, Z.}
\newblock \bibinfo{journal}{\bibinfo{title}{General relation between tensile strength and fatigue strength of metallic materials}}.
\newblock {\emph{\JournalTitle{Materials Science and Engineering: A}}} \textbf{\bibinfo{volume}{564}}, \bibinfo{pages}{331--341} (\bibinfo{year}{2013}).

\bibitem{Stinville2022}
\bibinfo{author}{Stinville, J.} \emph{et~al.}
\newblock \bibinfo{journal}{\bibinfo{title}{On the origins of fatigue strength in crystalline metallic materials}}.
\newblock {\emph{\JournalTitle{Science}}} \textbf{\bibinfo{volume}{377}}, \bibinfo{pages}{1065--1071} (\bibinfo{year}{2022}).

\bibitem{Charkaluk2002}
\bibinfo{author}{Charkaluk, {\'E}.}, \bibinfo{author}{Bignonnet, A.}, \bibinfo{author}{Constantinescu, A.} \& \bibinfo{author}{Dang~Van, K.}
\newblock \bibinfo{journal}{\bibinfo{title}{Fatigue design of structures under thermomechanical loadings}}.
\newblock {\emph{\JournalTitle{Fatigue \& Fracture of Engineering Materials \& Structures}}} \textbf{\bibinfo{volume}{25}}, \bibinfo{pages}{1199--1206} (\bibinfo{year}{2002}).

\bibitem{Cassagne2025}
\bibinfo{author}{Cassagne, A.~R.}, \bibinfo{author}{Lagoudas, D.~C.} \& \bibinfo{author}{le~Graverend, J.-B.}
\newblock \bibinfo{journal}{\bibinfo{title}{A multi-scale modeling of complex thermomechanical loading paths in high-temperature shape memory alloys using a crystal-plasticity framework}}.
\newblock {\emph{\JournalTitle{International Journal of Plasticity}}} \bibinfo{pages}{104598} (\bibinfo{year}{2025}).

\bibitem{Prouteau2022}
\bibinfo{author}{Prouteau, J.}, \bibinfo{author}{Villechaise, P.}, \bibinfo{author}{Signor, L.} \& \bibinfo{author}{Cormier, J.}
\newblock \bibinfo{journal}{\bibinfo{title}{Early stages of fatigue crack initiation in relation with twin boundaries in polycrystalline ni-based superalloy ad730{\texttrademark}}}.
\newblock {\emph{\JournalTitle{Metallurgical and Materials Transactions A}}} \doiprefix\url{10.1007/s11661-022-06916-7} (\bibinfo{year}{2022}).

\bibitem{Anjaria2024}
\bibinfo{author}{Anjaria, D.} \emph{et~al.}
\newblock \bibinfo{journal}{\bibinfo{title}{Plastic deformation delocalization at cryogenic temperatures in a nickel-based superalloy}}.
\newblock {\emph{\JournalTitle{Acta Materialia}}} \textbf{\bibinfo{volume}{276}}, \bibinfo{pages}{120106}, \doiprefix\url{https://doi.org/10.1016/j.actamat.2024.120106} (\bibinfo{year}{2024}).

\bibitem{Texier2024}
\bibinfo{author}{Texier, D.} \emph{et~al.}
\newblock \bibinfo{journal}{\bibinfo{title}{Strain localization in the alloy 718 ni-based superalloy: From room temperature to 650 °c}}.
\newblock {\emph{\JournalTitle{Acta Materialia}}} \textbf{\bibinfo{volume}{268}}, \bibinfo{pages}{119759}, \doiprefix\url{https://doi.org/10.1016/j.actamat.2024.119759} (\bibinfo{year}{2024}).

\bibitem{Yvinec2026}
\bibinfo{author}{Yvinec, T.} \emph{et~al.}
\newblock \bibinfo{journal}{\bibinfo{title}{Low stress grain boundary mediated plasticity and early fracture at basal twist grain boundaries in a titanium alloy}}.
\newblock {\emph{\JournalTitle{Communications Materials}}}  (\bibinfo{year}{2026}).

\bibitem{GIANOLA2023101090}
\bibinfo{author}{Gianola, D.~S.} \emph{et~al.}
\newblock \bibinfo{journal}{\bibinfo{title}{Advances and opportunities in high-throughput small-scale mechanical testing}}.
\newblock {\emph{\JournalTitle{Current Opinion in Solid State and Materials Science}}} \textbf{\bibinfo{volume}{27}}, \bibinfo{pages}{101090}, \doiprefix\url{https://doi.org/10.1016/j.cossms.2023.101090} (\bibinfo{year}{2023}).

\bibitem{Salzbrenner2017}
\bibinfo{author}{Salzbrenner, B.~C.} \emph{et~al.}
\newblock \bibinfo{journal}{\bibinfo{title}{High-throughput stochastic tensile performance of additively manufactured stainless steel}}.
\newblock {\emph{\JournalTitle{Journal of Materials Processing Technology}}} \textbf{\bibinfo{volume}{241}}, \bibinfo{pages}{1--12}, \doiprefix\url{https://doi.org/10.1016/j.jmatprotec.2016.10.023} (\bibinfo{year}{2017}).

\bibitem{Gu2021}
\bibinfo{author}{Gu, D.} \emph{et~al.}
\newblock \bibinfo{journal}{\bibinfo{title}{Material-structure-performance integrated laser-metal additive manufacturing}}.
\newblock {\emph{\JournalTitle{Science}}} \textbf{\bibinfo{volume}{372}}, \bibinfo{pages}{eabg1487} (\bibinfo{year}{2021}).

\bibitem{Ren2022}
\bibinfo{author}{Ren, J.} \emph{et~al.}
\newblock \bibinfo{journal}{\bibinfo{title}{Strong yet ductile nanolamellar high-entropy alloys by additive manufacturing}}.
\newblock {\emph{\JournalTitle{Nature}}} \textbf{\bibinfo{volume}{608}}, \bibinfo{pages}{62--68} (\bibinfo{year}{2022}).

\bibitem{Gao2023}
\bibinfo{author}{Gao, S.} \emph{et~al.}
\newblock \bibinfo{title}{Additive manufacturing of alloys with programmable microstructure and properties, nat. commun. 14 (2023) 6752}.

\bibitem{Zhu2026}
\bibinfo{author}{Zhu, T.} \& \bibinfo{author}{Chen, W.}
\newblock \bibinfo{journal}{\bibinfo{title}{Mechanical behaviour of additively manufactured metals}}.
\newblock {\emph{\JournalTitle{Nature Materials}}} \bibinfo{pages}{1--13} (\bibinfo{year}{2026}).

\bibitem{Borg2020}
\bibinfo{author}{Borg, C. K.~H.} \emph{et~al.}
\newblock \bibinfo{journal}{\bibinfo{title}{Expanded dataset of mechanical properties and observed phases of multi-principal element alloys}}.
\newblock {\emph{\JournalTitle{Scientific Data}}} \textbf{\bibinfo{volume}{7}}, \doiprefix\url{10.1038/s41597-020-00768-9} (\bibinfo{year}{2020}).

\bibitem{Lee2021}
\bibinfo{author}{Lee, J.-W.} \emph{et~al.}
\newblock \bibinfo{journal}{\bibinfo{title}{A machine-learning-based alloy design platform that enables both forward and inverse predictions for thermo-mechanically controlled processed (tmcp) steel alloys}}.
\newblock {\emph{\JournalTitle{Scientific Reports}}} \textbf{\bibinfo{volume}{11}}, \doiprefix\url{10.1038/s41598-021-90237-z} (\bibinfo{year}{2021}).

\bibitem{Hu2022}
\bibinfo{author}{Hu, Q.-M.} \& \bibinfo{author}{Yang, R.}
\newblock \bibinfo{journal}{\bibinfo{title}{The endless search for better alloys}}.
\newblock {\emph{\JournalTitle{Science}}} \textbf{\bibinfo{volume}{378}}, \bibinfo{pages}{26--27} (\bibinfo{year}{2022}).

\bibitem{Deng2022}
\bibinfo{author}{Deng, Y.} \emph{et~al.}
\newblock \bibinfo{journal}{\bibinfo{title}{An intelligent design for ni-based superalloy based on machine learning and multi-objective optimization}}.
\newblock {\emph{\JournalTitle{Materials and Design}}} \textbf{\bibinfo{volume}{221}}, \bibinfo{pages}{110935}, \doiprefix\url{https://doi.org/10.1016/j.matdes.2022.110935} (\bibinfo{year}{2022}).

\bibitem{Gu2024}
\bibinfo{author}{Gu, Y.}, \bibinfo{author}{Stiles, C.~D.} \& \bibinfo{author}{El-Awady, J.~A.}
\newblock \bibinfo{journal}{\bibinfo{title}{A statistical perspective for predicting the strength of metals: Revisiting the hall–petch relationship using machine learning}}.
\newblock {\emph{\JournalTitle{Acta Materialia}}} \textbf{\bibinfo{volume}{266}}, \bibinfo{pages}{119631}, \doiprefix\url{https://doi.org/10.1016/j.actamat.2023.119631} (\bibinfo{year}{2024}).

\bibitem{Sohail2025}
\bibinfo{author}{Sohail, Y.} \emph{et~al.}
\newblock \bibinfo{journal}{\bibinfo{title}{Machine-learning design of ductile fenicoalta alloys with high strength}}.
\newblock {\emph{\JournalTitle{Nature}}} \textbf{\bibinfo{volume}{643}}, \bibinfo{pages}{119--124} (\bibinfo{year}{2025}).

\bibitem{Black2023}
\bibinfo{author}{Black, R.~L.} \emph{et~al.}
\newblock \bibinfo{journal}{\bibinfo{title}{High-throughput high-resolution digital image correlation measurements by multi-beam sem imaging}}.
\newblock {\emph{\JournalTitle{Experimental Mechanics}}} \textbf{\bibinfo{volume}{63}}, \bibinfo{pages}{939–953}, \doiprefix\url{10.1007/s11340-023-00961-y} (\bibinfo{year}{2023}).

\bibitem{Deronja2024}
\bibinfo{author}{DeRonja, J.}, \bibinfo{author}{Nowell, M.}, \bibinfo{author}{Wright, S.} \& \bibinfo{author}{Kacher, J.}
\newblock \bibinfo{journal}{\bibinfo{title}{Generational assessment of ebsd detectors for cross-correlation-based analysis: From scintillators to direct detection}}.
\newblock {\emph{\JournalTitle{Ultramicroscopy}}} \textbf{\bibinfo{volume}{257}}, \bibinfo{pages}{113913} (\bibinfo{year}{2024}).

\bibitem{Calvat2026}
\bibinfo{author}{Calvat, M.} \emph{et~al.}
\newblock \bibinfo{journal}{\bibinfo{title}{Plasticity encoding and mapping during elementary loading for accelerated mechanical properties prediction}}.
\newblock {\emph{\JournalTitle{Scripta Materialia}}} \textbf{\bibinfo{volume}{273}}, \bibinfo{pages}{117082} (\bibinfo{year}{2026}).

\bibitem{Pantleon2008}
\bibinfo{author}{Pantleon, W.}
\newblock \bibinfo{journal}{\bibinfo{title}{Resolving the geometrically necessary dislocation content by conventional electron backscattering diffraction}}.
\newblock {\emph{\JournalTitle{Scripta Materialia}}} \textbf{\bibinfo{volume}{58}}, \bibinfo{pages}{994--997} (\bibinfo{year}{2008}).

\bibitem{STINVILLE2020110600}
\bibinfo{author}{Stinville, J.} \emph{et~al.}
\newblock \bibinfo{journal}{\bibinfo{title}{Measurement of elastic and rotation fields during irreversible deformation using heaviside-digital image correlation}}.
\newblock {\emph{\JournalTitle{Materials Characterization}}} \textbf{\bibinfo{volume}{169}}, \bibinfo{pages}{110600}, \doiprefix\url{https://doi.org/10.1016/j.matchar.2020.110600} (\bibinfo{year}{2020}).

\bibitem{Bourdin2018}
\bibinfo{author}{Bourdin, F.} \emph{et~al.}
\newblock \bibinfo{journal}{\bibinfo{title}{Measurements of plastic localization by heaviside-digital image correlation}}.
\newblock {\emph{\JournalTitle{Acta Materialia}}} \textbf{\bibinfo{volume}{157}}, \bibinfo{pages}{307 -- 325}, \doiprefix\url{https://doi.org/10.1016/j.actamat.2018.07.013} (\bibinfo{year}{2018}).

\bibitem{Anjaria2026}
\bibinfo{author}{Anjaria, D.} \emph{et~al.}
\newblock \bibinfo{journal}{\bibinfo{title}{Dynamic plastic deformation delocalization in fcc solid solution metals}}.
\newblock {\emph{\JournalTitle{Nature Communications}}}  (\bibinfo{year}{2026}).

\bibitem{Ribeiro2016}
\bibinfo{author}{Ribeiro, M.~T.}, \bibinfo{author}{Singh, S.} \& \bibinfo{author}{Guestrin, C.}
\newblock \bibinfo{title}{" why should i trust you?" explaining the predictions of any classifier}.
\newblock In \emph{\bibinfo{booktitle}{Proceedings of the 22nd ACM SIGKDD international conference on knowledge discovery and data mining}}, \bibinfo{pages}{1135--1144} (\bibinfo{year}{2016}).

\bibitem{VERMEIJ2023118502}
\bibinfo{author}{Vermeij, T.}, \bibinfo{author}{Peerlings, R.}, \bibinfo{author}{Geers, M.} \& \bibinfo{author}{Hoefnagels, J.}
\newblock \bibinfo{journal}{\bibinfo{title}{Automated identification of slip system activity fields from digital image correlation data}}.
\newblock {\emph{\JournalTitle{Acta Materialia}}} \textbf{\bibinfo{volume}{243}}, \bibinfo{pages}{118502}, \doiprefix\url{https://doi.org/10.1016/j.actamat.2022.118502} (\bibinfo{year}{2023}).

\bibitem{HU2023103618}
\bibinfo{author}{Hu, H.}, \bibinfo{author}{Briffod, F.}, \bibinfo{author}{Shiraiwa, T.} \& \bibinfo{author}{Enoki, M.}
\newblock \bibinfo{journal}{\bibinfo{title}{Automated slip system identification and strain analysis framework using high-resolution digital image correlation data: Application to a bimodal ti-6al-4v alloy}}.
\newblock {\emph{\JournalTitle{International Journal of Plasticity}}} \textbf{\bibinfo{volume}{166}}, \bibinfo{pages}{103618}, \doiprefix\url{https://doi.org/10.1016/j.ijplas.2023.103618} (\bibinfo{year}{2023}).

\bibitem{NI2024104119}
\bibinfo{author}{Ni, R.} \emph{et~al.}
\newblock \bibinfo{journal}{\bibinfo{title}{Automated analysis framework of strain partitioning and deformation mechanisms via multimodal fusion and computer vision}}.
\newblock {\emph{\JournalTitle{International Journal of Plasticity}}} \textbf{\bibinfo{volume}{182}}, \bibinfo{pages}{104119}, \doiprefix\url{https://doi.org/10.1016/j.ijplas.2024.104119} (\bibinfo{year}{2024}).

\bibitem{Bean2025a}
\bibinfo{author}{Bean, C.} \emph{et~al.}
\newblock \bibinfo{journal}{\bibinfo{title}{Statistical analyses of plastic deformation events via computer vision: Case study of additive manufactured microstructures}}.
\newblock {\emph{\JournalTitle{Materials Characterization}}} \bibinfo{pages}{115406} (\bibinfo{year}{2025}).

\bibitem{Stein2014}
\bibinfo{author}{Stein, C.~A.} \emph{et~al.}
\newblock \bibinfo{journal}{\bibinfo{title}{Fatigue crack initiation, slip localization and twin boundaries in a nickel-based superalloy}}.
\newblock {\emph{\JournalTitle{Current Opinion in Solid State and Materials Science}}} \textbf{\bibinfo{volume}{18}}, \bibinfo{pages}{244--252} (\bibinfo{year}{2014}).

\bibitem{Zhang2008}
\bibinfo{author}{Zhang, Z.} \& \bibinfo{author}{Wang, Z.}
\newblock \bibinfo{journal}{\bibinfo{title}{Grain boundary effects on cyclic deformation and fatigue damage}}.
\newblock {\emph{\JournalTitle{Progress in Materials Science}}} \textbf{\bibinfo{volume}{53}}, \bibinfo{pages}{1025--1099} (\bibinfo{year}{2008}).

\bibitem{Meric1991}
\bibinfo{author}{Me{\'{}}~ric, L.}, \bibinfo{author}{Poubanne, P.} \& \bibinfo{author}{Cailletaud, G.}
\newblock \bibinfo{journal}{\bibinfo{title}{Single crystal modeling for structural calculations: part 1—model presentation}}.
\newblock {\emph{\JournalTitle{Journal of Engineering materials and Technology}}}  (\bibinfo{year}{1991}).

\bibitem{Berbenni2007}
\bibinfo{author}{Berbenni, S.}, \bibinfo{author}{Favier, V.} \& \bibinfo{author}{Berveiller, M.}
\newblock \bibinfo{journal}{\bibinfo{title}{Impact of the grain size distribution on the yield stress of heterogeneous materials}}.
\newblock {\emph{\JournalTitle{International Journal of Plasticity}}} \textbf{\bibinfo{volume}{23}}, \bibinfo{pages}{114--142} (\bibinfo{year}{2007}).

\bibitem{Zhou2010}
\bibinfo{author}{Zhou, C.}, \bibinfo{author}{Biner, S.~B.} \& \bibinfo{author}{LeSar, R.}
\newblock \bibinfo{journal}{\bibinfo{title}{Discrete dislocation dynamics simulations of plasticity at small scales}}.
\newblock {\emph{\JournalTitle{Acta Materialia}}} \textbf{\bibinfo{volume}{58}}, \bibinfo{pages}{1565--1577} (\bibinfo{year}{2010}).

\bibitem{Zepeda2017}
\bibinfo{author}{Zepeda-Ruiz, L.~A.}, \bibinfo{author}{Stukowski, A.}, \bibinfo{author}{Oppelstrup, T.} \& \bibinfo{author}{Bulatov, V.~V.}
\newblock \bibinfo{journal}{\bibinfo{title}{Probing the limits of metal plasticity with molecular dynamics simulations}}.
\newblock {\emph{\JournalTitle{Nature}}} \textbf{\bibinfo{volume}{550}}, \bibinfo{pages}{492--495} (\bibinfo{year}{2017}).

\bibitem{Cormier2010}
\bibinfo{author}{Cormier, J.} \& \bibinfo{author}{Cailletaud, G.}
\newblock \bibinfo{journal}{\bibinfo{title}{Constitutive modeling of the creep behavior of single crystal superalloys under non-isothermal conditions inducing phase transformations}}.
\newblock {\emph{\JournalTitle{Materials Science and Engineering: A}}} \textbf{\bibinfo{volume}{527}}, \bibinfo{pages}{6300--6312} (\bibinfo{year}{2010}).

\bibitem{Castelluccio2017}
\bibinfo{author}{Castelluccio, G.~M.} \& \bibinfo{author}{McDowell, D.~L.}
\newblock \bibinfo{journal}{\bibinfo{title}{Mesoscale cyclic crystal plasticity with dislocation substructures}}.
\newblock {\emph{\JournalTitle{International Journal of Plasticity}}} \textbf{\bibinfo{volume}{98}}, \bibinfo{pages}{1--26} (\bibinfo{year}{2017}).

\bibitem{Zhu2023}
\bibinfo{author}{Zhu, D.}, \bibinfo{author}{Zhang, W.}, \bibinfo{author}{Ding, Z.} \& \bibinfo{author}{Kim, J.}
\newblock \bibinfo{journal}{\bibinfo{title}{Investigation of crack propagation driving force based on crystal plasticity and cyclic j-integral}}.
\newblock {\emph{\JournalTitle{Engineering Fracture Mechanics}}} \textbf{\bibinfo{volume}{289}}, \bibinfo{pages}{109362} (\bibinfo{year}{2023}).

\bibitem{Liu2017}
\bibinfo{author}{Liu, J.}, \bibinfo{author}{Xiong, W.}, \bibinfo{author}{Behera, A.}, \bibinfo{author}{Thompson, S.} \& \bibinfo{author}{To, A.~C.}
\newblock \bibinfo{journal}{\bibinfo{title}{Mean-field polycrystal plasticity modeling with grain size and shape effects for laser additive manufactured fcc metals}}.
\newblock {\emph{\JournalTitle{International Journal of Solids and Structures}}} \textbf{\bibinfo{volume}{112}}, \bibinfo{pages}{35--42} (\bibinfo{year}{2017}).

\bibitem{Healy2014}
\bibinfo{author}{Healy, C.~J.} \& \bibinfo{author}{Ackland, G.~J.}
\newblock \bibinfo{journal}{\bibinfo{title}{Molecular dynamics simulations of compression--tension asymmetry in plasticity of fe nanopillars}}.
\newblock {\emph{\JournalTitle{Acta Materialia}}} \textbf{\bibinfo{volume}{70}}, \bibinfo{pages}{105--112} (\bibinfo{year}{2014}).

\bibitem{Calvat2025a}
\bibinfo{author}{Calvat, M.}, \bibinfo{author}{Keller, C.} \& \bibinfo{author}{Barbe, F.}
\newblock \bibinfo{journal}{\bibinfo{title}{Micromechanical analysis of a unimodal and a bimodal polycrystal with paired microstructures of ultrafine grains, 2d \& 3d}}.
\newblock {\emph{\JournalTitle{European Journal of Mechanics-A/Solids}}} \textbf{\bibinfo{volume}{109}}, \bibinfo{pages}{105434} (\bibinfo{year}{2025}).

\bibitem{Aubin2003}
\bibinfo{author}{Aubin, V.}, \bibinfo{author}{Quaegebeur, P.} \& \bibinfo{author}{Degallaix, S.}
\newblock \bibinfo{journal}{\bibinfo{title}{Cyclic plasticity of a duplex stainless steel under non-proportional loading}}.
\newblock {\emph{\JournalTitle{Materials Science and Engineering: A}}} \textbf{\bibinfo{volume}{346}}, \bibinfo{pages}{208--215} (\bibinfo{year}{2003}).

\bibitem{Kumar2016}
\bibinfo{author}{Kumar, S.~S.}, \bibinfo{author}{Raghu, T.}, \bibinfo{author}{Bhattacharjee, P.~P.}, \bibinfo{author}{Rao, G.~A.} \& \bibinfo{author}{Borah, U.}
\newblock \bibinfo{journal}{\bibinfo{title}{Strain rate dependent microstructural evolution during hot deformation of a hot isostatically processed nickel base superalloy}}.
\newblock {\emph{\JournalTitle{Journal of Alloys and Compounds}}} \textbf{\bibinfo{volume}{681}}, \bibinfo{pages}{28--42} (\bibinfo{year}{2016}).

\bibitem{HULL2011171}
\bibinfo{author}{Hull, D.} \& \bibinfo{author}{Bacon, D.}
\newblock \bibinfo{title}{Chapter 9 - dislocation arrays and crystal boundaries}.
\newblock In \bibinfo{editor}{Hull, D.} \& \bibinfo{editor}{Bacon, D.} (eds.) \emph{\bibinfo{booktitle}{Introduction to Dislocations (Fifth Edition)}}, \bibinfo{pages}{171--204}, \doiprefix\url{https://doi.org/10.1016/B978-0-08-096672-4.00009-8} (\bibinfo{publisher}{Butterworth-Heinemann}, \bibinfo{address}{Oxford}, \bibinfo{year}{2011}), \bibinfo{edition}{fifth edition} edn.

\bibitem{Andani2020}
\bibinfo{author}{Andani, M.~T.} \emph{et~al.}
\newblock \bibinfo{journal}{\bibinfo{title}{A quantitative study of stress fields ahead of a slip band blocked by a grain boundary in unalloyed magnesium}}.
\newblock {\emph{\JournalTitle{Scientific Reports}}} \textbf{\bibinfo{volume}{10}}, \doiprefix\url{10.1038/s41598-020-59684-y} (\bibinfo{year}{2020}).

\bibitem{ANDANI2022117613}
\bibinfo{author}{Andani, M.~T.}, \bibinfo{author}{Lakshmanan, A.}, \bibinfo{author}{Sundararaghavan, V.}, \bibinfo{author}{Allison, J.} \& \bibinfo{author}{Misra, A.}
\newblock \bibinfo{journal}{\bibinfo{title}{Estimation of micro-hall-petch coefficients for prismatic slip system in mg-4al as a function of grain boundary parameters}}.
\newblock {\emph{\JournalTitle{Acta Materialia}}} \textbf{\bibinfo{volume}{226}}, \bibinfo{pages}{117613}, \doiprefix\url{https://doi.org/10.1016/j.actamat.2021.117613} (\bibinfo{year}{2022}).

\bibitem{doi:https://doi.org/10.1002/9781119296126.ch173}
\bibinfo{author}{Guo, Y.}, \bibinfo{author}{Britton, T.~B.} \& \bibinfo{author}{Wilkinson, A.~J.}
\newblock \emph{\bibinfo{title}{Stress Concentrations, Slip Bands and Grain Boundaries In Commercially Pure Titanium}}, chap. \bibinfo{chapter}{173}, \bibinfo{pages}{1017--1021} (\bibinfo{publisher}{John Wiley And Sons, Ltd}, \bibinfo{year}{2016}).
\newblock \eprint{https://onlinelibrary.wiley.com/doi/pdf/10.1002/9781119296126.ch173}.

\bibitem{doi:10.1126/sciadv.abo5735}
\bibinfo{author}{Edwards, T. E.~J.}, \bibinfo{author}{Maeder, X.}, \bibinfo{author}{Ast, J.}, \bibinfo{author}{Berger, L.} \& \bibinfo{author}{Michler, J.}
\newblock \bibinfo{journal}{\bibinfo{title}{Mapping pure plastic strains against locally applied stress: Revealing toughening plasticity}}.
\newblock {\emph{\JournalTitle{Science Advances}}} \textbf{\bibinfo{volume}{8}}, \bibinfo{pages}{eabo5735}, \doiprefix\url{10.1126/sciadv.abo5735} (\bibinfo{year}{2022}).
\newblock \eprint{https://www.science.org/doi/pdf/10.1126/sciadv.abo5735}.

\bibitem{CLEVERINGA1999837}
\bibinfo{author}{Cleveringa, H.}, \bibinfo{author}{{Van der Giessen}, E.} \& \bibinfo{author}{Needleman, A.}
\newblock \bibinfo{journal}{\bibinfo{title}{A discrete dislocation analysis of bending}}.
\newblock {\emph{\JournalTitle{International Journal of Plasticity}}} \textbf{\bibinfo{volume}{15}}, \bibinfo{pages}{837--868}, \doiprefix\url{https://doi.org/10.1016/S0749-6419(99)00013-3} (\bibinfo{year}{1999}).

\bibitem{CHAKRAVARTHY2010625}
\bibinfo{author}{Chakravarthy, S.~S.} \& \bibinfo{author}{Curtin, W.}
\newblock \bibinfo{journal}{\bibinfo{title}{Effect of source and obstacle strengths on yield stress: A discrete dislocation study}}.
\newblock {\emph{\JournalTitle{Journal of the Mechanics and Physics of Solids}}} \textbf{\bibinfo{volume}{58}}, \bibinfo{pages}{625 -- 635}, \doiprefix\url{https://doi.org/10.1016/j.jmps.2010.03.004} (\bibinfo{year}{2010}).

\bibitem{refId0}
\bibinfo{author}{{Cleveringa, H.}}, \bibinfo{author}{{Van der Giessen, E.}} \& \bibinfo{author}{{Needleman, A.}}
\newblock \bibinfo{journal}{\bibinfo{title}{Discrete dislocation simulations and size dependent hardening in single slip}}.
\newblock {\emph{\JournalTitle{J. Phys. IV France}}} \textbf{\bibinfo{volume}{08}}, \bibinfo{pages}{Pr4--83--Pr4--92}, \doiprefix\url{10.1051/jp4:1998410} (\bibinfo{year}{1998}).

\bibitem{Hall1951}
\bibinfo{author}{Hall, E.}
\newblock \bibinfo{journal}{\bibinfo{title}{The deformation and ageing of mild steel: Iii discussion of results}}.
\newblock {\emph{\JournalTitle{Proceedings of the Physical Society. Section B}}} \textbf{\bibinfo{volume}{64}}, \bibinfo{pages}{747} (\bibinfo{year}{1951}).

\bibitem{Petch1953}
\bibinfo{author}{Petch, N.~J.}
\newblock \bibinfo{journal}{\bibinfo{title}{The cleavage strength of polycrystals}}.
\newblock {\emph{\JournalTitle{J. Iron Steel Inst.}}} \textbf{\bibinfo{volume}{174}}, \bibinfo{pages}{25--28} (\bibinfo{year}{1953}).

\bibitem{Salem2006}
\bibinfo{author}{Salem, A.~A.}, \bibinfo{author}{Kalidindi, S.}, \bibinfo{author}{Doherty, R.} \& \bibinfo{author}{Semiatin, S.}
\newblock \bibinfo{journal}{\bibinfo{title}{Strain hardening due to deformation twinning in $\alpha$-titanium: Mechanisms}}.
\newblock {\emph{\JournalTitle{Metallurgical and Materials Transactions A}}} \textbf{\bibinfo{volume}{37}}, \bibinfo{pages}{259--268} (\bibinfo{year}{2006}).

\bibitem{Bean2022}
\bibinfo{author}{Bean, C.} \emph{et~al.}
\newblock \bibinfo{journal}{\bibinfo{title}{Heterogeneous slip localization in an additively manufactured 316l stainless steel}}.
\newblock {\emph{\JournalTitle{International Journal of Plasticity}}} \textbf{\bibinfo{volume}{159}}, \bibinfo{pages}{103436} (\bibinfo{year}{2022}).

\bibitem{Yvinec2026b}
\bibinfo{author}{Yvinec, T.}, \bibinfo{author}{Valle, V.}, \bibinfo{author}{Hamon, F.} \& \bibinfo{author}{H{\'e}mery, S.}
\newblock \bibinfo{journal}{\bibinfo{title}{A criterion for slip transfer in ti-6al-4 v determined using hr-dic: toward a quantitative assessment of slip length in microtextured regions}}.
\newblock {\emph{\JournalTitle{Materials \& Design}}} \bibinfo{pages}{115974} (\bibinfo{year}{2026}).

\bibitem{Mughrabi2009}
\bibinfo{author}{Mughrabi, H.}
\newblock \bibinfo{journal}{\bibinfo{title}{Cyclic slip irreversibilities and the evolution of fatigue damage}}.
\newblock {\emph{\JournalTitle{Metallurgical and Materials Transactions A}}} \textbf{\bibinfo{volume}{40}}, \bibinfo{pages}{1257--1279}, \doiprefix\url{10.1007/s11661-009-9839-8} (\bibinfo{year}{2009}).

\bibitem{STINVILLE2020172}
\bibinfo{author}{Stinville, J.} \emph{et~al.}
\newblock \bibinfo{journal}{\bibinfo{title}{Direct measurements of slip irreversibility in a nickel-based superalloy using high resolution digital image correlation}}.
\newblock {\emph{\JournalTitle{Acta Materialia}}} \textbf{\bibinfo{volume}{186}}, \bibinfo{pages}{172--189}, \doiprefix\url{https://doi.org/10.1016/j.actamat.2019.12.009} (\bibinfo{year}{2020}).

\bibitem{Pan2026}
\bibinfo{author}{Pan, Q.} \& \bibinfo{author}{Lu, L.}
\newblock \bibinfo{journal}{\bibinfo{title}{Fatigue in metals and alloys}}.
\newblock {\emph{\JournalTitle{Nature Materials}}} \textbf{\bibinfo{volume}{25}}, \bibinfo{pages}{357--365} (\bibinfo{year}{2026}).

\bibitem{STINVILLE201816}
\bibinfo{author}{Stinville, J.~C.} \emph{et~al.}
\newblock \bibinfo{journal}{\bibinfo{title}{Fatigue deformation in a polycrystalline nickel base superalloy at intermediate and high temperature: Competing failure modes}}.
\newblock {\emph{\JournalTitle{Acta Materialia}}} \textbf{\bibinfo{volume}{152}}, \bibinfo{pages}{16 -- 33}, \doiprefix\url{https://doi.org/10.1016/j.actamat.2018.03.035} (\bibinfo{year}{2018}).

\bibitem{annurev:/content/journals/10.1146/annurev-matsci-080921-102621}
\bibinfo{author}{Stinville, J.} \emph{et~al.}
\newblock \bibinfo{journal}{\bibinfo{title}{Insights into plastic localization by crystallographic slip from emerging experimental and numerical approaches}}.
\newblock {\emph{\JournalTitle{Annual Review of Materials Research}}} \textbf{\bibinfo{volume}{53}}, \bibinfo{pages}{275--317}, \doiprefix\url{https://doi.org/10.1146/annurev-matsci-080921-102621} (\bibinfo{year}{2023}).

\bibitem{Vermeij2023}
\bibinfo{author}{Vermeij, T.}, \bibinfo{author}{Peerlings, R.}, \bibinfo{author}{Geers, M.} \& \bibinfo{author}{Hoefnagels, J.}
\newblock \bibinfo{journal}{\bibinfo{title}{Automated identification of slip system activity fields from digital image correlation data}}.
\newblock {\emph{\JournalTitle{Acta Materialia}}} \textbf{\bibinfo{volume}{243}}, \bibinfo{pages}{118502}, \doiprefix\url{https://doi.org/10.1016/j.actamat.2022.118502} (\bibinfo{year}{2023}).

\bibitem{Mornout2025}
\bibinfo{author}{Mornout, C.}, \bibinfo{author}{Slokker, G.}, \bibinfo{author}{Vermeij, T.}, \bibinfo{author}{König, D.} \& \bibinfo{author}{Hoefnagels, J.}
\newblock \bibinfo{journal}{\bibinfo{title}{Slide: Automated identification and quantification of grain boundary sliding and opening in 3d}}.
\newblock {\emph{\JournalTitle{Scripta Materialia}}} \textbf{\bibinfo{volume}{268}}, \bibinfo{pages}{116861}, \doiprefix\url{https://doi.org/10.1016/j.scriptamat.2025.116861} (\bibinfo{year}{2025}).

\bibitem{POLAK20031}
\bibinfo{author}{Polák, J.}
\newblock \bibinfo{title}{4.01 - cyclic deformation, crack initiation, and low-cycle fatigue}.
\newblock In \bibinfo{editor}{Milne, I.}, \bibinfo{editor}{Ritchie, R.} \& \bibinfo{editor}{Karihaloo, B.} (eds.) \emph{\bibinfo{booktitle}{Comprehensive Structural Integrity}}, \bibinfo{pages}{1 -- 39}, \doiprefix\url{https://doi.org/10.1016/B0-08-043749-4/04060-X} (\bibinfo{publisher}{Pergamon}, \bibinfo{address}{Oxford}, \bibinfo{year}{2003}).

\bibitem{MUGHRABI20131197}
\bibinfo{author}{Mughrabi, H.}
\newblock \bibinfo{journal}{\bibinfo{title}{Cyclic slip irreversibility and fatigue life: A microstructure-based analysis}}.
\newblock {\emph{\JournalTitle{Acta Materialia}}} \textbf{\bibinfo{volume}{61}}, \bibinfo{pages}{1197 -- 1203}, \doiprefix\url{https://doi.org/10.1016/j.actamat.2012.10.029} (\bibinfo{year}{2013}).

\bibitem{CHARPAGNE2021117037}
\bibinfo{author}{Charpagne, M.} \emph{et~al.}
\newblock \bibinfo{journal}{\bibinfo{title}{Slip localization in inconel 718: A three-dimensional and statistical perspective}}.
\newblock {\emph{\JournalTitle{Acta Materialia}}} \textbf{\bibinfo{volume}{215}}, \bibinfo{pages}{117037}, \doiprefix\url{https://doi.org/10.1016/j.actamat.2021.117037} (\bibinfo{year}{2021}).

\bibitem{HONG19901581}
\bibinfo{author}{Hong, S.~I.} \& \bibinfo{author}{Laird, C.}
\newblock \bibinfo{journal}{\bibinfo{title}{Mechanisms of slip mode modification in f.c.c. solid solutions}}.
\newblock {\emph{\JournalTitle{Acta Metallurgica et Materialia}}} \textbf{\bibinfo{volume}{38}}, \bibinfo{pages}{1581--1594}, \doiprefix\url{https://doi.org/10.1016/0956-7151(90)90126-2} (\bibinfo{year}{1990}).

\bibitem{CHARPAGNE2022143291}
\bibinfo{author}{Charpagne, M.}, \bibinfo{author}{Stinville, J.}, \bibinfo{author}{Wang, F.}, \bibinfo{author}{Philips, N.} \& \bibinfo{author}{Pollock, T.}
\newblock \bibinfo{journal}{\bibinfo{title}{Orientation dependent plastic localization in the refractory high entropy alloy hfnbtatizr at room temperature}}.
\newblock {\emph{\JournalTitle{Materials Science and Engineering: A}}} \textbf{\bibinfo{volume}{848}}, \bibinfo{pages}{143291}, \doiprefix\url{https://doi.org/10.1016/j.msea.2022.143291} (\bibinfo{year}{2022}).

\bibitem{Goodfellow2016}
\bibinfo{author}{Goodfellow, I.}, \bibinfo{author}{Bengio, Y.} \& \bibinfo{author}{Courville, A.}
\newblock \emph{\bibinfo{title}{Deep Learning}} (\bibinfo{publisher}{MIT Press}, \bibinfo{year}{2016}).
\newblock \bibinfo{note}{\url{http://www.deeplearningbook.org}}.

\bibitem{Pei2021}
\bibinfo{author}{Pei, Z.} \emph{et~al.}
\newblock \bibinfo{journal}{\bibinfo{title}{Machine-learning microstructure for inverse material design}}.
\newblock {\emph{\JournalTitle{Advanced Science}}} \textbf{\bibinfo{volume}{8}}, \bibinfo{pages}{2101207} (\bibinfo{year}{2021}).

\bibitem{Calvat2025b}
\bibinfo{author}{Calvat, M.} \emph{et~al.}
\newblock \bibinfo{journal}{\bibinfo{title}{Learning metal microstructural heterogeneity through spatial mapping of diffraction latent space features}}.
\newblock {\emph{\JournalTitle{npj Computational Materials}}} \textbf{\bibinfo{volume}{11}}, \bibinfo{pages}{284} (\bibinfo{year}{2025}).

\bibitem{Islam2026}
\bibinfo{author}{Islam, M.~M.}, \bibinfo{author}{Belal, T.}, \bibinfo{author}{Anik, M. A. H.~C.} \& \bibinfo{author}{Sharif, A.}
\newblock \bibinfo{journal}{\bibinfo{title}{Predicting materials using variational autoencoders: A systematic literature review}}.
\newblock {\emph{\JournalTitle{Archives of Computational Methods in Engineering}}} \bibinfo{pages}{1--52} (\bibinfo{year}{2026}).

\bibitem{Mcinnes2018}
\bibinfo{author}{McInnes, L.}, \bibinfo{author}{Healy, J.} \& \bibinfo{author}{Melville, J.}
\newblock \bibinfo{journal}{\bibinfo{title}{Umap: Uniform manifold approximation and projection for dimension reduction}}.
\newblock {\emph{\JournalTitle{arXiv preprint arXiv:1802.03426}}}  (\bibinfo{year}{2018}).

\bibitem{Becht2019}
\bibinfo{author}{Becht, E.} \emph{et~al.}
\newblock \bibinfo{journal}{\bibinfo{title}{Dimensionality reduction for visualizing single-cell data using umap}}.
\newblock {\emph{\JournalTitle{Nature biotechnology}}} \textbf{\bibinfo{volume}{37}}, \bibinfo{pages}{38--44} (\bibinfo{year}{2019}).

\bibitem{Bobbio2017}
\bibinfo{author}{Bobbio, L.~D.} \emph{et~al.}
\newblock \bibinfo{journal}{\bibinfo{title}{Additive manufacturing of a functionally graded material from ti-6al-4v to invar: Experimental characterization and thermodynamic calculations}}.
\newblock {\emph{\JournalTitle{Acta Materialia}}} \textbf{\bibinfo{volume}{127}}, \bibinfo{pages}{133--142} (\bibinfo{year}{2017}).

\bibitem{BEAN2025114115}
\bibinfo{author}{Bean, C.} \emph{et~al.}
\newblock \bibinfo{journal}{\bibinfo{title}{Accelerated fatigue strength prediction via additive manufactured functionally graded materials and high-throughput plasticity quantification}}.
\newblock {\emph{\JournalTitle{Materials \& Design}}} \textbf{\bibinfo{volume}{256}}, \bibinfo{pages}{114115}, \doiprefix\url{https://doi.org/10.1016/j.matdes.2025.114115} (\bibinfo{year}{2025}).

\bibitem{Jullien2025}
\bibinfo{author}{Jullien, M.}, \bibinfo{author}{Legros, M.}, \bibinfo{author}{Calvat, M.}, \bibinfo{author}{Stinville, J.-C.} \& \bibinfo{author}{Texier, D.}
\newblock \bibinfo{journal}{\bibinfo{title}{Quantifying the impact of oxidation on the mechanical properties of alloy 718 using local mechanical testing techniques}}.
\newblock {\emph{\JournalTitle{Materials \& Design}}} \bibinfo{pages}{114669} (\bibinfo{year}{2025}).

\bibitem{Gu2018}
\bibinfo{author}{Gu, G.~X.} \& \bibinfo{author}{Buehler, M.~J.}
\newblock \bibinfo{journal}{\bibinfo{title}{Tunable mechanical properties through texture control of polycrystalline additively manufactured materials using adjoint-based gradient optimization: Gx gu, mj buehler}}.
\newblock {\emph{\JournalTitle{Acta Mechanica}}} \textbf{\bibinfo{volume}{229}}, \bibinfo{pages}{4033--4044} (\bibinfo{year}{2018}).

\bibitem{Keller2021}
\bibinfo{author}{Keller, C.}, \bibinfo{author}{Mokhtari, M.}, \bibinfo{author}{Vieille, B.}, \bibinfo{author}{Briatta, H.} \& \bibinfo{author}{Bernard, P.}
\newblock \bibinfo{journal}{\bibinfo{title}{Influence of a rescanning strategy with different laser powers on the microstructure and mechanical properties of hastelloy x elaborated by powder bed fusion}}.
\newblock {\emph{\JournalTitle{Materials Science and Engineering: A}}} \textbf{\bibinfo{volume}{803}}, \bibinfo{pages}{140474} (\bibinfo{year}{2021}).

\bibitem{Kammers_2013a}
\bibinfo{author}{Kammers, A.} \& \bibinfo{author}{Daly, S.}
\newblock \bibinfo{journal}{\bibinfo{title}{Self-assembled nanoparticle surface patterning for improved digital image correlation in a scanning electron microscope}}.
\newblock {\emph{\JournalTitle{Experimental Mechanics}}} \textbf{\bibinfo{volume}{53}}, \bibinfo{pages}{1333--1341}, \doiprefix\url{10.1007/s11340-013-9734-5} (\bibinfo{year}{2013}).

\bibitem{Montgomery2019}
\bibinfo{author}{Montgomery, C.}, \bibinfo{author}{Koohbor, B.} \& \bibinfo{author}{Sottos, N.}
\newblock \bibinfo{journal}{\bibinfo{title}{A robust patterning technique for electron microscopy-based digital image correlation at sub-micron resolutions}}.
\newblock {\emph{\JournalTitle{Experimental Mechanics}}} \textbf{\bibinfo{volume}{59}}, \bibinfo{pages}{1063--1073}, \doiprefix\url{10.1007/s11340-019-00487-2} (\bibinfo{year}{2019}).

\bibitem{Besson2009}
\bibinfo{author}{Besson, J.}, \bibinfo{author}{Cailletaud, G.}, \bibinfo{author}{Chaboche, J.-L.} \& \bibinfo{author}{Forest, S.}
\newblock \emph{\bibinfo{title}{Non-linear mechanics of materials}}, vol. \bibinfo{volume}{167} (\bibinfo{publisher}{Springer Science \& Business Media}, \bibinfo{year}{2009}).

\bibitem{Kammers_2013b}
\bibinfo{author}{Kammers, A.} \& \bibinfo{author}{Daly, S.}
\newblock \bibinfo{journal}{\bibinfo{title}{Digital image correlation under scanning electron microscopy: Methodology and validation}}.
\newblock {\emph{\JournalTitle{Experimental Mechanics}}} \textbf{\bibinfo{volume}{53}}, \bibinfo{pages}{1743--1761}, \doiprefix\url{10.1007/s11340-013-9782-x} (\bibinfo{year}{2013}).

\bibitem{Stinville_2015a}
\bibinfo{author}{Stinville, J.} \emph{et~al.}
\newblock \bibinfo{journal}{\bibinfo{title}{Sub-grain scale digital image correlation by electron microscopy for polycrystalline materials during elastic and plastic deformation}}.
\newblock {\emph{\JournalTitle{Experimental Mechanics}}} \bibinfo{pages}{1--20}, \doiprefix\url{10.1007/s11340-015-0083-4} (\bibinfo{year}{2015}).

\bibitem{Mello2017}
\bibinfo{author}{Mello, A.~W.} \emph{et~al.}
\newblock \bibinfo{journal}{\bibinfo{title}{Distortion correction protocol for digital image correlation after scanning electron microscopy: Emphasis on long duration and ex-situ experiments}}.
\newblock {\emph{\JournalTitle{Experimental Mechanics}}} \doiprefix\url{10.1007/s11340-017-0303-1} (\bibinfo{year}{2017}).

\bibitem{Valle2017}
\bibinfo{author}{Valle, V.} \& \bibinfo{author}{Hedan, S.}
\newblock \bibinfo{journal}{\bibinfo{title}{{Crack Analysis in Mudbricks under Compression Using Specific Development of Stereo-Digital Image Correlation}}}.
\newblock {\emph{\JournalTitle{Experimental Mechanics}}} \doiprefix\url{10.1007/s11340-017-0363-2} (\bibinfo{year}{2017}).

\bibitem{CVAT2022}
\bibinfo{author}{{CVAT.ai Corporation}}.
\newblock \bibinfo{title}{Computer vision annotation tool (cvat)} (\bibinfo{year}{2022}).

\bibitem{Bachmann2010}
\bibinfo{author}{Bachmann, F.}, \bibinfo{author}{Hielscher, R.} \& \bibinfo{author}{Schaeben, H.}
\newblock \bibinfo{journal}{\bibinfo{title}{Texture analysis with mtex--free and open source software toolbox}}.
\newblock {\emph{\JournalTitle{Solid state phenomena}}} \textbf{\bibinfo{volume}{160}}, \bibinfo{pages}{63--68} (\bibinfo{year}{2010}).

\bibitem{Bergmann2015}
\bibinfo{author}{Bergmann, R.}, \bibinfo{author}{Chan, R.~H.}, \bibinfo{author}{Hielscher, R.}, \bibinfo{author}{Persch, J.} \& \bibinfo{author}{Steidl, G.}
\newblock \bibinfo{journal}{\bibinfo{title}{Restoration of manifold-valued images by half-quadratic minimization}}.
\newblock {\emph{\JournalTitle{arXiv preprint arXiv:1505.07029}}}  (\bibinfo{year}{2015}).

\bibitem{Glorot2010}
\bibinfo{author}{Glorot, X.} \& \bibinfo{author}{Bengio, Y.}
\newblock \bibinfo{title}{Understanding the difficulty of training deep feedforward neural networks}.
\newblock In \emph{\bibinfo{booktitle}{Proceedings of the thirteenth international conference on artificial intelligence and statistics}}, \bibinfo{pages}{249--256} (\bibinfo{organization}{JMLR Workshop and Conference Proceedings}, \bibinfo{year}{2010}).

\bibitem{Kingma2014}
\bibinfo{author}{Kingma, D.~P.}
\newblock \bibinfo{journal}{\bibinfo{title}{Adam: A method for stochastic optimization}}.
\newblock {\emph{\JournalTitle{arXiv preprint arXiv:1412.6980}}}  (\bibinfo{year}{2014}).

\bibitem{Smith2023}
\bibinfo{author}{Smith, T.~M.} \emph{et~al.}
\newblock \bibinfo{journal}{\bibinfo{title}{A 3d printable alloy designed for extreme environments}}.
\newblock {\emph{\JournalTitle{Nature}}} \textbf{\bibinfo{volume}{617}}, \bibinfo{pages}{513--518} (\bibinfo{year}{2023}).

\bibitem{He2016}
\bibinfo{author}{He, K.}, \bibinfo{author}{Zhang, X.}, \bibinfo{author}{Ren, S.} \& \bibinfo{author}{Sun, J.}
\newblock \bibinfo{title}{Deep residual learning for image recognition}.
\newblock In \emph{\bibinfo{booktitle}{Proceedings of the IEEE conference on computer vision and pattern recognition}}, \bibinfo{pages}{770--778} (\bibinfo{year}{2016}).

\bibitem{Lu2019}
\bibinfo{author}{Lu, L.}, \bibinfo{author}{Shin, Y.}, \bibinfo{author}{Su, Y.} \& \bibinfo{author}{Karniadakis, G.~E.}
\newblock \bibinfo{journal}{\bibinfo{title}{Dying relu and initialization: Theory and numerical examples}}.
\newblock {\emph{\JournalTitle{arXiv preprint arXiv:1903.06733}}}  (\bibinfo{year}{2019}).

\end{thebibliography}

\noindent\textbf{Acknowledgments} \\

\justify M.C., C.B., H.W., K.V. and J.C.S. are grateful for financial support from the Defense Advanced Research Projects Agency (DARPA - HR001124C0394). C.B., D.A. and J.C.S. acknowledge the NSF (award \#2338346) for financial support. This work was carried out in the Materials Research Laboratory Central Research Facilities, University of Illinois. EBSD measurements were conducted at the AFRL/RX Materials Characterization Facility (MCF), which is supported by Air Force contract FA2394-23-C-B028. G.S. was supported by the DARPA METALS program through the on-site contract (FA8650-20-F-5212) at Air Force Research Laboratory, WPAFB, OH. A.M.B acknowledge the NSF (award CMMI \#2334678) for financial support. ATI, Noah Philips and Ming Li are acknowledged for providing the W\_Invar material. Carpenter Technology is acknowledged for providing the W\_RX\_In718 material. Acknowledgment is extended to Valery Valle for providing the Heaviside-DIC code. \\

\noindent\textbf{CRediT authorship contribution statement} \\

\justify \textbf{M.C.}:  Conceptualization, Data curation, Formal analysis, Investigation, Methodology, Writing – original draft, Writing – review \& editing. \textbf{D.A.}: Investigation, Data curation, Writing – review \& editing. \textbf{C.B.}: Methodology, Data curation. \textbf{G.S.}: Resources \textbf{H.W.}: Resources. \textbf{P.G.}: Resources. \textbf{T.M.S.}: Resources. \textbf{A.M.B.}: Resources. \textbf{G.D.}: Resources. \textbf{K.V.}: Resources, Writing – review \& editing. \textbf{M.B.}: Methodology, Writing – review \& editing. \textbf{J.C.S.}: Conceptualization, Funding acquisition, Methodology, Project administration, Resources, Supervision, Writing – original draft, Writing – review \& editing. \\

\noindent\textbf{Competing interests} \\

\justify The authors declare that they have no known competing financial interests or personal relationships that could have appeared to influence the work reported in this paper. \\

\noindent\textbf{Code availability} \\

\justify The architectures used in this study were implemented on MATLAB and the associated code will be made available upon request to the corresponding author. \\

\newpage

\thispagestyle{empty}

\begin{large}
    \noindent \textbf{SUPPLEMENTARY MATERIALS}
\end{large}

\renewcommand{\thefigure}{S\arabic{figure}}
\renewcommand{\thetable}{S\arabic{table}}
\setcounter{figure}{0}

\section{Materials}

\justify In this study, a total of 10 FCC alloys were characterized including Invar, Nickel 600, Steel 330, a Copper-Nickel-Tin alloy (Cu77NiSn), stainless steel 316L (Steel316), a Cobalt-based alloy (Co76A), three variants of the Nickel-based superalloy Inconel 718 (In718) and an ODS Nickel-based alloy (GRX-810)\cite{Smith2023}. W\_RX\_In718 was annealed at $1050^{\circ}\mathrm{C}$ for 30 minutes to generate a nearly random texture, followed by 8 hours at $720^{\circ}\mathrm{C}$, forming $\gamma$' precipitates. AM\_AB\_In718, AM\_AB\_Steel316 and AM\_AB\_Co76A were all produced with a Formalloy L2 Directed Energy Deposition (DED) unit utilizing a 650 W Nuburu 450 nm blue laser capable of achieving a 400~$\mu$m laser spot size. During the additive manufacturing process, Argon was employed as shielding and carrier gas. AM\_HIP\_GRX810 was produced thanks to a EOS M290 printer (L-PBF) with a laser power of $275\:\mathrm{W}$, a scan speed of $1000\:\mathrm{mm/s}$, a layer thickness of $40\:\mathrm{\mu m}$ and an energy density in the range of $90-110\:\mathrm{J/mm^3}$. AM\_HIP\_GRX810 was then subjected to a HIP treatment at $1163^{\circ}\mathrm{C}$ and $100\:\mathrm{MPa}$ for 3.5 hours. These alloys are referred to as \textbf{W}, \textbf{AM}, \textbf{RX}, \textbf{AB} and \textbf{HIP}, corresponding to Wrought, Additively Manufactured, Recrystallized, As-Built, and Hot Isostatic Pressing, respectively. The chemical composition of these alloys is listed in Supplementary Table \ref{tab:compo}. Supplementary Fig. \ref{fig:materials} provides an overview of the investigated materials, showing their microstructure obtained from Electron BackScatter Diffraction (EBSD) along with the associated plastic localization depicted as longitudinal strain fields obtained from High-Resolution Digital Image Correlation (HR-DIC). Additionally, Supplementary Figs. \ref{fig:FTEMPLATE_W_Invar}, \ref{fig:FTEMPLATE_W_Nickel600}, \ref{fig:FTEMPLATE_W_Steel330}, \ref{fig:FTEMPLATE_W_In718}, \ref{fig:FTEMPLATE_W_Cu77NiSn}, \ref{fig:FTEMPLATE_W_RX_In718}, \ref{fig:FTEMPLATE_AM_AB_In718}, \ref{fig:FTEMPLATE_AM_AB_Steel316}, \ref{fig:FTEMPLATE_AM_AB_Co76A}, \ref{fig:FTEMPLATE_AM_HIP_GRX810} offer detailed visualizations of various modalities of the multimodal dataset for each of the ten alloys investigated in this study.


\begin{supptable}[h!]
    \centering
    \setlength{\tabcolsep}{3pt}
    \caption{Chemical compositions, in wt.\%, of the 10 FCC materials considered in this study. Values denoted by a \textsuperscript{m} correspond to maximum values according to specifications.}
    \label{tab:compo}
        \resizebox{\textwidth}{!}{%
            \begin{tabular}{|c||c|c|c|c|c|c|c|c|c|c|c|c|c|c|c|c|c|}
            \hline
            \textbf{Alloy} & Fe & Ni & Cu & Co & Cr & Nb & Mo & Mn & Ti & Al & Si & Sn & W & Re & C & N & Other \\
            \hline
            \textbf{Denomination} & ($\%$) & ($\%$) & ($\%$) & ($\%$) & ($\%$) & ($\%$) & ($\%$) & ($\%$) & ($\%$) & ($\%$) & ($\%$) & ($\%$) & ($\%$) & ($\%$) & ($\%$) & ($\%$) & ($\%$) \\
            \hline
            W\_Invar & bal. & 38 & / & 0.5 & 0.25 & / & / & 0.6 & / & 0.1 & / & / & / & / & 0.05 & / & 0.1 \\
            \hline
            W\_Nickel600 & 6 - 10 & bal. & / & / & 14 - 17 & / & / & 1\textsuperscript{m} & / & / & 0.5\textsuperscript{m} & / & / & / & / & / & / \\
            \hline
            W\_Steel330 & bal. & 34 - 37 & 1\textsuperscript{m} & / & 18 - 20 & / & / & 2\textsuperscript{m} & / & / & 1.5\textsuperscript{m} & / & / & / & 0.08\textsuperscript{m} & / & / \\
            \hline
            W\_In718 & bal. & 23.5 & / & / & 17 - 21 & 4.75 - 5.5 & 2.8 - 3.3 & 0.35\textsuperscript{m} & 0.65 - 1.15 & 0.2 - 0.8 & 0.35\textsuperscript{m} & / & / & / & 0.08\textsuperscript{m} & / & / \\
            \hline
            W\_Cu77NiSn & 0.5\textsuperscript{m} & 14.5 - 15.5 & bal. & / & / & 0.1\textsuperscript{m} & / & / & / & / & / & 7.5 - 8.5 & / & / & / & / & 0.7 \\
            \hline
            W\_RX\_In718 & 17.31 & bal. & / & 0.14 & 17.97 & 5.4 & / & / & 1 & 0.56 & / & / & / & / & 0.023 & 0.0062 & 0 \\
            \hline
            AM\_AB\_In718 & 18.77 & bal. & 0.02 & 0.07 & 18.88 & 5.08 & 3.04 & 0.04 & 0.96 & / & 0.08 & / & / & / & 0.036 & / & 0 \\
            \hline
            AM\_AB\_Steel316 & bal. & 12 & / & / & 18 & / & 2 & / & / & / & / & / & / & / & 0.03 & / & 1 \\
            \hline
            AM\_AB\_Co76A & / & / & / & bal. & 29 & / & 6 & / & / & / & / & / & / & / & 0.12 & / & 0.1 \\
            \hline
            AM\_HIP\_GRX810 & / & bal. & / & 33 & 29 & 0.75 & / & / & 0.25 & 0.3 & / & / & 3 & 1.5 & 0.05 & / & 0 \\
            \hline
        \end{tabular}
        }
\end{supptable}

\begin{suppfigure}
    \centering
    \includegraphics[width=1\linewidth]{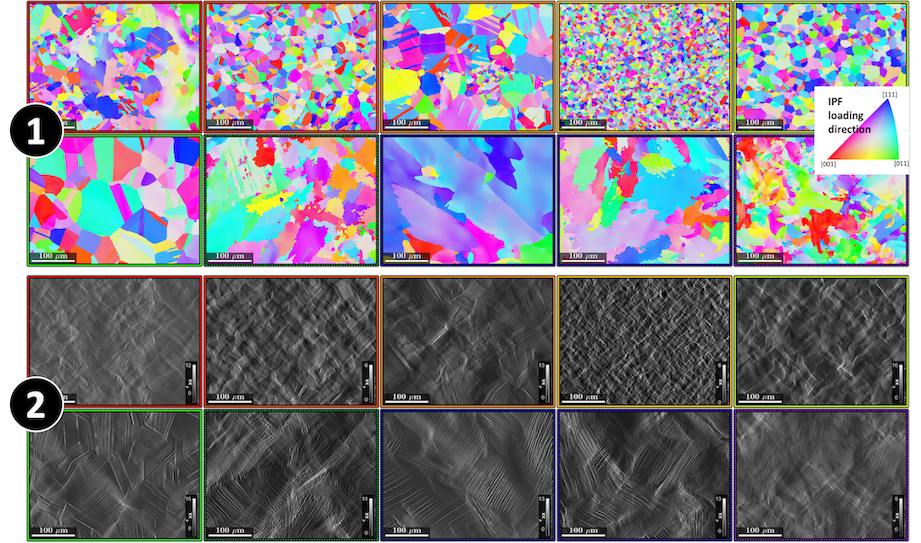}
    \caption{\textbf{Overview of the investigated materials}. \textbf{(1)} inverse pole figure (IPF)-colored orientation maps and \textbf{(2)} associated longitudinal strain fields $\varepsilon_{XX}$.}
    \label{fig:materials}
\end{suppfigure}

\newpage

\begin{suppfigure}
    \centering
    \includegraphics[width=1\linewidth]{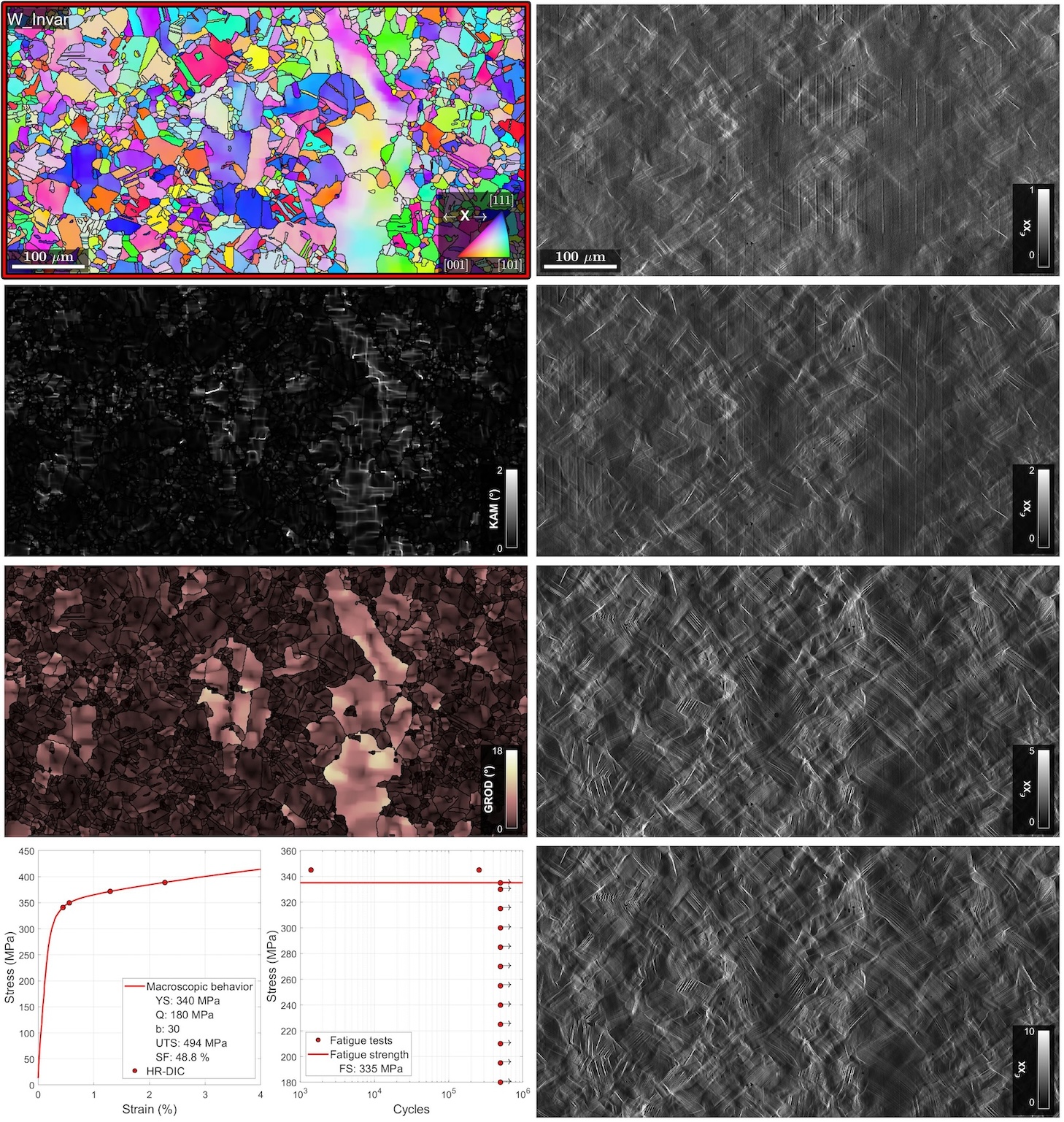}
    \caption{\textbf{Multimodal dataset for the conventionally wrought Invar alloy.} The left column presents experimentally measured microstructural descriptors, including Inverse Pole Figure (IPF)-colored orientation maps, Kernel Average Misorientation (KAM), and Grain Reference Orientation Deviation (GROD) maps obtained from EBSD measurements. The bottom row shows the associated macroscopic mechanical behavior, including the monotonic tensile response and fatigue properties. The right column presents spatially resolved deformation maps, longitudinal strain $\varepsilon_{XX}$, measured by HR-DIC at increasing applied deformation levels, highlighting the development and spatial organization of plastic deformation.}
    \label{fig:FTEMPLATE_W_Invar}
\end{suppfigure}

\newpage

\begin{suppfigure}
    \centering
    \includegraphics[width=1\linewidth]{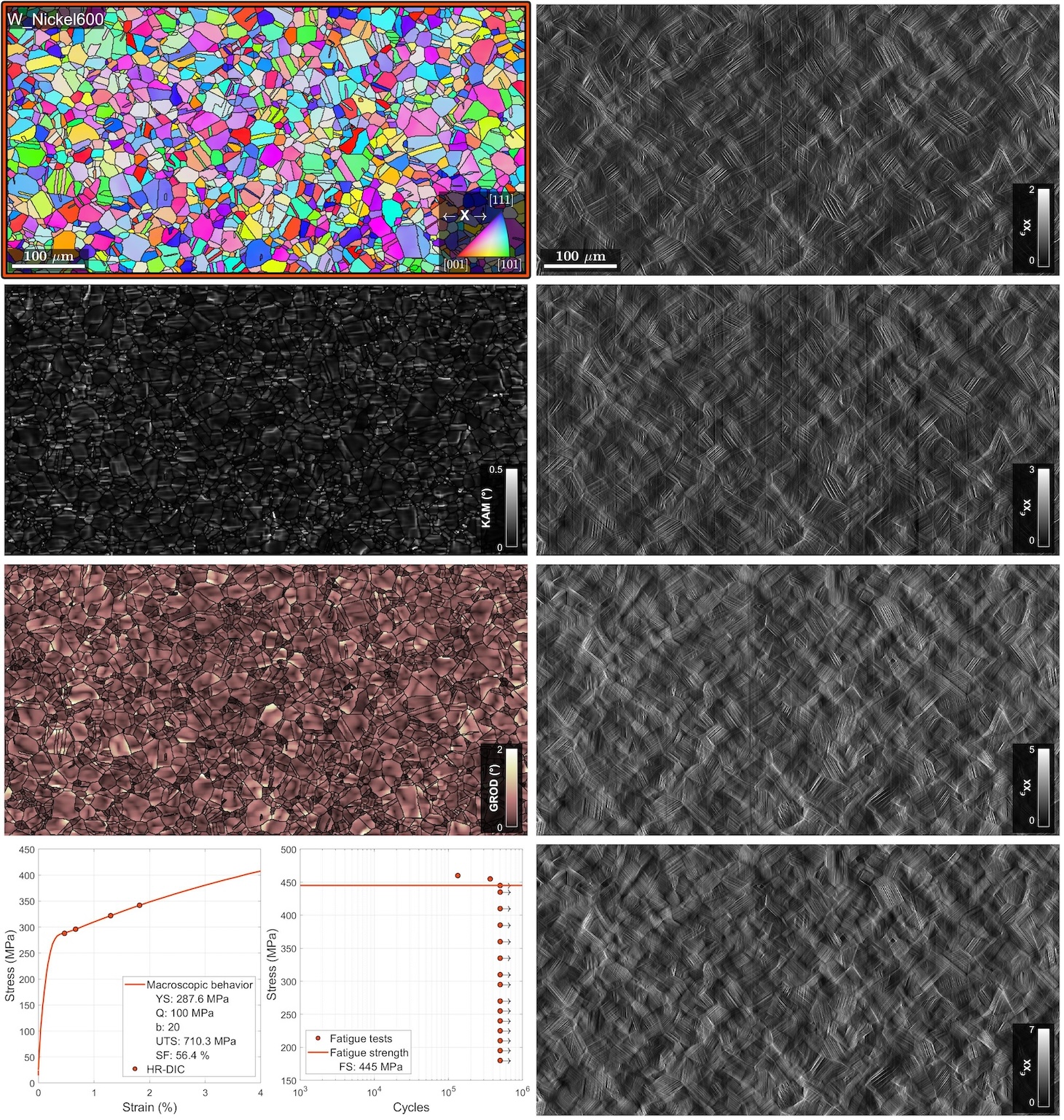}
    \caption{\textbf{Multimodal dataset for the conventionally wrought Nickel600 alloy.} The left column presents experimentally measured microstructural descriptors, including Inverse Pole Figure (IPF)-colored orientation maps, Kernel Average Misorientation (KAM), and Grain Reference Orientation Deviation (GROD) maps obtained from EBSD measurements. The bottom row shows the associated macroscopic mechanical behavior, including the monotonic tensile response and fatigue properties. The right column presents spatially resolved deformation maps, longitudinal strain $\varepsilon_{XX}$, measured by HR-DIC at increasing applied deformation levels, highlighting the development and spatial organization of plastic deformation.}
    \label{fig:FTEMPLATE_W_Nickel600}
\end{suppfigure}

\newpage

\begin{suppfigure}
    \centering
    \includegraphics[width=1\linewidth]{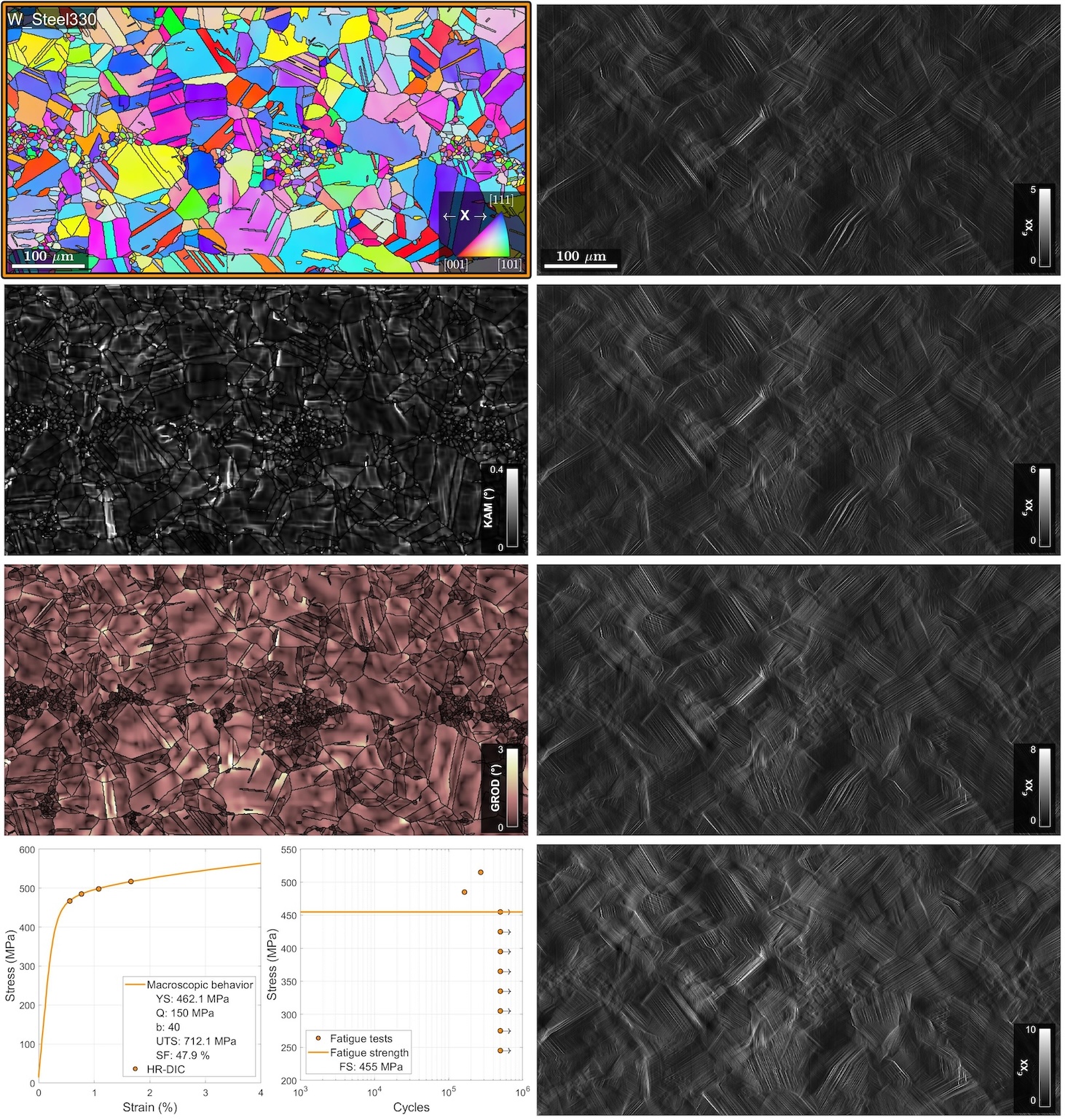}
    \caption{\textbf{Multimodal dataset for the conventionally wrought Steel330 alloy.} The left column presents experimentally measured microstructural descriptors, including Inverse Pole Figure (IPF)-colored orientation maps, Kernel Average Misorientation (KAM), and Grain Reference Orientation Deviation (GROD) maps obtained from EBSD measurements. The bottom row shows the associated macroscopic mechanical behavior, including the monotonic tensile response and fatigue properties. The right column presents spatially resolved deformation maps, longitudinal strain $\varepsilon_{XX}$, measured by HR-DIC at increasing applied deformation levels, highlighting the development and spatial organization of plastic deformation.}
    \label{fig:FTEMPLATE_W_Steel330}
\end{suppfigure}

\newpage

\begin{suppfigure}
    \centering
    \includegraphics[width=1\linewidth]{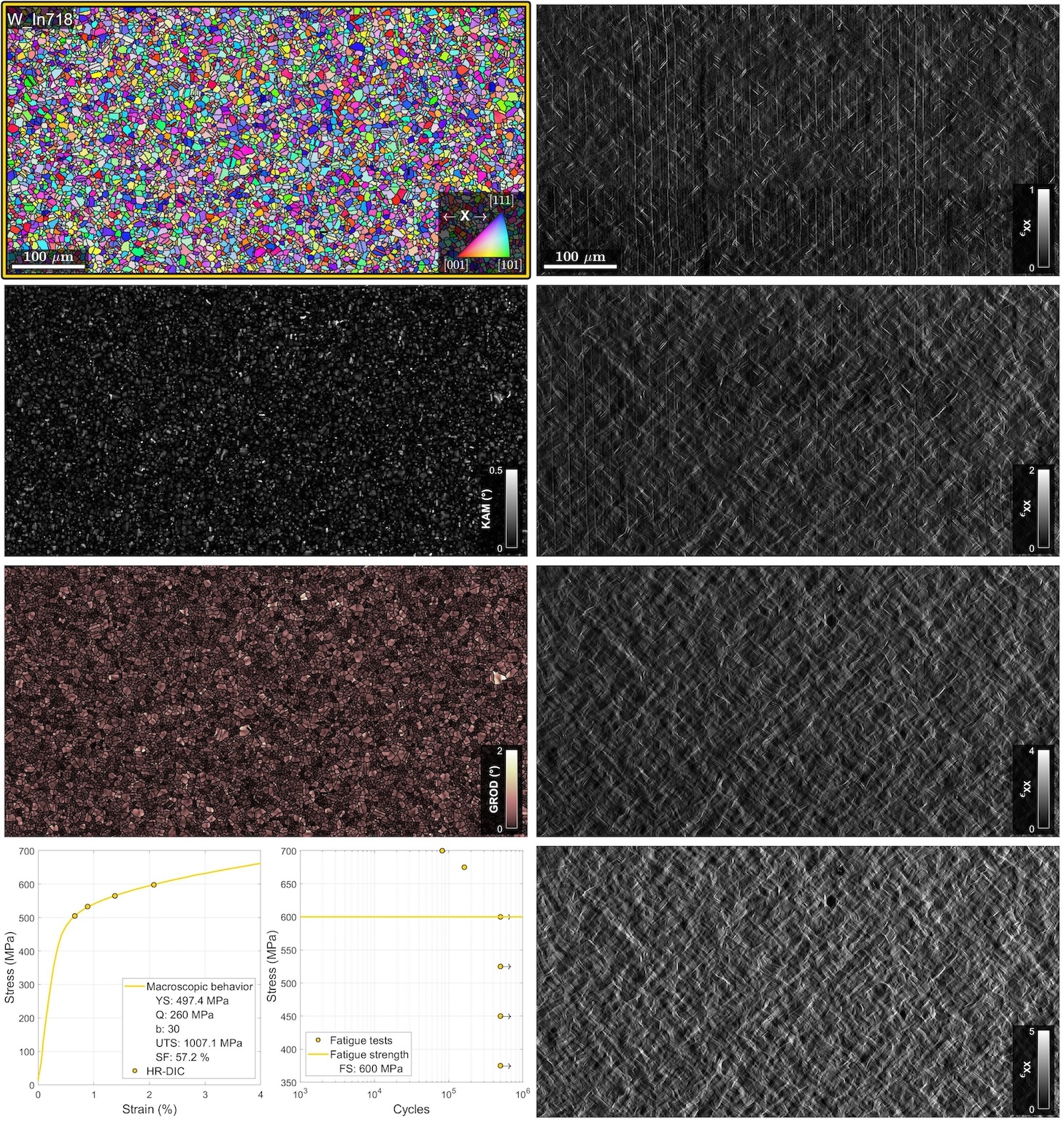}
    \caption{\textbf{Multimodal dataset for the conventionally wrought In718 alloy.} The left column presents experimentally measured microstructural descriptors, including Inverse Pole Figure (IPF)-colored orientation maps, Kernel Average Misorientation (KAM), and Grain Reference Orientation Deviation (GROD) maps obtained from EBSD measurements. The bottom row shows the associated macroscopic mechanical behavior, including the monotonic tensile response and fatigue properties. The right column presents spatially resolved deformation maps, longitudinal strain $\varepsilon_{XX}$, measured by HR-DIC at increasing applied deformation levels, highlighting the development and spatial organization of plastic deformation.}
    \label{fig:FTEMPLATE_W_In718}
\end{suppfigure}

\newpage

\begin{suppfigure}
    \centering
    \includegraphics[width=1\linewidth]{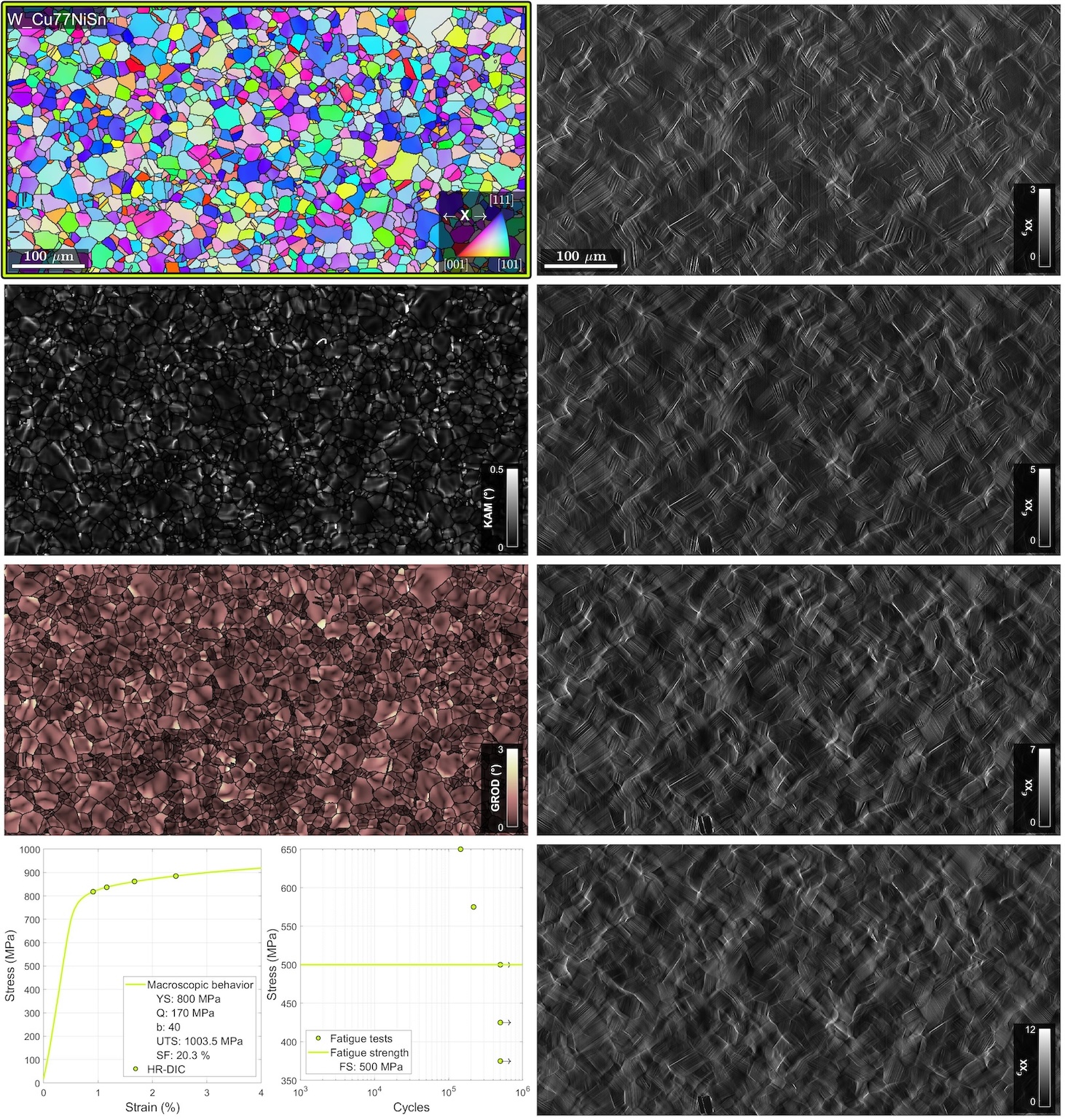}
    \caption{\textbf{Multimodal dataset for the conventionally wrought Cu77NiSn alloy.} The left column presents experimentally measured microstructural descriptors, including Inverse Pole Figure (IPF)-colored orientation maps, Kernel Average Misorientation (KAM), and Grain Reference Orientation Deviation (GROD) maps obtained from EBSD measurements. The bottom row shows the associated macroscopic mechanical behavior, including the monotonic tensile response and fatigue properties. The right column presents spatially resolved deformation maps, longitudinal strain $\varepsilon_{XX}$, measured by HR-DIC at increasing applied deformation levels, highlighting the development and spatial organization of plastic deformation.}
    \label{fig:FTEMPLATE_W_Cu77NiSn}
\end{suppfigure}

\newpage

\begin{suppfigure}
    \centering
    \includegraphics[width=1\linewidth]{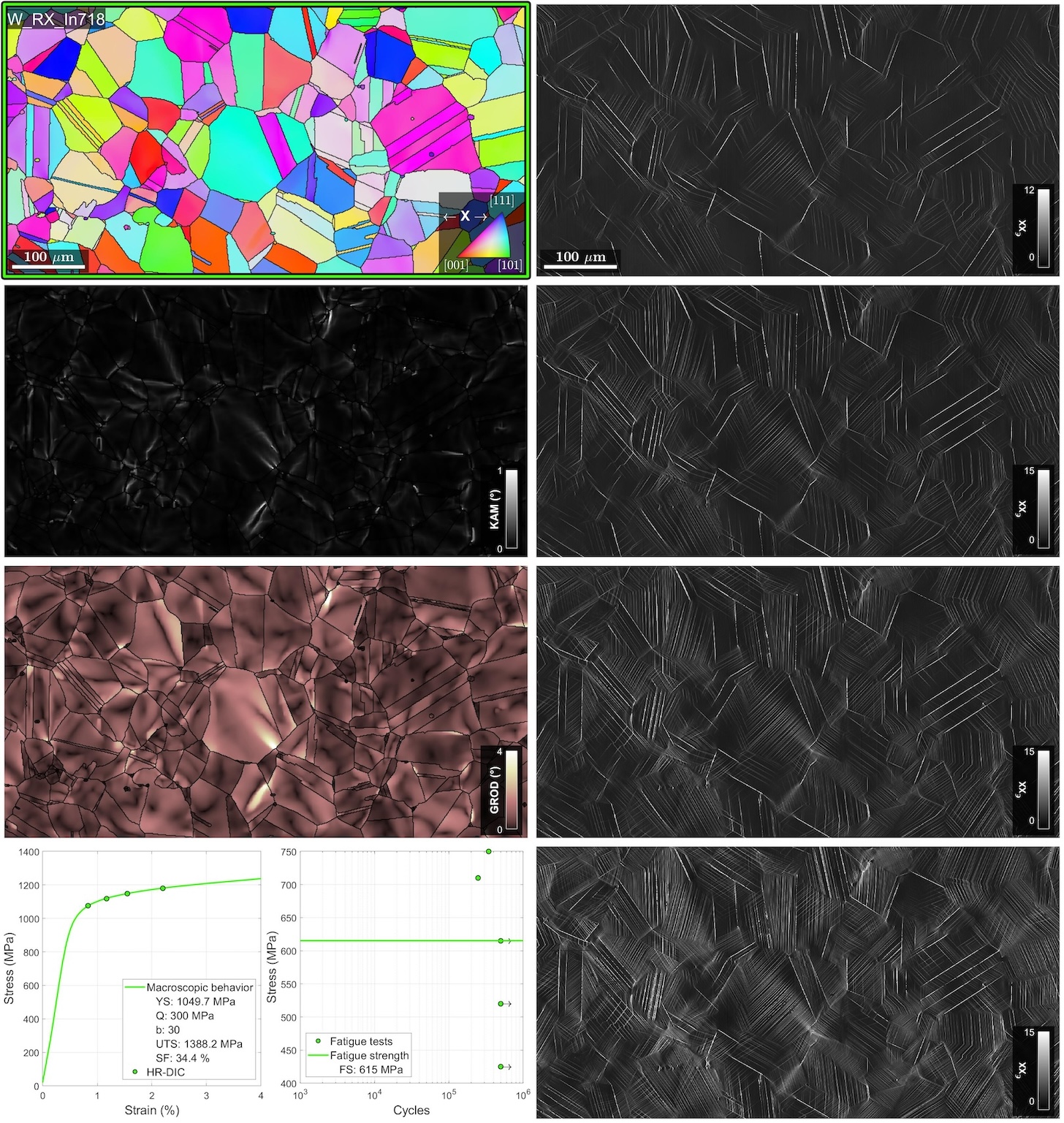}
    \caption{\textbf{Multimodal dataset for the conventionally wrought and recrystallized In718 alloy.} The left column presents experimentally measured microstructural descriptors, including Inverse Pole Figure (IPF)-colored orientation maps, Kernel Average Misorientation (KAM), and Grain Reference Orientation Deviation (GROD) maps obtained from EBSD measurements. The bottom row shows the associated macroscopic mechanical behavior, including the monotonic tensile response and fatigue properties. The right column presents spatially resolved deformation maps, longitudinal strain $\varepsilon_{XX}$, measured by HR-DIC at increasing applied deformation levels, highlighting the development and spatial organization of plastic deformation.}
    \label{fig:FTEMPLATE_W_RX_In718}
\end{suppfigure}

\newpage

\begin{suppfigure}
    \centering
    \includegraphics[width=1\linewidth]{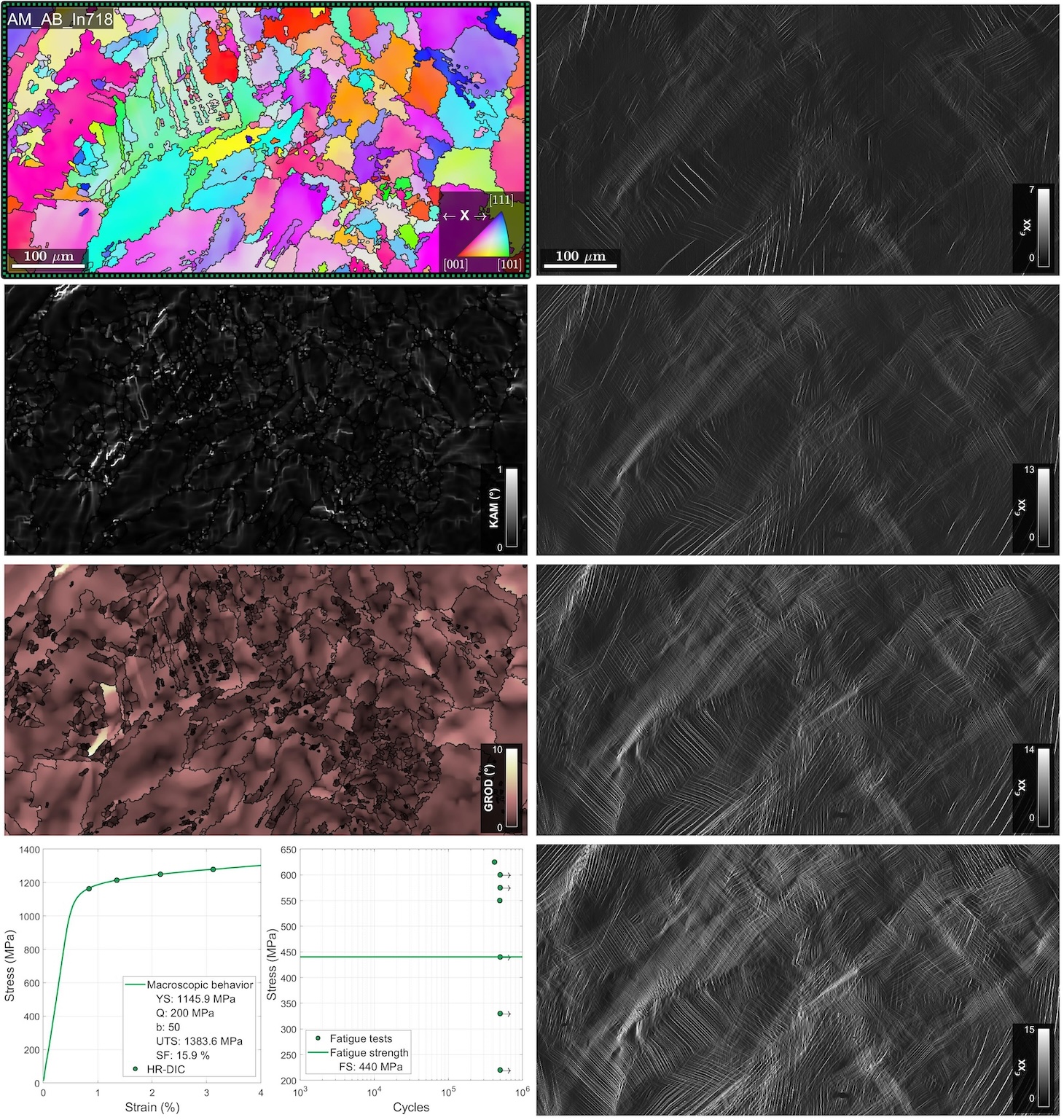}
    \caption{\textbf{Multimodal dataset for the additively manufactured In718 alloy.} The left column presents experimentally measured microstructural descriptors, including Inverse Pole Figure (IPF)-colored orientation maps, Kernel Average Misorientation (KAM), and Grain Reference Orientation Deviation (GROD) maps obtained from EBSD measurements. The bottom row shows the associated macroscopic mechanical behavior, including the monotonic tensile response and fatigue properties. The right column presents spatially resolved deformation maps, longitudinal strain $\varepsilon_{XX}$, measured by HR-DIC at increasing applied deformation levels, highlighting the development and spatial organization of plastic deformation.}
    \label{fig:FTEMPLATE_AM_AB_In718}
\end{suppfigure}

\newpage

\begin{suppfigure}
    \centering
    \includegraphics[width=1\linewidth]{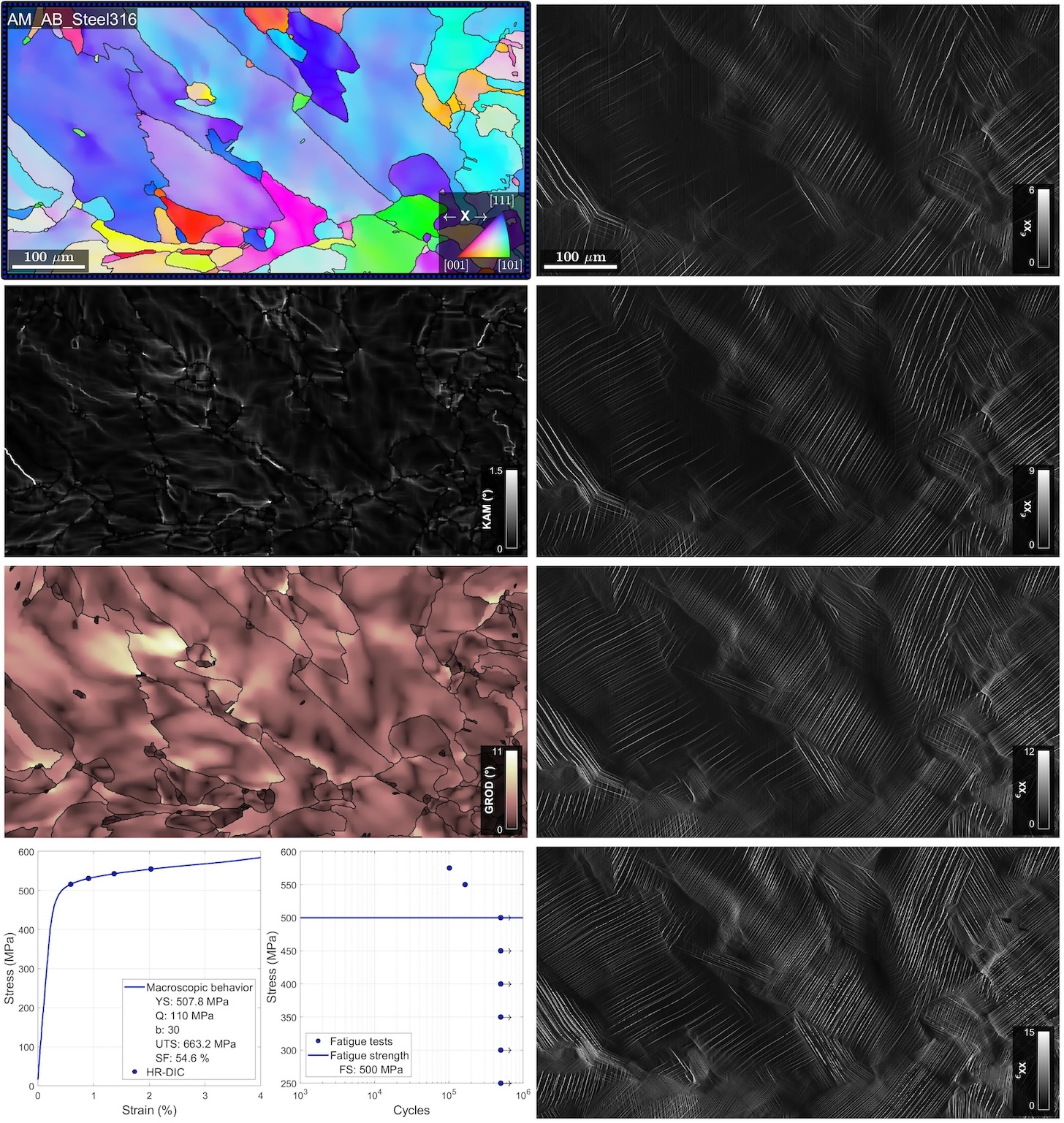}
    \caption{\textbf{Multimodal dataset for the additively manufactured Steel316 alloy.} The left column presents experimentally measured microstructural descriptors, including Inverse Pole Figure (IPF)-colored orientation maps, Kernel Average Misorientation (KAM), and Grain Reference Orientation Deviation (GROD) maps obtained from EBSD measurements. The bottom row shows the associated macroscopic mechanical behavior, including the monotonic tensile response and fatigue properties. The right column presents spatially resolved deformation maps, longitudinal strain $\varepsilon_{XX}$, measured by HR-DIC at increasing applied deformation levels, highlighting the development and spatial organization of plastic deformation.}
    \label{fig:FTEMPLATE_AM_AB_Steel316}
\end{suppfigure}
\newpage

\begin{suppfigure}
    \centering
    \includegraphics[width=1\linewidth]{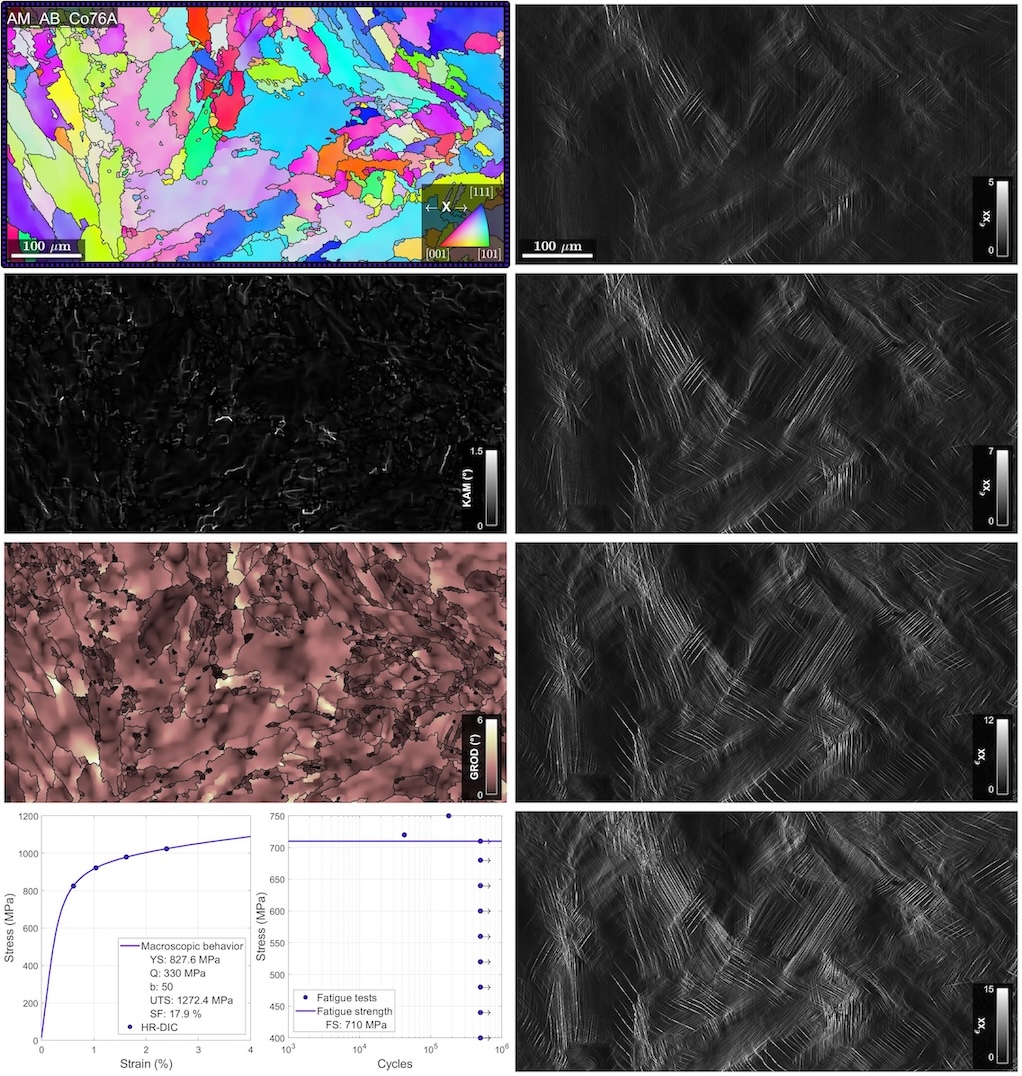}
    \caption{\textbf{Multimodal dataset for the additively manufactured Co76A alloy.} The left column presents experimentally measured microstructural descriptors, including Inverse Pole Figure (IPF)-colored orientation maps, Kernel Average Misorientation (KAM), and Grain Reference Orientation Deviation (GROD) maps obtained from EBSD measurements. The bottom row shows the associated macroscopic mechanical behavior, including the monotonic tensile response and fatigue properties. The right column presents spatially resolved deformation maps, longitudinal strain $\varepsilon_{XX}$, measured by HR-DIC at increasing applied deformation levels, highlighting the development and spatial organization of plastic deformation.}
    \label{fig:FTEMPLATE_AM_AB_Co76A}
\end{suppfigure}

\newpage

\begin{suppfigure}
    \centering
    \includegraphics[width=1\linewidth]{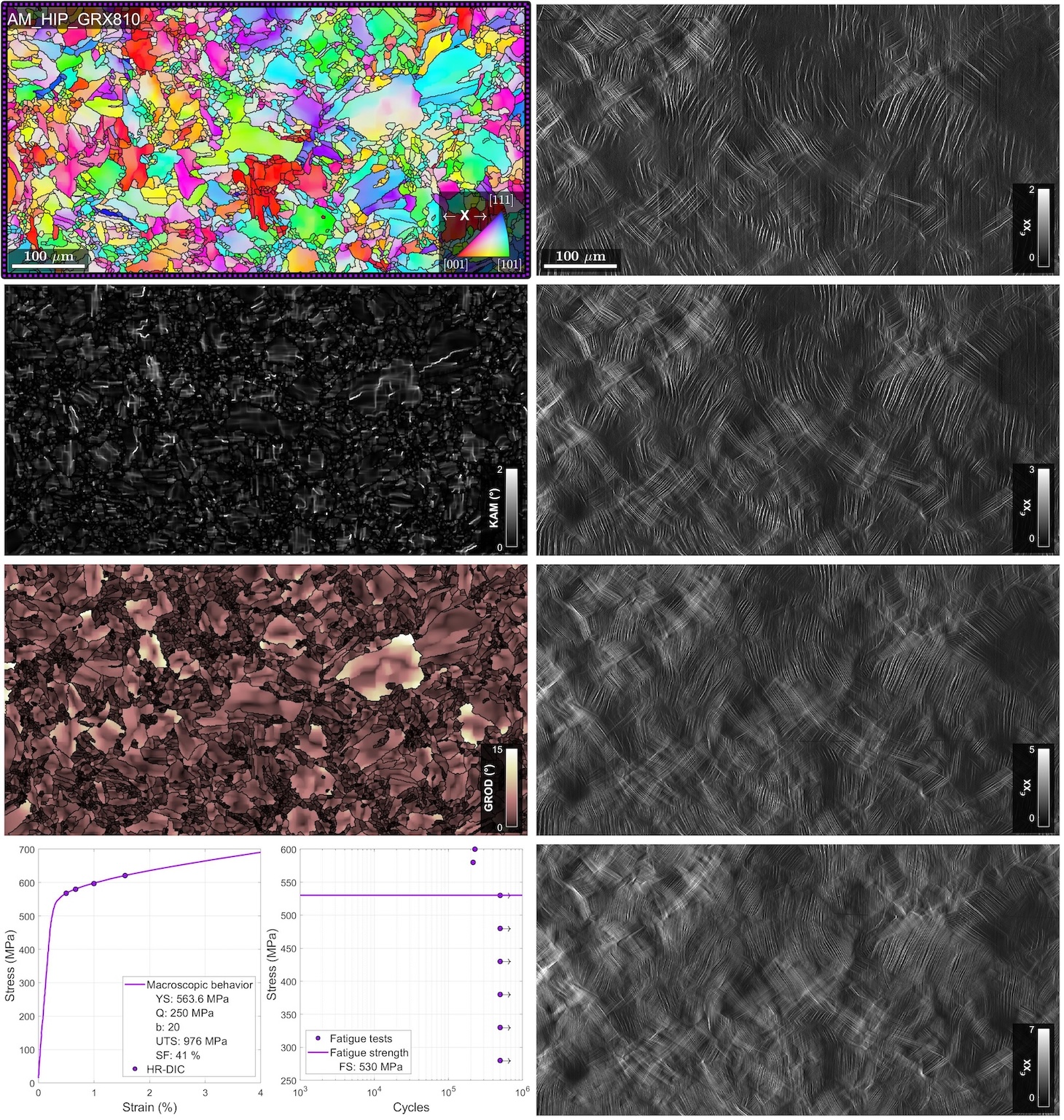}
    \caption{\textbf{Multimodal dataset for the additively manufactured GRX810 alloy.} The left column presents experimentally measured microstructural descriptors, including Inverse Pole Figure (IPF)-colored orientation maps, Kernel Average Misorientation (KAM), and Grain Reference Orientation Deviation (GROD) maps obtained from EBSD measurements. The bottom row shows the associated macroscopic mechanical behavior, including the monotonic tensile response and fatigue properties. The right column presents spatially resolved deformation maps, longitudinal strain $\varepsilon_{XX}$, measured by HR-DIC at increasing applied deformation levels, highlighting the development and spatial organization of plastic deformation.}
    \label{fig:FTEMPLATE_AM_HIP_GRX810}
\end{suppfigure}

\section{Sample preparation}

\justify For each material considered in this study, three different types of mechanical testing were performed: tensile testing for HR-DIC, tensile testing for optical DIC and conventional fatigue testing, using the geometries depicted in Supplementary Fig. \ref{fig:testing}(A), \ref{fig:testing}(C) and \ref{fig:testing}(F), respectively. The speckles used for optical DIC and HR-DIC measurements are provided in Supplementary Fig. \ref{fig:testing}(B) and \ref{fig:testing}(D), respectively.

\begin{suppfigure}
    \centering
    \includegraphics[width=1\linewidth]{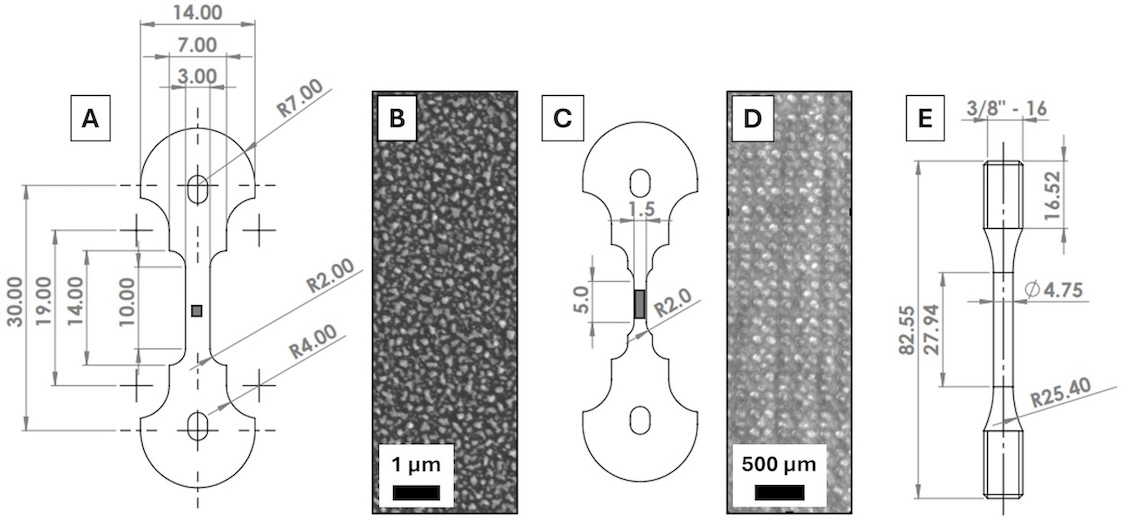}
    \caption{Specimen geometries used for \textbf{(A)} HR-DIC and \textbf{(B)} its associated speckle pattern, \textbf{(C)} optical DIC and \textbf{(D)} its associated speckle pattern generated by laser engraving, \textbf{(E)} macroscopic testing.}
    \label{fig:testing}
\end{suppfigure}

\section{Mechanical properties}

\justify Supplementary Table \ref{tab:props} contains the strain and stress applied to the HR-DIC specimens to obtain the longitudinal strain fields illustrated in Supplementary Fig. \ref{fig:materials}, along with the mechanical properties, including yield Strength (YS), two hardening parameters (Q and b), Ultimate Tensile Strength (UTS), Strain at Failure (SF), and Fatigue Strength (FS).

\begin{supptable}
    \centering
    \setlength{\tabcolsep}{6pt}
    \caption{Plastic strain values obtained by HR-DIC and mechanical properties of the 10 FCC materials considered in this study. All stress and strain values are engineering values.}
    \label{tab:props}
    \begin{tabular}{|c||c|c|c|c||c|c|c|c|c|c|}
        \hline
        \textbf{Alloy} & \multicolumn{4}{c||}{HR-DIC plastic strains} & \multicolumn{6}{c|}{Mechanical properties} \\
        \textbf{Denomination} & step 1 & step 2 & step 3 & step 4 & YS & Q & b & UTS & FS & FS \\
         & (\%) & (\%) & (\%) & (\%) & (MPa) & (MPa) & & (MPa) & (\%) & (MPa) \\
        \hline \hline
        W\_Invar & 0.18 & 0.29 & 1.00 & 1.98 & 340 & 180 & 30 & 494 & 48.8 & 335 \\
        \hline
        W\_Nickel600 & 0.34 & 0.51 & 1.16 & 1.63 & 287.6 & 100 & 20 & 710.3 & 56.4 & 445 \\
        \hline
        W\_Steel330 & 0.35 & 0.56 & 0.86 & 1.43 & 462.1 & 150 & 40 & 712.1 & 47.9 & 455 \\
        \hline
        W\_In718 & 0.13 & 0.36 & 0.83 & 1.48 & 497.4 & 260 & 30 & 1007.1 & 57.2 & 600 \\
        \hline
        W\_Cu77NiSn & 0.28 & 0.53 & 1.02 & 1.75 & 800 & 170 & 40 & 1003.5 & 20.3 & 500 \\
        \hline
        W\_RX\_In718 & 0.31 & 0.65 & 0.97 & 1.61 & 1049.7 & 300 & 30 & 1388.2 & 34.4 & 615 \\
        \hline
        AM\_AB\_In718 & 0.27 & 0.75 & 1.53 & 2.46 & 1145.9 & 200 & 50 & 1383.6 & 15.9 & 440 \\
        \hline
        AM\_AB\_Steel316 & 0.35 & 0.64 & 1.10 & 1.77 & 507.8 & 110 & 30 & 663.2 & 54.6 & 500 \\
        \hline
        AM\_AB\_Co76A & 0.27 & 0.61 & 1.14 & 1.85 & 827.6 & 330 & 50 & 1272.4 & 17.9 & 710 \\
        \hline
        AM\_HIP\_GRX810 & 0.23 & 0.39 & 0.72 & 1.28 & 563.6 & 250 & 20 & 887.6 & 41 & 530 \\
        \hline
    \end{tabular}
\end{supptable}

\section{Electron backscatter diffraction measurements}

\justify EBSD measurements were processed using MTEX \cite{Bachmann2010} to extract the microstructure characteristics provided in Supplementary Table \ref{tab:ebsd} that include the grain size (10\textsuperscript{th}, 50\textsuperscript{th} and 90\textsuperscript{th} quantiles), average Grain Reference Orientation Deviation (GROD), Multiples of Uniform Distribution (MUD) and twin area fraction.

\begin{supptable}
    \centering
    \setlength{\tabcolsep}{6pt}
    \caption{Microstructure characteristics of the 10 FCC materials considered in this study.}
    \label{tab:ebsd}
    \begin{tabular}{|c||c|c|c|c|}
        \hline
        \textbf{Alloy} & \multicolumn{4}{c|}{Microstructure characteristics} \\
        \textbf{Denomination} & GS & GROD & MUD & Twin fraction \\
        & ($\mu m$) & (${}^{\circ}$) & & (\%) \\
        \hline \hline
        W\_Invar & 4.37 / 8.29 / 21.02 & 1.81 & 1.71 & 41.00 \\
        \hline
        W\_Nickel600 & 4.79 / 9.71 / 19.96 & 0.41 & 1.31 & 30.69 \\
        \hline
        W\_Steel330 & 4.51 / 7.98 / 26.22 & 0.55 & 1.70 & 39.37 \\
        \hline
        W\_In718 & 4.24 / 5.97 / 9.17 & 0.27 & 1.28 & 22.32 \\
        \hline
        W\_Cu77NiSn & 4.65 / 10.28 / 20.65 & 0.48 & 1.70 & 17.14 \\
        \hline
        W\_RX\_In718 & 5.41 / 21.11 / 53.76 & 0.64 & 1.46 & 31.55 \\
        \hline
        AM\_AB\_In718 & 4.37 / 9.24 / 44.92 & 1.46 & 3.90 & 0.47 \\
        \hline
        AM\_AB\_Steel316 & 6.08 / 21.56 / 83.14 & 2.32 & 2.51 & 0.18 \\
        \hline
        AM\_AB\_Co76A & 4.51 / 9.44 / 38.89 & 1.21 & 2.31 & 0.84 \\
        \hline
        AM\_HIP\_GRX810 & 4.51 / 10.70 / 34.69 & 4.33 & 3.16 & 0.61 \\
        \hline
    \end{tabular}
\end{supptable}

\section{CNN architectures}

\justify The global CNN architecture is illustrated in Supplementary Fig. \ref{fig:cnn} and adapted from Ref. \cite{Calvat2026}. This architecture integrates two encoding head variants for processing EBSD or HR-DIC maps (shown in Supplementary Fig. \ref{fig:cnn}(A) and \ref{fig:cnn}(B), which allow these maps to be projected onto a latent representation. These new representations are sequentially processed by a prediction head (given in Supplementary Fig. \ref{fig:cnn}(C)) to predict six mechanical properties of interest in a simultaneous manner. These encoders have been developed to be compatible with various resolutions: $846 \times 846$, $1410 \times 1410$, $1974 \times 1974$ or $2538 \times 2538$ for HR-DIC, and $57 \times 57$, $95 \times 95$, $133 \times 133$ or $171 \times 171$ for EBSD maps. Both encoders use "shortcut connections" to accelerate training \cite{He2016}, and are composed by large convolution kernels and integrate strides of two or three for HR-DIC and EBSD, respectively. Both encoding heads and prediction heads employ the leaky ReLU activation function to avoid the "dying ReLU" issue \cite{Lu2019}. Due to the limited amount of data, the prediction head includes 15\% dropout layers to prevent overfitting during training. Conversely to Ref. \cite{Calvat2026}, no macroscopic state input is employed to limit their contribution and consequently forces the model in extracting governing features from the map rather than drawing relationships from macroscopic states. All these architectures were implemented and trained on MATLAB using the Deep Learning Toolbox.

\begin{suppfigure}
    \centering
    \includegraphics[width=1\linewidth]{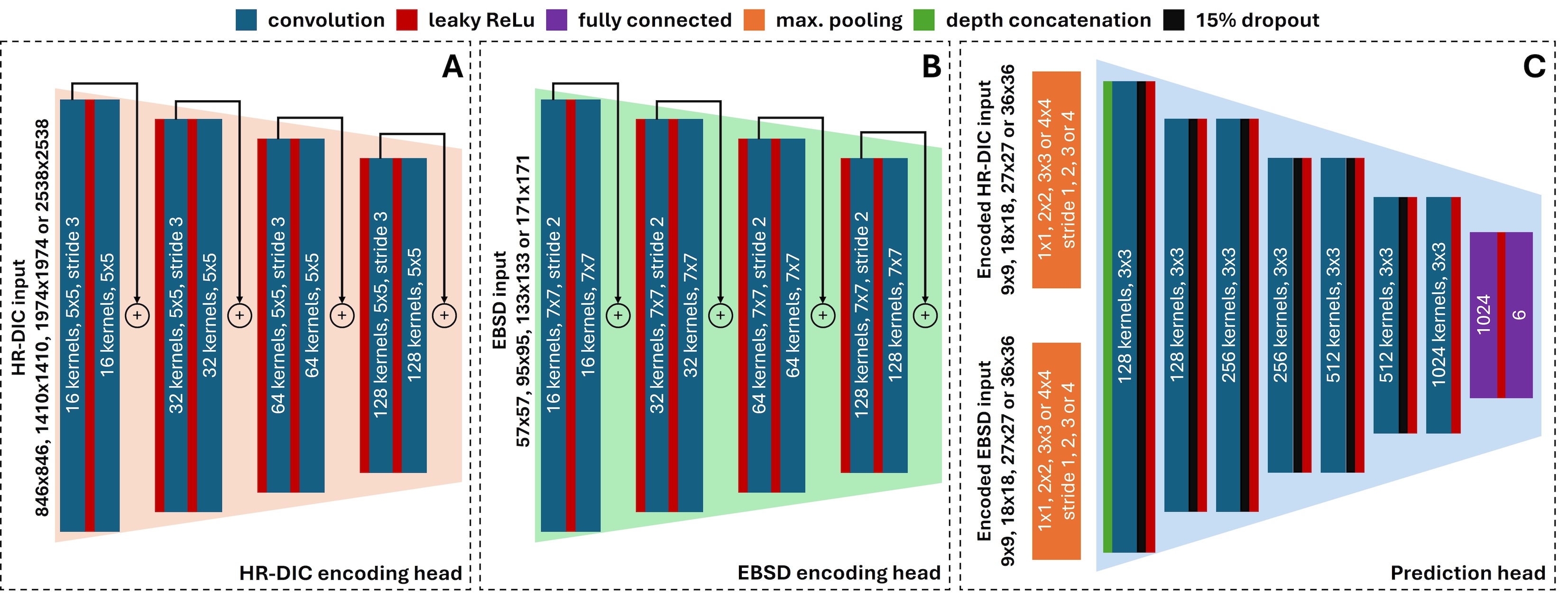}
    \caption{\textbf{(A)} HR-DIC encoding head, \textbf{(B)} EBSD encoding head and \textbf{(C)} prediction head.}
    \label{fig:cnn}
\end{suppfigure}

\section{Training data and preprocessing} 

\justify Supplementary Table \ref{tab:tiles} provides the number of tiles used for training the various CNN architectures with the four different dimensions considered in this study: $57 \times 57\:\mathrm{\mu m}$, $95 \times 95\:\mathrm{\mu m}$, $133 \times 133\:\mathrm{\mu m}$ or $171 \times 171\:\mathrm{\mu m}$.

\begin{supptable}[]
    \centering
    \setlength{\tabcolsep}{6pt}
    \caption{Number of tiles for the different dimensions employed to train the CNN architectures.}
    \label{tab:tiles}
    \begin{tabular}{|c||c|c|c|c|}
        \hline
        \textbf{Alloy} & \multicolumn{4}{c|}{Training / Validation} \\
        \textbf{Denomination} & $846 \times 846$ & $1410 \times 1410$ & $1974 \times 1974$ & $2538 \times 2538$ \\
        \hline \hline
        W\_Invar & 1008 / 280 & 320 / 64  & 140 / 20 & 80 / 16 \\
        \hline
        W\_Nickel600 & 988 / 260 & 352 / 96  & 140 / 20 & 80 / 16 \\
        \hline
        W\_Steel330 & 1064 / 280 & 352 / 96  & 140 / 20 & 80 / 16 \\
        \hline
        W\_In718 & 748 / 220 & 240 / 48  & 112 / 16 & 60 / 12 \\
        \hline
        W\_Cu77NiSn & 1140 / 300 & 396 / 108  & 168 / 124 & 80 / 16 \\
        \hline
        W\_RX\_In718 & 1216 / 320 & 396 / 108  & 168 / 124 & 100 / 20 \\
        \hline
        AM\_AB\_In718 & 1080 / 300 & 360 / 72  & 168 / 124 & 80 / 16 \\
        \hline
        AM\_AB\_Steel316 & 1216 / 320 & 396 / 108  & 168 / 124 & 80 / 16 \\
        \hline
        AM\_AB\_Co76A & 1152 / 320 & 360 / 108  & 168 / 124 & 100 / 20 \\
        \hline
        AM\_HIP\_GRX810 & 1140 / 300 & 396 / 108  & 168 / 124 & 80 / 16 \\
        \hline \hline
        \textbf{Total} & 820 / 164 & 1540 / 220  & 3568 / 916 & 10752 / 2900 \\
        \hline
    \end{tabular}
\end{supptable}

\section{CNN trainings}

\justify A total of 440 different models were trained to predict six mechanical properties from EBSD modalities (crystallographic orientation described by quaternion, KAM, GAM, GROD, GOS and GND) or HR-DIC modalities (longitudinal strain, effective strain, plastic deformation intensity and lattice rotation). For each of the considered modality, individual models were trained using 4 different tile sizes with custom CNN encoding heads designed for each tile size dimension (see Supplementary Fig. \ref{fig:cnn}). All EBSD-based and HR-DIC-based models were trained for 60 epochs corresponding to about 5,000 iterations for the smallest tile size. Apart from the number of epochs utilized, all other training parameters remain consistent across all models.

\justify The resulting training losses are given in Supplementary Fig. \ref{fig:loss}(A.1) and \ref{fig:loss}(A.2) for models predicting properties from EBSD and HR-DIC modalities, respectively. Throughout training, validation metrics were computed from four additional maps not included in the training set to evaluate the models' ability to predict mechanical properties as well as overfitting with respect to the training set. These metrics were estimated without data augmentation using a CNN with dropout layers removed. The resulting validation losses and the average absolute relative error are shown in Supplementary Fig. \ref{fig:loss}(B.1) and \ref{fig:loss}(C.1) for the EBSD modalities and Supplementary Fig. \ref{fig:loss}(B.2) and \ref{fig:loss}(C.2) for the HR-DIC modalities. For visualization purposes, the different metrics were smoothed over a 100-iteration window. Only the metrics associated with the models predicting from $57\times 57\:\mathrm{\mu m}$ tiles are shown in these figures. All training losses associated with EBSD modalities show a similar decrease. However, significant differences are observed in terms of validation loss and absolute relative error, for which neither quaternions nor GND show improvement. While a limited to no improvement in validation loss seems to be observed, a decrease in the average absolute relative error is observed for all modalities, demonstrating convergence of the CNN models. This difference in predictive ability is primarily explained by the nature of the modalities. While quaternions inherently contain all the information of the other modalities (KAM, GAM, GROD, GOS, and GND), they depend on the quaternion manifold. In contrast, KAM acts as distilled information about local misorientation facilitating prediction and identification of governing states. In this study, GND was employed using a logarithmic scale to facilitate utilization of this modality; however, the difference between low and high dislocation densities appears compressed compared to the original scale, which limits the model's predictive accuracy. Regarding the HR-DIC-derived models, plastic deformation intensity shows the fastest improvement in training loss, followed by longitudinal and effective strain (similar evolution) and lattice rotation showing a significantly slower decrease. This difference in training loss evolution stems from the nature of the preprocessing employed for the strain components and lattice rotation. The strain components and lattice rotation were preprocessed using a Gaussian blur with a standard deviation of 3 and 10, respectively. As with quaternions or GND for EBSD-derived models, no improvement in validation loss is observed for the lattice rotation model. However, all HR-DIC-derived models demonstrate an improvement in average absolute relative error.

\begin{suppfigure}
    \centering
    \includegraphics[width=1\linewidth]{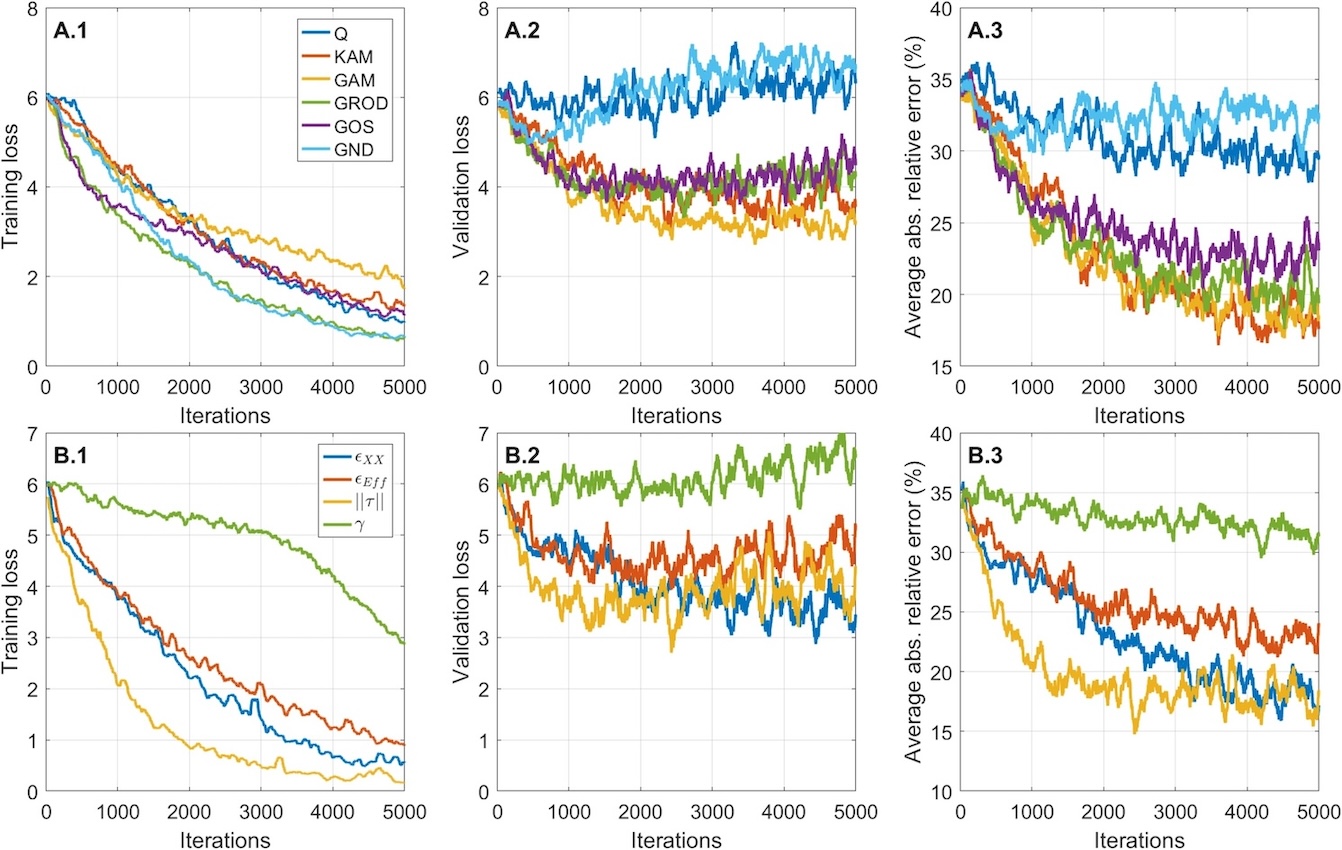}
    \caption{\textbf{Training and validation metrics}. \textbf{(A)} Training losses, \textbf{(B)} validation losses and \textbf{(C)} average absolute relative errors associated with \textbf{(.1)} EBSD modalities and \textbf{(.2)} HR-DIC modalities.}
    \label{fig:loss}
\end{suppfigure}

\section{Mechanical properties prediction performance}

\justify After training the various architectures, they were subsequently used to predict mechanical properties across the validation regions (which were not employed during training). The distributions of predicted properties are summarized in Supplementary Fig. \ref{fig:preds} using boxes, each extending from the first to the third quartile, with whiskers corresponding to the 5\textsuperscript{th} and 95\textsuperscript{th} percentiles. In addition, the centered line depicts the average of the distribution and circle markers designate the minimum and maximum extreme predicted values. Here, the mechanical properties were only computed from the validation regions using $57\times 57\:\mathrm{\mu m}$ tiles. The boxes associated with the HR-DIC predictions represent the overall distributions estimated from the four deformation steps. Lastly, solid lines indicate a perfect match between experimental and predicted mechanical properties, while the dashed and dotted lines delimit 10\% and 20\% errors, respectively. Supplementary Fig. \ref{fig:preds}(A -- B) first provide predictions of the YS, UTS, and FS using either quaternions or KAM. Depending on the mechanical property examined and the modality employed, materials exhibit different levels of variability that are directly related to microstructure. For example, W\_Invar, characterized by its partially recrystallized microstructure, demonstrates significant variability. Conversely, W\_In718 and AM\_HIP\_GRX810 demonstrate low variability regardless of the considered modality or predicted property. Supplementary Fig. \ref{fig:preds}(C -- D) provide predictions for the same mechanical properties using longitudinal strain and lattice rotation, demonstrating similar accuracy to that achieved using EBSD modalities.

\begin{suppfigure}
    \centering
    \includegraphics[width=1\linewidth]{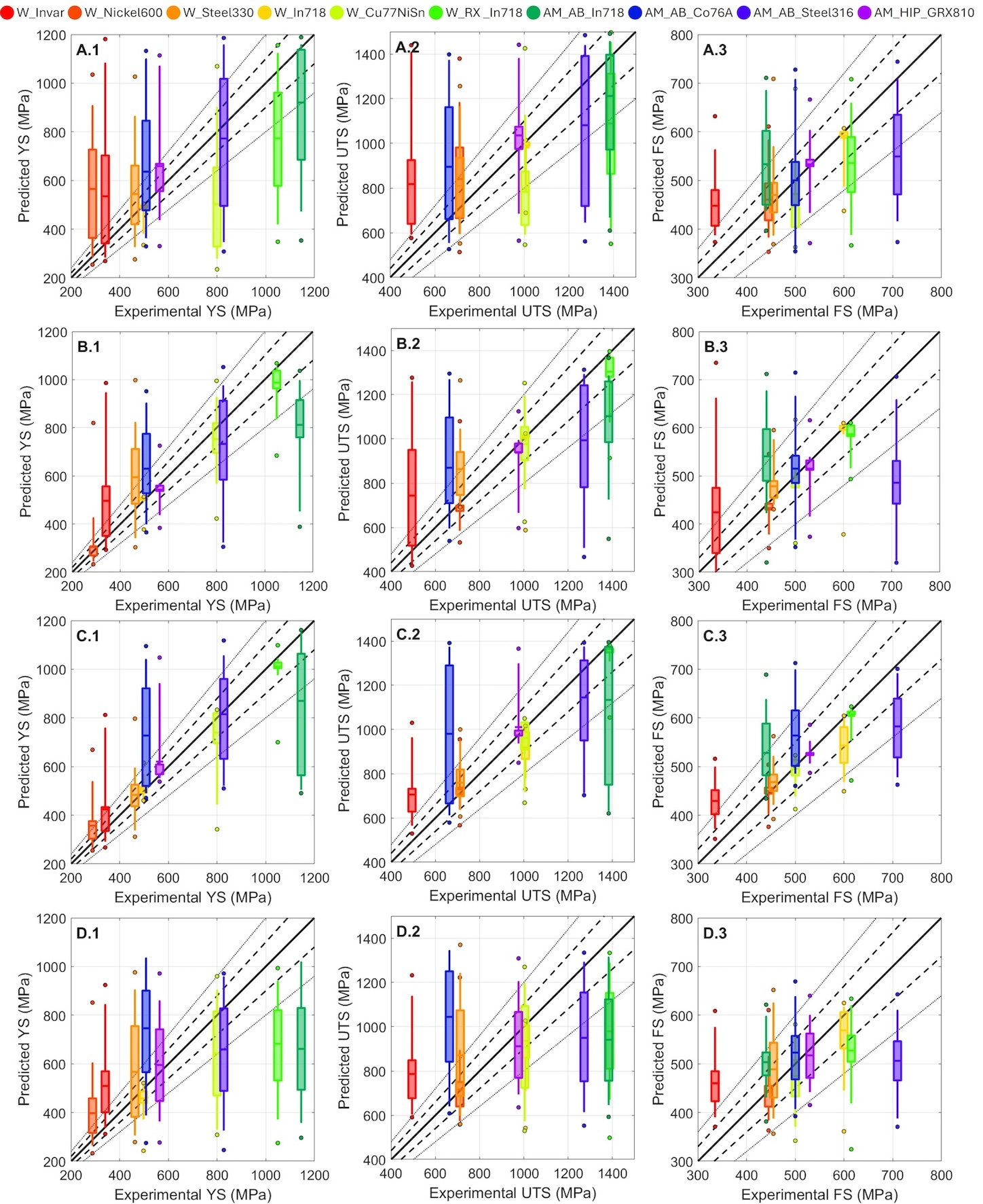}
    \caption{\textbf{Mechanical properties prediction.} Predictions from \textbf{(A)} quaternions, \textbf{(B)} kernel Average Misorientation (KAM), \textbf{(C)} longitudinal strain $\varepsilon_{XX}$ and \textbf{(D)} lattice rotation $\gamma$, \textbf{(.1)} Yield Strength (YS), \textbf{(.2)} Ultimate Tensile Strength (UTS) and \textbf{(.3)} Fatigue Strength (FS) predicted from $57\times 57\:\mathrm{\mu m}$ tiles.}
    \label{fig:preds}
\end{suppfigure}

\justify In addition to examples of predicted mechanical properties shown in Supplementary Fig. \ref{fig:preds}, Supplementary Fig. \ref{fig:accuracy_preds} illustrates the accuracy and repeatability of the predictions associated with the different microstructure or plasticity localization modalities and were evaluated from $57\times 57\:\mathrm{\mu m}$ tiles within the validation regions. The color of each marker indicates the accuracy of the prediction (using the average predicted value). Vertically-elongated markers indicate low repeatability, while horizontally-elongated markers indicate high repeatability. For fair comparison between the different predicted properties, the repeatability has been inversely scaled by the variability (standard deviation) of each experimental mechanical property. Each row corresponds to a single material and the predictions are grouped by models (separated by solid vertical black lines). The repeatability is directly related to the grain size and grain size distribution. Significantly high repeatability is observed for both W\_In718 and AM\_HIP\_GRX810 when considering EBSD modalities due to their relatively small grain size. Similarly, repeatability is high for W\_In718 when considering HR-DIC modalities due to the uniformity of the observed plastic localization (see Supplementary Fig. \ref{fig:FTEMPLATE_W_In718}). On the contrary, the alloys produced by DED additive manufacturing, considered in this study, are associated with a large grain size, which leads to lower repeatability.

\begin{suppfigure}
    \centering
    \includegraphics[width=1\linewidth]{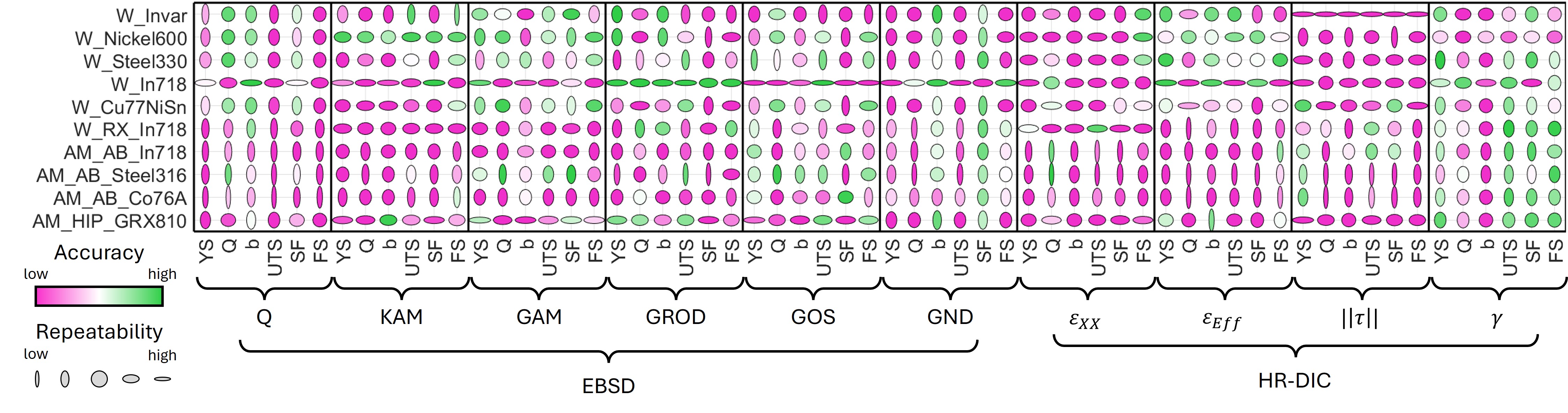}
    \caption{Accuracy and repeatability of predicted mechanical properties.}
    \label{fig:accuracy_preds}
\end{suppfigure}

\justify Figure \ref{fig:corr} provides a correlation matrix that illustrates the eventual relationships between the various models' predictions and their associated errors. The correlation coefficients have been calculated using the Pearson definition using only predicted values from $57\times 57\:\mathrm{\mu m}$ tiles within validation regions. The correlations associated with the HR-DIC modalities were estimated using the four deformation steps. The lower-left half provides the correlations between the mechanical properties predicted by the different models, arranged by mechanical property (delimited by solid black lines). The upper-right half contains the correlations computed between the absolute relative errors associated with the predictions, arranged by model (delimited by solid black lines). For the sake of visualization, HR-DIC modalities and their associated correlations are depicted using a gray background. The size of the markers illustrates the strength of the correlation, and the color indicates whether the correlation is positive (red) or negative (blue). 

\justify In the lower-left half are given the correlations between the mechanical properties predicted by the different models. In the context of property prediction, a strong positive correlation means that if a high value is predicted for a tile, then a high value will also be predicted for that tile. Conversely, a strong negative correlation means that if a high value is predicted for one tile, a low value will be predicted for that same tile. The diagonal is always made up of ones because it compares the same property predicted by the same model. Most of the predicted mechanical properties exhibit strong positive or negative correlation coefficients when comparing predictions from each model individually (diagonal values in each square). Predictions from the four different HR-DIC modalities are associated with comparable correlation coefficients, even when comparing predictions from different models. These strong correlation values are directly inherited from the accuracy of the predictions (see Supplementary Fig. \ref{fig:preds}(C -- D) and Supplementary Fig. \ref{fig:accuracy_preds}) and tied to the experimental values and eventual trade-offs. Q-hardening and b-hardening parameters, as well as Strain at Failure (SF) and Q-hardening, only exhibit weak correlations. Specifically, SF shows strong negative correlation coefficients with the other mechanical properties, particularly with Yield Strength (YS), the b-hardening parameter, and Ultimate Tensile Strength (UTS). In other words, when SF is predicted to be high, these properties are predicted to be low, and vice versa. For properties predicted from EBSD modalities (often associated with variability), the presence of weak correlations when comparing different properties and/or models indicates that these properties are likely derived from different microstructural features. 

\justify In the upper-right half are given the correlations between the absolute errors corresponding to the predictions made by the different models. Conversely to the lower-left half, a strong positive correlation indicates that a prediction associated with a large error from a given tile will also lead to a large error for another prediction. For each individual model, most errors exhibit strong positive correlations for the different mechanical properties. Except for the GND-based model, all EBSD-based models exhibit medium to strong correlations when comparing errors across different mechanical properties. Conversely, most of the HR-DIC-based models show lower correlation coefficients, except for the model based on the plastic deformation intensity. Interestingly, regardless of the model considered (employing either EBSD or HR-DIC modalities), FS shows lower correlations with respect to b-hardening and SF. Depending on the modality under consideration and the model's associated accuracy to predict (see Supplementary Fig. \ref{fig:accuracy_preds}), either no trade-off exists between FS and b-hardening or SF, or different features are employed within the maps to predict these properties. Additionally, when comparing the predicted properties one by one, GOS shows a medium positive correlation with GAM and GROD since all three depend on grain structure, and their associated models learn similar features for predicting properties. Strong correlations between several properties may hinder the accuracy of the remaining properties if the extracted features are not suitable for accurately predicting them. Thus, reducing overall loss can be achieved at the expense of predictions' accuracy if it benefits a greater number of properties.

\begin{suppfigure}
    \centering
    \includegraphics[width=1\linewidth]{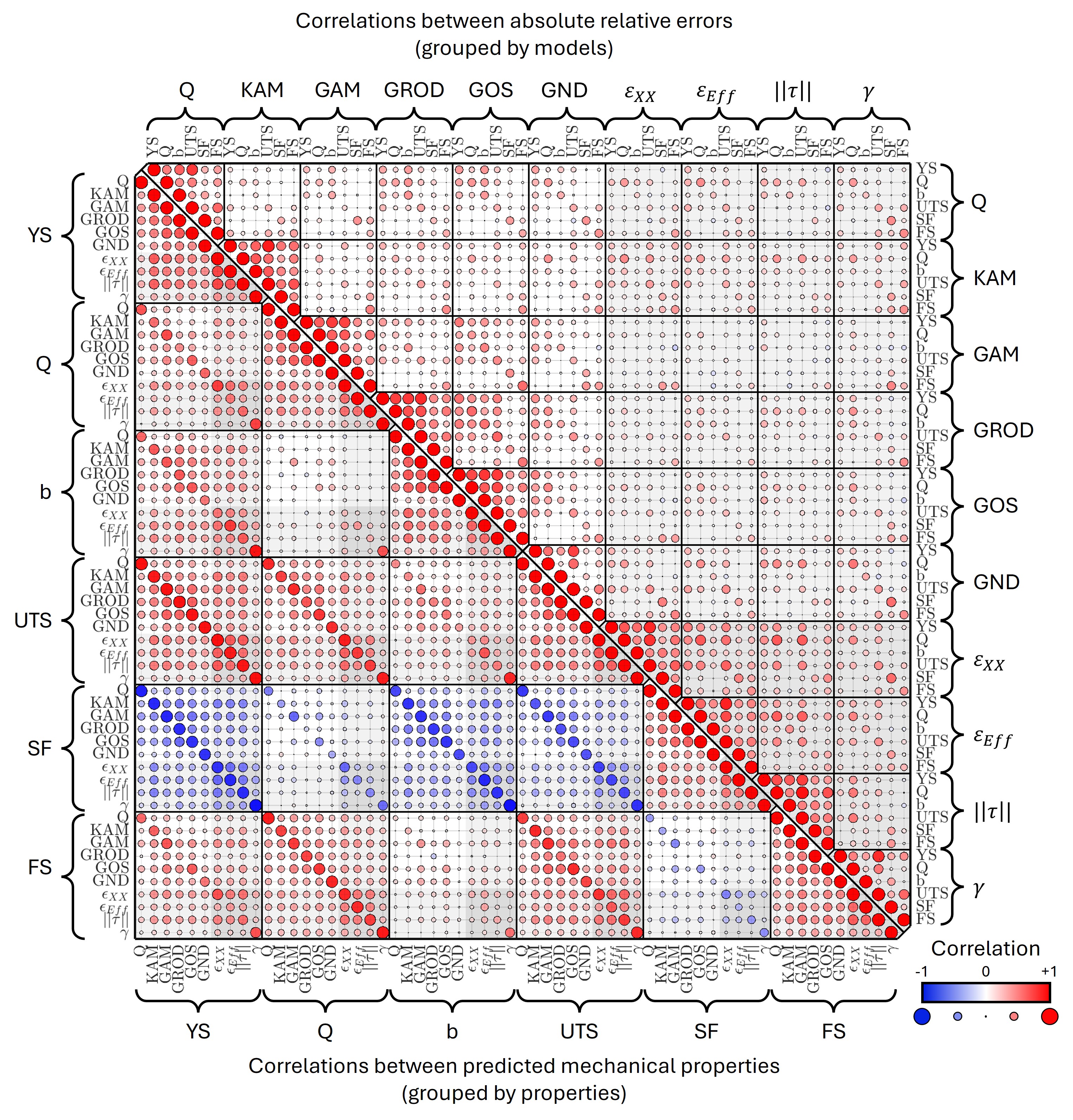}
    \caption{Correlation matrix between predicted mechanical properties from the different EBSD-based and HR-DIC-based models using only $57\times 57\:\mathrm{\mu m}$ tiles.}
    \label{fig:corr}
\end{suppfigure}

\section{Structure of the latent space}

\justify Compared to our previous study\cite{Calvat2026}, the structure of the latent space in this study is not directly constrained by a loss nor requires the input to be reconstructed from it. Instead, the latent space is being iteratively optimized to provide the best predicting ability over 10 materials and 6 mechanical properties. Low-dimensional representations were extracted as 128-dimensional vectors from the HR-DIC and EBSD encoding heads, as well as from the predicting heads as 1024-dimensional vectors. They were obtained from the processing of $57\times 57\:\mathrm{\mu m}$ tiles and transformed into 2D representations using uniform manifold approximation and projection (UMAP) \cite{Mcinnes2018,Becht2019}. The same parameters were employed for all the projections: minimum distance of 0.02, 50 neighbors and spread of 10. The resulting projections are shown in Supplementary Fig. \ref{fig:latent}. For all results, a projection template was first constructed from latent space representations of training tiles, and the resulting projections are depicted with transparent markers. Validation tiles were projected directly using the trained template and are shown as markers with solid black borders. Each material is represented by a single color to highlight the structure of the latent space and the heterogeneities captured. 

\justify Supplementary Fig. \ref{fig:latent}(A.1 -- A.4) and Supplementary Fig. \ref{fig:latent}(B.1 -- B.4) provide projections of various low-dimensional EBSD representations obtained from encoding and predicting heads, respectively. While the topology obtained from the outputs of the encoding heads appears rather continuous, deeper activations in the network enable a higher degree of abstraction (identical UMAP parameters), resulting in a better material separation, here optimized for predicting mechanical properties. The low-dimensional representation of quaternions captures the grain structure but also local dispersion of crystallographic orientations (see Supplementary Fig. \ref{fig:latent}(A.1, B.1)). Both quaternions and GND provide local measurements, while both GAM and GOS depend on the grain structure. In principle, quaternions contain all the information necessary to construct local maps (such as KAM and GND) as well as non-local maps, depending on the grain structure (such as GAM, GROD, and GOS). 

\justify Regarding materials, W\_Invar and W\_Steel330 feature dispersed low-dimensional representations due to their partially recrystallized microstructures, composed by large grains with significant orientation spread and smaller grains with uniform crystallographic orientation. This notably results in projections distant from the main group for W\_Invar (identified by the red arrows in Supplementary Fig. \ref{fig:latent}(A.1 -- A.3)). For quaternions, GAM, and GOS, W\_Nickel600, W\_Cu77NiSn, and W\_RX\_In718 are connected due to their equiaxed microstructures (see Supplementary Fig. \ref{fig:latent}(A.1 -- A.3)). However, despite its equiaxed grain structure, W\_In718 appears disjoint from the other materials due to its significantly finer grain size, regardless of the considered EBSD modality. While W\_RX\_In718 is characterized by an equiaxed grain structure, it exhibits crystallographic orientation variation (see Supplementary Fig. \ref{fig:FTEMPLATE_W_RX_In718}) due to the presence of microvolumes and intense plasticity localization. Although both GAM and GOS capture local variations within grains, GAM provides an average of local variations, whereas GOS describes the largest misorientation per grain. According to GOS low-dimensional representations, W\_RX\_In718 is connected to W\_Invar and W\_Steel330 due to their similar crystallographic orientation spreads within grains (see Supplementary Fig. \ref{fig:latent}(A.3)), despite their different natures: non-recrystallized grains for W\_Invar and W\_Steel330, and intense plastic localization for W\_RX\_In718. Regarding GAM, W\_RX\_In718 is associated with significantly variable latent representation in relation to the variability of misorientation from one grain to another. As expected, AM\_AB\_In718, AM\_AB\_Steel316, and AM\_AB\_Co76A, all of which produced using the same additive manufacturing technique (DED), appear closely related for the different EBSD modalities. Interestingly, although AM\_HIP\_GRX810 was produced using additive manufacturing (L-PBF), it is shown as either distant from the other AM materials (see Supplementary Fig. \ref{fig:latent}(A.1,A.2)) or as an intermediate between wrought and AM materials (see Supplementary Fig. \ref{fig:latent}(A.3)).

\justify Additionally, Supplementary Fig. \ref{fig:latent}(C.1 -- C.4) and Supplementary Fig. \ref{fig:latent}(D.1 -- D.4) provide projections of various low-dimensional HR-DIC representations obtained from encoding and predicting heads, respectively. The projections associated with the encoding heads demonstrate a continuous manifold, except for plastic deformation intensity, where W\_Invar appears distant from the other materials (see Supplementary Fig. \ref{fig:latent}(C.3)). Similarly to EBSD-derived latent spaces, HR-DIC-derived latent spaces mostly group local states representation by materials when produced by the predicting heads. Conversely, lattice rotation is associated with a rather continuous manifold of latent representations, from which W\_In718 appears distant from the other materials (see Supplementary Fig. \ref{fig:latent}(D.4)).

\begin{suppfigure}
    \centering
    \includegraphics[width=1\linewidth]{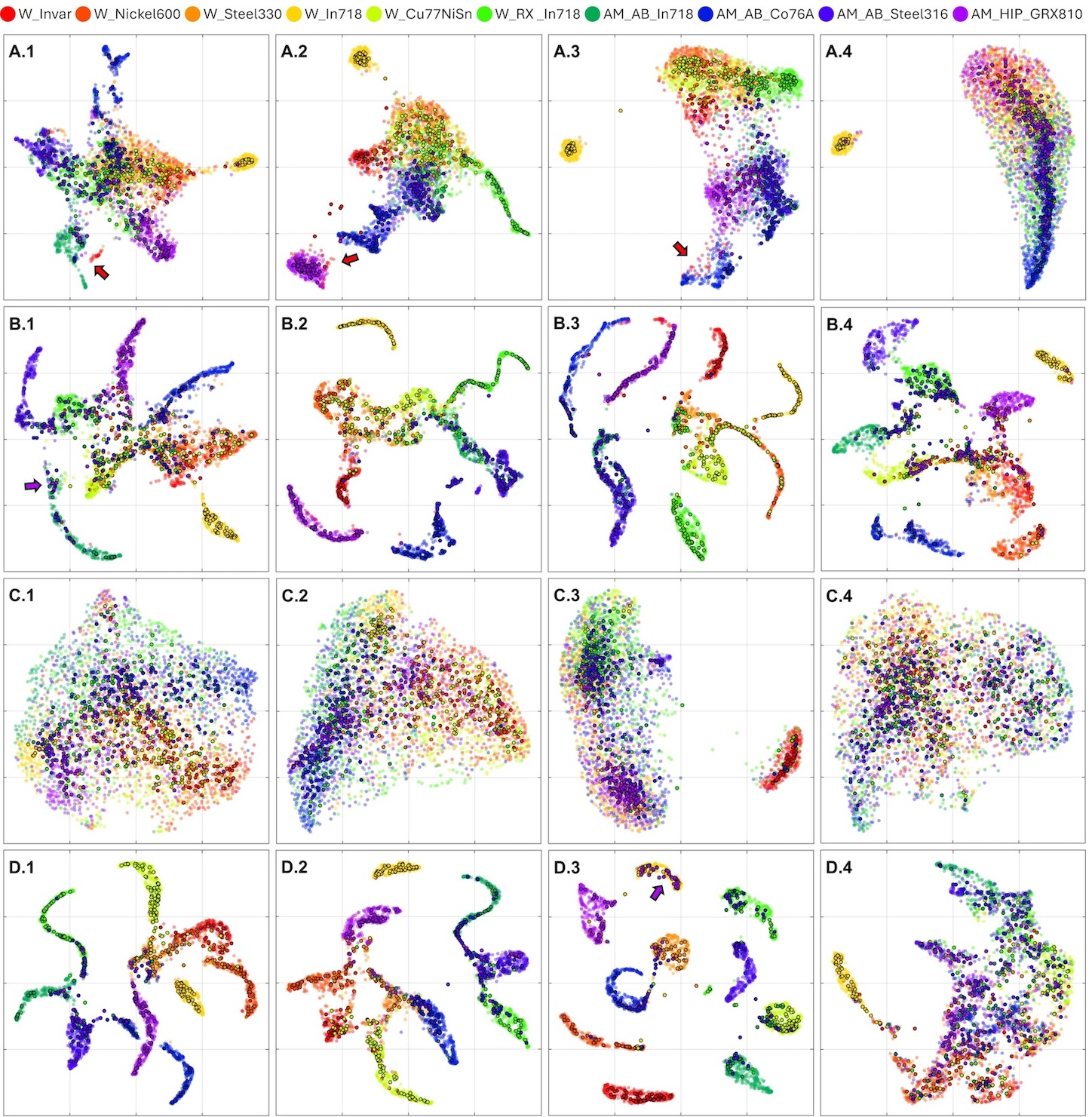}
    \caption{Uniform manifold approximation and projection associated with low-dimensional representations from EBSD maps: \textbf{(A)} from encoding heads and \textbf{(B)} from predicting heads (penultimate layer) and, from HR-DIC maps: \textbf{(C)} from encoding heads and \textbf{(D)} from predicting heads (penultimate layer). Structure of the latent space associated with \textbf{(A.1 -- B.1)} quaternions, \textbf{(A.2 -- B.2)} Grain Average Misorientation (GAM), \textbf{(A.3 -- B.3)} Grain Orientation Spread (GOS) and \textbf{(A.4 -- B.4)} Geometrically Necessary Dislocation (GND), \textbf{(C.1 -- D.1)} longitudinal strain $\varepsilon_{XX}$, \textbf{(C.2 -- D.2)} effective strain $\varepsilon_{eff}$, \textbf{(C.3 -- D.3)} plastic deformation intensity $||\tau||$ and \textbf{(C.4 -- D.4)} lattice rotation $\gamma$.}
    \label{fig:latent}
\end{suppfigure}

\justify Figure \ref{fig:latent_withpreds} illustrates the predicted yield and fatigue strengths overlaid on the various manifolds learned by the prediction heads. Since the latent spaces were optimized for property prediction, the structure of the latent space produced by the various models depends on microstructural variability and similarity, as well as the mechanical properties to be predicted. The relative position of a material or local state within the latent space directly controls the prediction accuracy, especially when materials with close latent space representations have significantly different mechanical properties. 

\begin{suppfigure}
    \centering
    \includegraphics[width=1\linewidth]{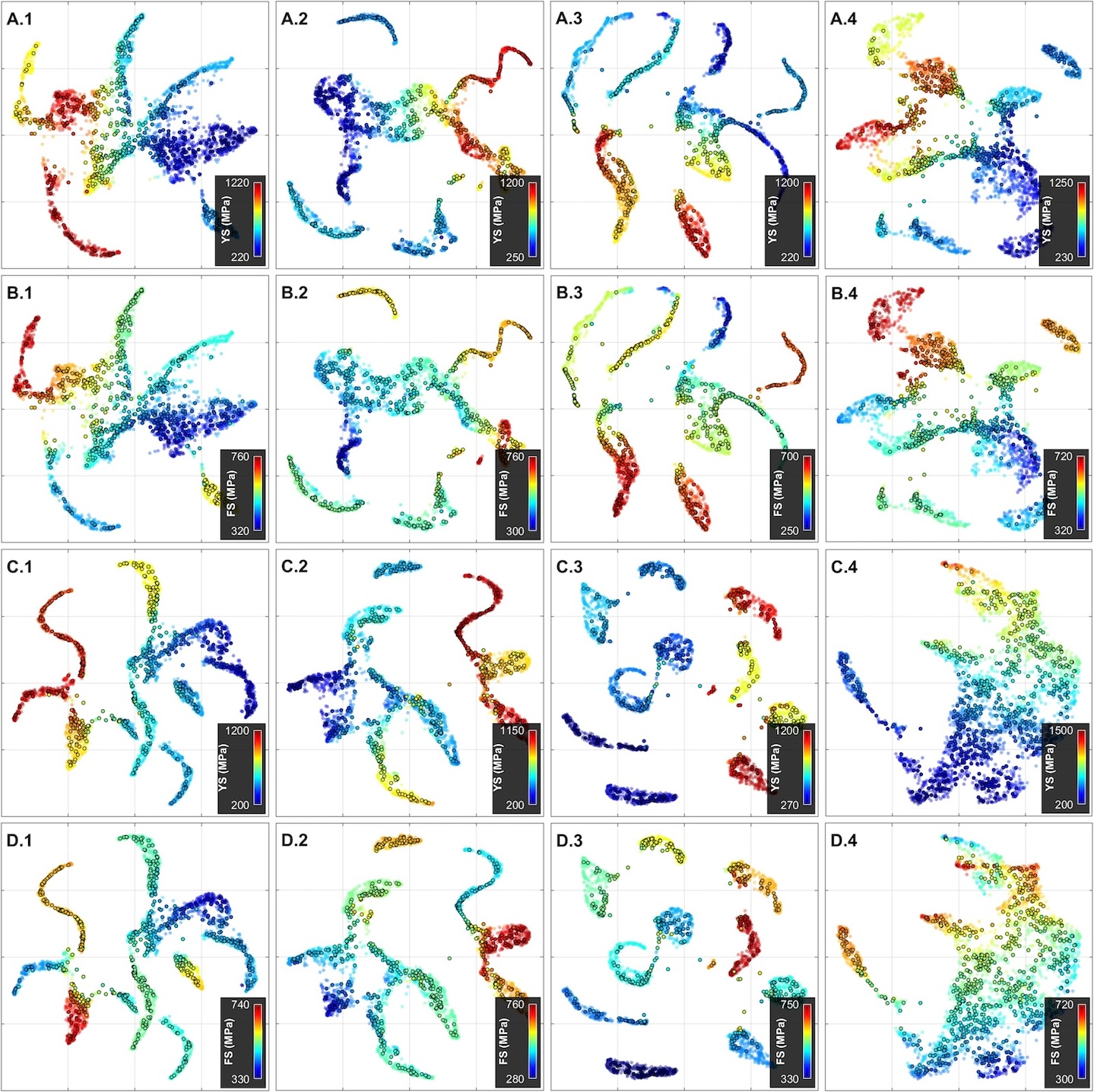}
    \caption{Uniform manifold approximation and projection associated with low-dimensional representations produced by \textbf{(A -- B)} EBSD predicting heads (penultimate layer) and \textbf{(C -- D)} HR-DIC predicting heads (penultimate layer). Illustration of the predicted mechanical properties: \textbf{(A -- C)} Yield Strength (YS) and \textbf{(B -- D)} Fatigue Strength (FS), predicted from \textbf{(A.1 -- B.1)} quaternions, \textbf{(A.2 -- B.2)} Grain Average Misorientation (GAM), \textbf{(A.3 -- B.3)} Grain Orientation Spread (GOS) and \textbf{(A.4 -- B.4)} Geometrically Necessary Dislocation (GND), \textbf{(C.1 -- D.1)} longitudinal strain $\varepsilon_{XX}$, \textbf{(C.2 -- D.2)} effective strain $\varepsilon_{eff}$, \textbf{(C.3 -- D.3)} plastic deformation intensity $||\tau||$ and \textbf{(C.4 -- D.4)} lattice rotation $\gamma$.}
    \label{fig:latent_withpreds}
\end{suppfigure}

\section{Local mechanical properties mapping}

\justify Following training of the various architectures, these various machine learning models were subsequently employed for predicting properties and their associated trade-offs over large fields of view. Maps were constructed from these predictions using a sliding window approach with a $5\:\mathrm{\mu m}$ step size in both horizontal and vertical directions. To conserve the same resolution as the original microstructure or plasticity maps, the predicted values were averaged over the windows. Several mechanical properties can be combined to illustrate trade-offs through a trade-off breaking index, as described in the Main Manuscript.

\section{Tile size influence on prediction performance} \label{sec:size}

\justify For all modalities, various models were trained to predict mechanical properties from different tile sizes: $57\times 57\:\mathrm{\mu m}$, $95\times 95\:\mathrm{\mu m}$, $133\times 133\:\mathrm{\mu m}$ and $171\times 171\:\mathrm{\mu m}$. The models have been trained for 60 epochs, corresponding to 5,000, 1,650, 700 and 390 iterations for $57\times 57\:\mathrm{\mu m}$, $95\times 95\:\mathrm{\mu m}$, $133\times 133\:\mathrm{\mu m}$ and $171\times 171\:\mathrm{\mu m}$ tile sizes, respectively. The loss metrics associated with the training from different tile sizes are provided in Supplementary Fig. \ref{fig:loss_withsize}. Only a few metrics were plotted for the sake of readability; quaternions and KAM as EBSD modalities, longitudinal strain and lattice rotation as HR-DIC modalities. The different metrics have been smoothed over a 100-iteration window for visualization purposes. Concerning first the training losses, quaternions and KAM are associated with similar training losses for the different tiles sizes considered as shown in Supplementary Fig. \ref{fig:loss_withsize}(A.1). Interestingly, they are associated with comparable final losses despite the lower number of iterations with increasingly larger tiles (and smaller datasets). Regarding the training losses associated with HR-DIC models depicted in Supplementary Fig. \ref{fig:loss_withsize}(B.1), the plastic deformation intensity model shows a faster decrease in training loss than the longitudinal strain model, similarly to what was previously observed (see Supplementary Fig. \ref{fig:loss}). Compared to EBSD modalities, HR-DIC modalities show a greater increase in final loss when larger tile sizes are considered. While the validation losses shown in Supplementary Fig. \ref{fig:loss_withsize}(A.2 -- B.2) demonstrate a limited decrease throughout training, the average absolute relative error shows a pronounced decrease, indicating convergence of the CNN models' prediction ability. When comparing EBSD and HR-DIC modalities, reducing the tile size for EBSD results in a faster decrease in validation loss and average absolute relative error than HR-DIC models, for which the trend appears globally similar regardless of tile size. 

\begin{suppfigure}
    \centering
    \includegraphics[width=1\linewidth]{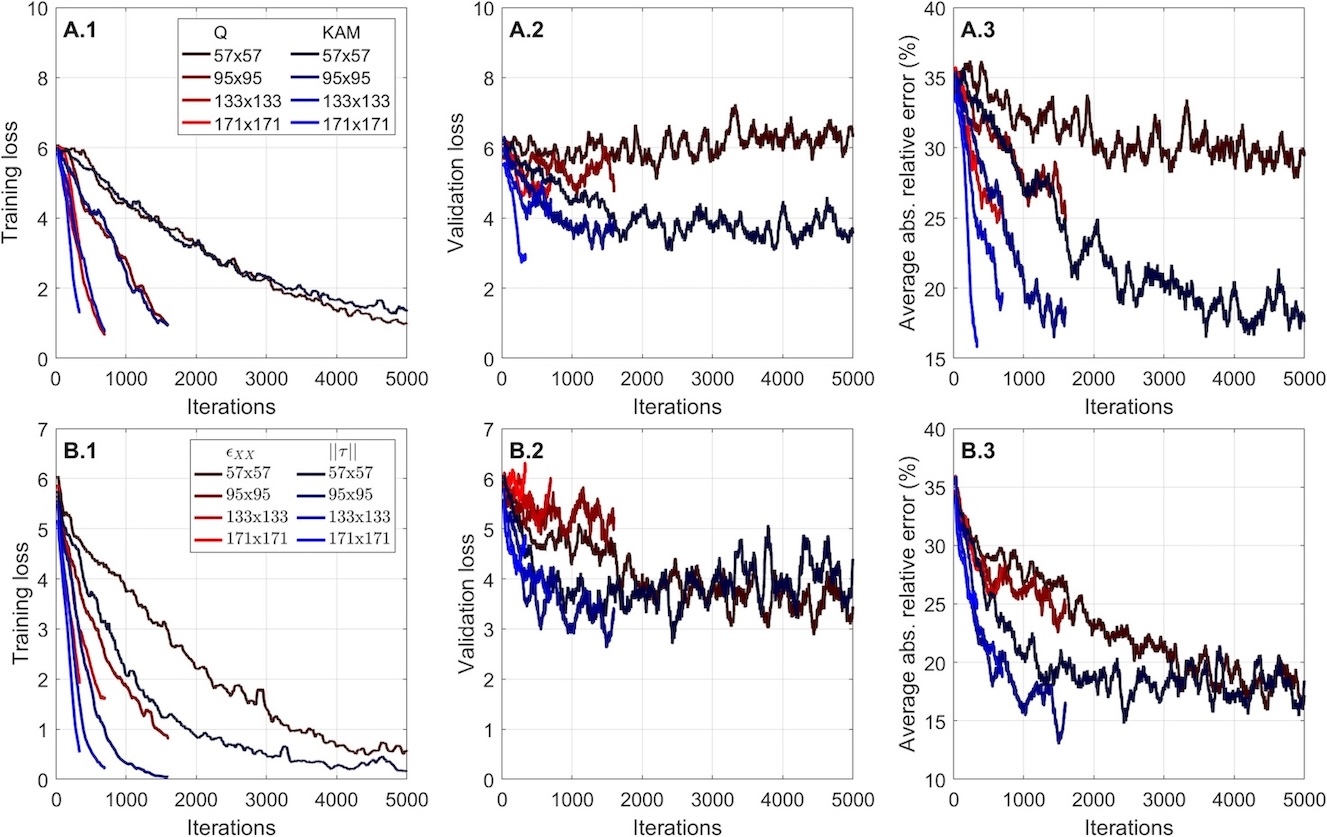}
    \caption{\textbf{Training and validation metrics with various tile sizes.} \textbf{(A)} Training losses, \textbf{(B)} validation losses and \textbf{(C)} average absolute relative errors associated with \textbf{(.1)} quaternion and Kernel Average Misorientation (KAM) and \textbf{(.2)} longitudinal strain $\varepsilon_{XX}$ and lattice rotation $\gamma$.}
    \label{fig:loss_withsize}
\end{suppfigure}

\justify The accuracies of these models are provided in Supplementary Fig. \ref{fig:preds_withsize} for the Yield Strength (YS) and Fatigue Strength (FS) predicted from KAM, GND, longitudinal strain and lattice rotation maps. These predictions were solely computed from validation regions. Each material is depicted by a different color and the distribution of predicted properties are summarized by four boxes, arranged from the smallest to the largest tile size (from left to right). Each box extends from the first quartile to the third quartile of the distribution, and the whiskers correspond to the 5\textsuperscript{th} and 95\textsuperscript{th} percentiles. For each of the ten materials, the solid black line denotes the experimental mechanical properties. The dashed and dotted lines delimit 10\% and 20\% error margins, respectively. 

\justify Regardless of the modality or tile size, yield strength is associated with greater variability than fatigue strength in relation to the range of experimental properties. Most materials demonstrate substantial variability decrease in their predictions with increasing tile sizes, encompassing both wrought and additively manufactured materials, since larger tiles can contain more microstructural or plasticity localization features. However, convergence to a given value does not necessarily mean convergence to the experimental value. Instead, it reveals a trade-off between map representativity (HR-DIC or EBSD) and the quantity of data associated with different tile sizes. Interestingly, while some materials demonstrate significant changes in the variability of their predictions, W\_In718 does not show significant changes due to its fine grain size. A large number of grains are already contained in the smaller tile size ($57\times 57\:\mathrm{\mu m}$), which is consistent with the results shown primarily in Supplementary Fig. \ref{fig:accuracy_preds}.

\begin{suppfigure}
    \centering
    \includegraphics[width=1\linewidth]{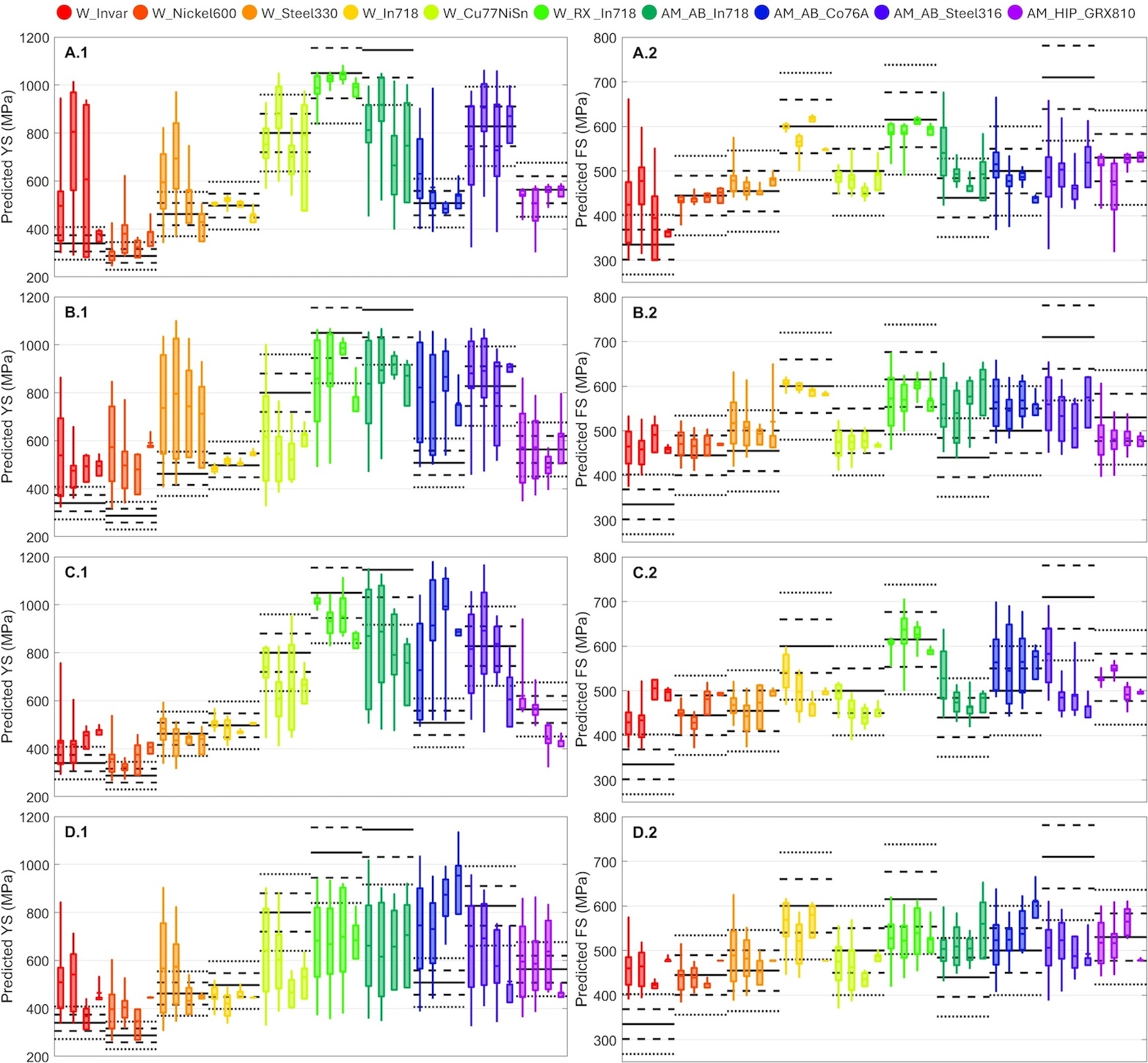}
    \caption{\textbf{Prediction accuracy from different tile sizes}. Predictions from \textbf{(A)} Kernel Average Misorientation (KAM), \textbf{(B)} Geometrically Necessary Dislocations (GND), \textbf{(B)} longitudinal strain $\varepsilon_{XX}$ and \textbf{(D)} lattice rotation $\gamma$, \textbf{(.1)} Yield Strength (YS), and \textbf{(.2)} Fatigue Strength (FS) predicted from validation regions and using increasingly large tile sizes.}
    \label{fig:preds_withsize}
\end{suppfigure}

\justify Similarly to Supplementary Fig. \ref{fig:accuracy_preds}, Supplementary Fig. \ref{fig:accuracy_withsize} provides a comprehensive comparison of the accuracy and repeatability of the mechanical properties predicted from the different modalities and using different tile sizes. Each row corresponds to predictions for one material and tile size. The predictions made by the different models are grouped and separated by solid vertical black lines. For EBSD-based models, increasing the tile size improves repeatability for most materials, including partially recrystallized materials (W\_Invar and W\_Steel330), equiaxed-grained materials (W\_Nickel600, W\_Cu77NiSn, and W\_RX\_In718), and additively manufactured materials (AM\_AB\_In718, AM\_AB\_316L, and AM\_AB\_Co76A). While partially recrystallized and equiaxed-grained materials demonstrate low variability in their predictions using the larger tile size, additively manufactured materials show improvement but still have higher variability due to their significantly larger grain size. In contrast, W\_In718 and AM\_HIP\_GRX810 are already associated with low variability using the smaller tile size due to their small grain size. Overall, increasing the tile size improves repeatability by enabling windows to contain more microstructural features, thus providing more robust predictions. However, increased repeatability does not necessarily mean higher-accuracy predictions (see Supplementary Fig. \ref{fig:preds_withsize}). Increasing the tile size may in general affect the accuracy of some models' predictions (see Supplementary Fig. \ref{fig:loss_withsize}) which can lead to some materials being affected significantly more than others (see Supplementary Fig. \ref{fig:accuracy_withsize}).

\begin{suppfigure}
    \centering
    \includegraphics[width=1\linewidth]{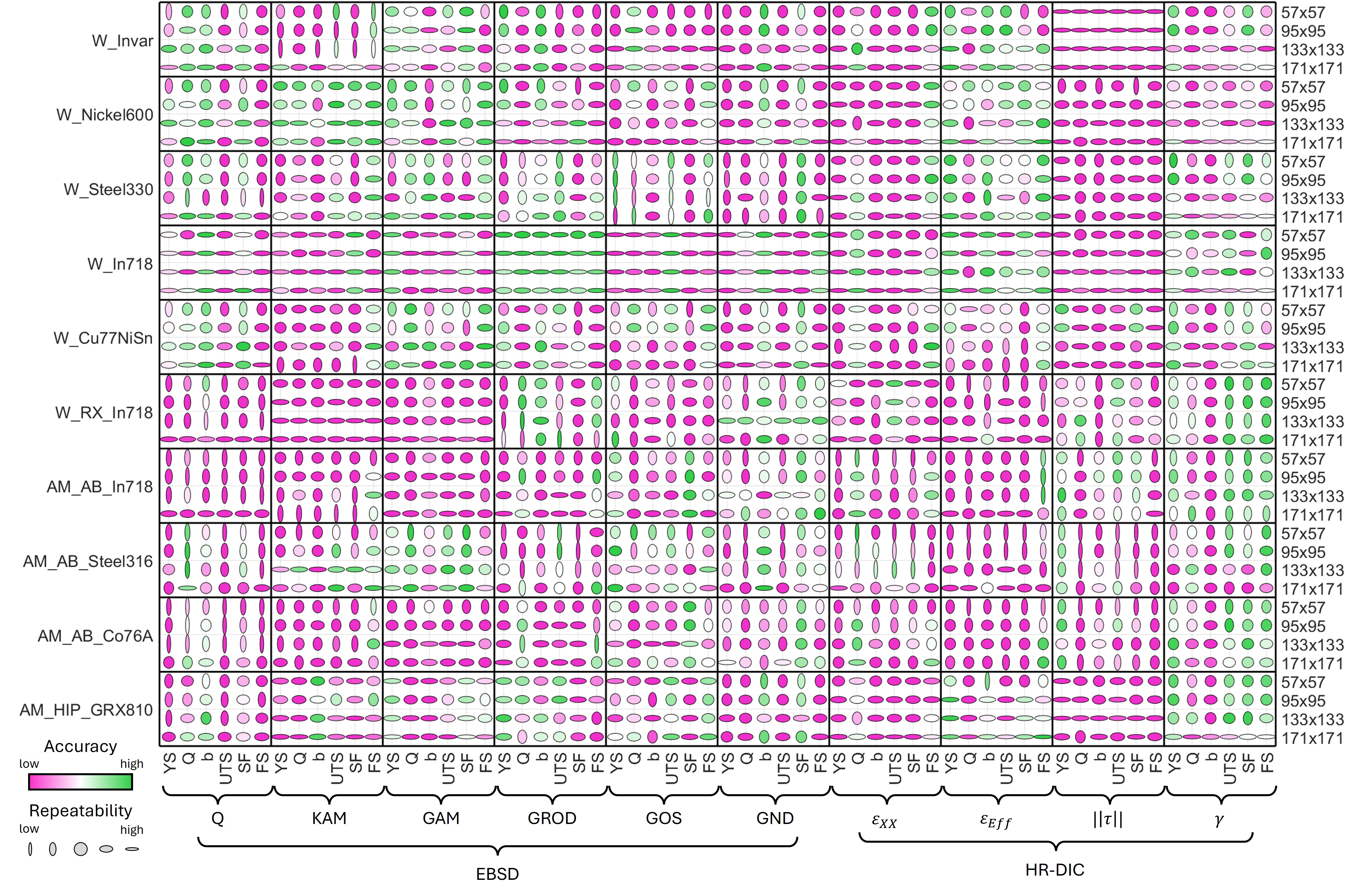}
    \caption{Accuracy and repeatability of predicted mechanical properties from the four different tile sizes.}
    \label{fig:accuracy_withsize}
\end{suppfigure}

\justify Additionally, predicted mechanical property maps for different tile sizes are shown in Supplementary Fig. \ref{fig:maps_withsize}. Since all the maps were generated using the same stride of $5\:\mathrm{\mu m}$, the value at a given pixel is obtained from more predictions than with the smaller tile size. As the tile size increases, the produced maps become smoother, and variability decreases since individual windows capture more microstructure or plasticity features. Using the larger tile size ($171\times 171\:\mathrm{\mu m}$) makes identifying governing features significantly more complex compared to maps produced with the smaller size ($57\times 57\:\mathrm{\mu m}$).

\begin{suppfigure}
    \centering
    \includegraphics[width=1\linewidth]{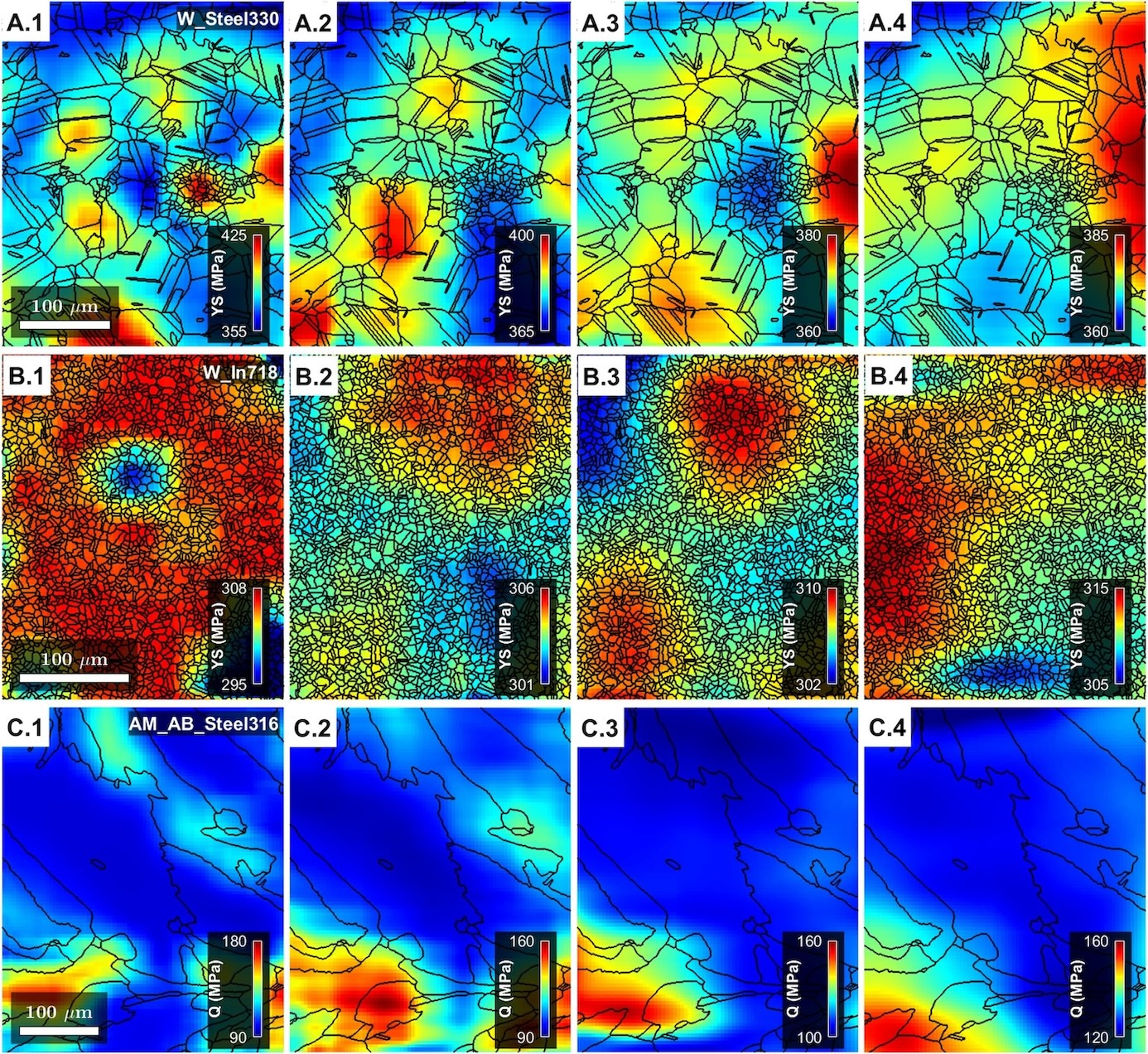}
    \caption{\textbf{(A)} W\_Steel330 Fatigue Strength (YS) predicted, \textbf{(B)} W\_In718 Yield Strength (YS), \textbf{(C)} AM\_AB\_Steel316 Q hardening parameter, all predicted from KAM using \textbf{(.1)} $57\times 57\:\mathrm{\mu m}$, \textbf{(.2)} $95\times 95\:\mathrm{\mu m}$, \textbf{(.3)} $133\times 133\:\mathrm{\mu m}$ and \textbf{(.4)} $171\times 171\:\mathrm{\mu m}$}
    \label{fig:maps_withsize}
\end{suppfigure}

 \section{A Path to Rapid Qualification of Materials and Processes} \label{sec:qualif}

\justify Qualification pathways for safety-critical systems are time-intensive, hierarchical, and application-specific. For instance, human-rated aerospace hardware must satisfy foundational material and process requirements (\textit{e.g.}, NASA-STD-6016), additive-manufacturing-specific standards (\textit{e.g.}, NASA-STD-6030), and process-specific qualification requirements (\textit{e.g.}, MSFC-STD-3716/3717 for L-PBF metals). Qualification typically relies on extensive destructive testing -- including tensile, fracture toughness, and fatigue testing (\textit{e.g.}, ASTM E8/E21, E399/E1820, and E466/E606) -- performed on witness coupons, representative geometries, qualification units and flight hardware, in order to establish material allowables, quantify manufacturing variability and verify performance under mission-relevant conditions. For additively manufactured metallic materials, a single build-test iteration can take months, and full qualification campaigns often span years.

\justify MSI-based prediction of material properties can directly accelerate this process by linking high-throughput, multiscale characterization under simple loading conditions to long-term engineering performance metrics such as fatigue life and creep resistance. By correlating nanometer-scale microstructural and deformation signatures with macroscopic properties measured after millions to billions of fatigue cycles or hundreds to thousands of hours of creep exposure, such approaches can reduce uncertainty during the early stages of material and process development. This could substantially lower testing costs and accelerate the deployment of advanced materials and manufacturing technologies by reducing the number of pre-qualification iterations.

\section{Perspectives and Next Steps} \label{sec:perspectives}

\justify Although the present MSI framework demonstrates promising predictive and interpretability capabilities, it should primarily be viewed as an initial step toward establishing materials spatial intelligence as a long-term research paradigm for microstructure-driven materials science. Future developments involving multimodal data integration, physics-informed learning, temporal representations, and larger experimental datasets will be required to fully capture the complexity of microstructure–deformation–property relationships across materials systems and loading conditions.

\justify An important limitation of the present MSI framework concerns its dependency on the experimental dataset used for training. The current dataset only includes a limited number of microstructural and deformation states, which restricts the broader applicability of the framework and the accuracy of property predictions across unexplored material systems and loading conditions. Consequently, the present work should primarily be viewed as a proof-of-concept demonstrating the potential of materials spatial intelligence rather than a fully generalized predictive framework. Application to entirely new material systems or loading conditions may therefore require substantial additional experimental data. Nevertheless, acquiring such large-scale, multimodal datasets is becoming more achievable due to recent advances in automated microscopy, high-throughput characterization, experimental robotics, and accelerated alloy processing and testing.

\justify In addition, the learned latent representations and associated property relationships are influenced by the diversity of alloys, processing routes, loading conditions, spatial resolutions, and experimental modalities included in the dataset. Consequently, the latent spaces may partially reflect dataset imbalance, alloy over-representation, microscopy artifacts, or acquisition-specific features. Although the framework demonstrates promising generalization capability across the investigated alloys, its transferability to entirely unseen material systems, characterization conditions, or deformation regimes remains limited by the representativeness of the available experimental data. For example, changes in loading conditions such as temperature may activate entirely different deformation mechanisms, including grain boundary sliding, phase transformations, or diffusion-assisted plasticity, requiring the framework to learn their influence on the resulting mechanical properties. Future developments will therefore require larger, more diverse, and standardized multimodal datasets to improve the robustness, transferability, and physical generalization capability of the MSI framework.

\justify In this study, the prediction models were primarily trained using a single microstructural or deformation modality at a time, enabling direct evaluation of the influence of individual experimentally measured features on the predicted mechanical properties and associated trade-offs. Although such an approach provides a first demonstration of the MSI framework and facilitates interpretability, it also introduces limitations regarding the representation of the full complexity of microstructure–deformation interactions governing macroscopic behavior. In practice, mechanical properties emerge from the coupled interaction of multiple microstructural and deformation features across length scales. Crystallographic orientation, grain morphology, lattice rotation, dislocation density, phase distribution, defects, and deformation localization are intrinsically correlated and evolve simultaneously during loading. Considering these modalities independently therefore limits the ability of the framework to capture higher-order correlations and coupled physical mechanisms. Future developments of MSI will consequently focus on multimodal artificial intelligence strategies capable of jointly encoding multiple experimental modalities into unified latent representations.

\justify In the present study, the MSI framework primarily considers microstructural and deformation states, while chemical composition is only implicitly incorporated through the investigated alloy systems and not directly encoded as a spatial modality. Future multimodal implementations will therefore aim to explicitly integrate spatially resolved chemical information together with microstructure and deformation measurements. 

\justify The present implementation also relies primarily on convolutional neural networks, which remain limited in their ability to capture long-range spatial relationships and interactions between distant microstructural regions. Future work will therefore investigate transformer-based and attention-based architectures capable of learning non-local interactions and collective spatial behavior within heterogeneous microstructures. Such approaches could enable improved representation of microstructural and deformation connectivity, long-range strain localization, and collective deformation mechanisms.

\justify Another important limitation concerns the static representation of deformation states. In the current framework, deformation is characterized using elementary loading conditions and discrete deformation states. Future MSI implementations will aim to incorporate temporal and sequential information through video-based or recurrent representations capable of learning the evolution of deformation processes during loading. Such an extension would enable direct representation of the dynamics of deformation and microstructure evolution.

\justify Finally, while the synthetic microstructure optimization framework introduced in this work demonstrates the possibility of virtually modifying governing features to improve predicted properties, these optimized states are not necessarily experimentally accessible through existing processing routes. Future developments will therefore focus on integrating MSI with processing-aware generative frameworks and physics-informed constraints to establish direct relationships between processing conditions, resulting spatial microstructure organization, and mechanical performance. Such integration could ultimately enable closed-loop microstructure design and autonomous materials optimization guided by experimentally grounded spatial intelligence.

\end{document}